\pdfoutput=1
\documentclass{CSML}

\def\dOi{13(1:7)2017}
\lmcsheading%
{\dOi}
{1--23}
{}
{}
{Aug.~15, 2014}
{Mar.~17, 2017}
{}

\usepackage{hyperref} 
\usepackage{asymptote}
\usepackage{graphicx}
\usepackage{amsmath}

\makeatletter
\@ifundefined{lhs2tex.lhs2tex.sty.read}%
  {\@namedef{lhs2tex.lhs2tex.sty.read}{}%
   \newcommand\SkipToFmtEnd{}%
   \newcommand\EndFmtInput{}%
   \long\def\SkipToFmtEnd#1\EndFmtInput{}%
  }\SkipToFmtEnd

\newcommand\ReadOnlyOnce[1]{\@ifundefined{#1}{\@namedef{#1}{}}\SkipToFmtEnd}
\usepackage{amstext}
\usepackage{amssymb}
\usepackage{stmaryrd}
\DeclareFontFamily{OT1}{cmtex}{}
\DeclareFontShape{OT1}{cmtex}{m}{n}
  {<5><6><7><8>cmtex8
   <9>cmtex9
   <10><10.95><12><14.4><17.28><20.74><24.88>cmtex10}{}
\DeclareFontShape{OT1}{cmtex}{m}{it}
  {<-> ssub * cmtt/m/it}{}

\DeclareFontShape{OT1}{cmtt}{bx}{n}
  {<5><6><7><8>cmtt8
   <9>cmbtt9
   <10><10.95><12><14.4><17.28><20.74><24.88>cmbtt10}{}
\DeclareFontShape{OT1}{cmtex}{bx}{n}
  {<-> ssub * cmtt/bx/n}{}

\newcommand{\Conid}[1]{\mathit{#1}}
\newcommand{\Varid}[1]{\mathit{#1}}
\newcommand{\anonymous}{\kern0.06em \vbox{\hrule\@width.5em}}

\newcommand{\bind}{\mathbin{>\!\!\!>\mkern-6.7mu=}}

\renewcommand{\leq}{\leqslant}
\renewcommand{\geq}{\geqslant}
\usepackage{polytable}

\@ifundefined{mathindent}%
  {\newdimen\mathindent\mathindent\leftmargini}%
  {}%

\def\resethooks{%
  \global\let\SaveRestoreHook\empty
  \global\let\ColumnHook\empty}
\newcommand*{\savecolumns}[1][default]%
  {\g@addto@macro\SaveRestoreHook{\savecolumns[#1]}}
\newcommand*{\restorecolumns}[1][default]%
  {\g@addto@macro\SaveRestoreHook{\restorecolumns[#1]}}
\newcommand*{\aligncolumn}[2]%
  {\g@addto@macro\ColumnHook{\column{#1}{#2}}}

\resethooks

\newcommand{\onelinecommentchars}{\quad-{}- }
\newcommand{\commentbeginchars}{\enskip\{-}
\newcommand{\commentendchars}{-\}\enskip}

\newcommand{\visiblecomments}{%
  \let\onelinecomment=\onelinecommentchars
  \let\commentbegin=\commentbeginchars
  \let\commentend=\commentendchars}

\newcommand{\invisiblecomments}{%
  \let\onelinecomment=\empty
  \let\commentbegin=\empty
  \let\commentend=\empty}

\visiblecomments

\newlength{\blanklineskip}
\newcommand{\hsindent}[1]{\quad}
\let\hspre\empty
\let\hspost\empty

\EndFmtInput
\makeatother
\ReadOnlyOnce{polycode.fmt}%
\makeatletter

\newcommand{\hsnewpar}[1]%
  {{\parskip=0pt\parindent=0pt\par\vskip #1\noindent}}

\newcommand{\hscodestyle}{}

\newcommand{\sethscode}[1]%
  {\expandafter\let\expandafter\hscode\csname #1\endcsname
   \expandafter\let\expandafter\endhscode\csname end#1\endcsname}

  {\par\noindent
   \advance\leftskip\mathindent
   \hscodestyle
   \let\\=\@normalcr
   \let\hspre\(\let\hspost\)%
   \pboxed}%
  {\endpboxed\)%
   \par\noindent
   \ignorespacesafterend}

  {\hsnewpar\abovedisplayskip
   \advance\leftskip\mathindent
   \hscodestyle
   \let\hspre\(\let\hspost\)%
   \pboxed}%
  {\endpboxed%
   \hsnewpar\belowdisplayskip
   \ignorespacesafterend}

  {\hsnewpar\abovedisplayskip
   \advance\leftskip\mathindent
   \hscodestyle
   \let\\=\@normalcr
   \(\pboxed}%
  {\endpboxed\)%
   \hsnewpar\belowdisplayskip
   \ignorespacesafterend}

\newcommand{\plainhs}{\sethscode{plainhscode}}

\plainhs

  {\hsnewpar\abovedisplayskip
   \advance\leftskip\mathindent
   \hscodestyle
   \let\\=\@normalcr
   \(\parray}%
  {\endparray\)%
   \hsnewpar\belowdisplayskip
   \ignorespacesafterend}

  {\parray}{\endparray}

  {\(\parray}{\endparray\)}

\def\codeframewidth{\arrayrulewidth}
\RequirePackage{calc}

  {\parskip=\abovedisplayskip\par\noindent
   \hscodestyle
   \arrayrulewidth=\codeframewidth
   \tabular{@{}|p{\linewidth-2\arraycolsep-2\arrayrulewidth-2pt}|@{}}%
   \hline\framedhslinecorrect\\{-1.5ex}%
   \let\endoflinesave=\\
   \let\\=\@normalcr
   \(\pboxed}%
  {\endpboxed\)%
   \framedhslinecorrect\endoflinesave{.5ex}\hline
   \endtabular
   \parskip=\belowdisplayskip\par\noindent
   \ignorespacesafterend}

\newcommand{\framedhslinecorrect}[2]%
  {#1[#2]}

  {\(\def\column##1##2{}%
   \let\>\undefined\let\<\undefined\let\\\undefined
   \newcommand\>[1][]{}\newcommand\<[1][]{}\newcommand\\[1][]{}%
   \def\fromto##1##2##3{##3}%
   }{\) }%

  {\let\orighscode=\hscode
   \let\origendhscode=\endhscode
   \def\endhscode{\def\hscode{\endgroup\def\@currenvir{hscode}\\}\begingroup}
   \orighscode\def\hscode{\endgroup\def\@currenvir{hscode}}}%
  {\origendhscode
   \global\let\hscode=\orighscode
   \global\let\endhscode=\origendhscode}%

\makeatother
\EndFmtInput

\def\commentbegin{\quad\{\ }
\def\commentend{\}}

\title[Sequential decision problems, dependent types and generic solutions]{Sequential decision problems, dependent types and generic solutions}

\author[N.~Botta]{Nicola Botta\rsuper a}
\address{{\lsuper a}Potsdam Institute for Climate Impact Research (PIK), Germany}
\email{botta@pik-potsdam.de}

\author[P.~Jansson]{Patrik Jansson\rsuper b}
\address{{\lsuper{b,c}}Chalmers University of Technology \& University of
Gothenburg, Sweden}
\email{\{patrikj, cezar\}@chalmers.se}
\thanks{{\lsuper{b,c}}Jansson and Ionescu were supported by the projects GRACeFUL
(grant agreement No 640954) and CoeGSS (grant agreement No 676547), which
have received funding from the European Union’s Horizon 2020 research and
innovation programme.}

\author[C.~Ionescu]{Cezar Ionescu\rsuper c}
\address{\vspace{-18 pt}}

\author[D.~R.~Christiansen]{David R. Christiansen\rsuper d}
\address{{\lsuper d}Indiana University, USA}
\email{davidchr@indiana.edu}
\thanks{{\lsuper d}Christiansen was funded by the Danish Advanced
Technology Foundation (Højteknologifonden) grant 17-2010-3.}

\author[E.~Brady]{Edwin Brady\rsuper e}
\address{{\lsuper e}University of St Andrews, Scotland}
\email{ecb10@st-andrews.ac.uk}

\begin{document}

\begin{abstract}
We present a computer-checked generic implementation for solving
finite-horizon sequential decision problems. This is a wide class of
problems, including inter-temporal optimizations, knapsack, optimal
bracketing, scheduling, etc. The implementation can handle time-step
dependent control and state spaces, and monadic representations of
uncertainty (such as stochastic, non-deterministic, fuzzy, or
combinations thereof). This level of genericity is achievable in a
programming language with dependent types (we have used both Idris and
Agda). Dependent types are also the means that allow us to obtain a
formalization and computer-checked proof of the central component of our
implementation: Bellman's principle of optimality and the associated
backwards induction algorithm. The formalization clarifies certain
aspects of backwards induction and, by making explicit notions such as
viability and reachability, can serve as a starting point for a theory
of controllability of monadic dynamical systems, commonly encountered
in, e.g., climate impact research.
\end{abstract}

\maketitle

\vfill
\renewcommand{\hscodestyle}{%
   \setlength\leftskip{1em}%
}

\section{Introduction}
\label{section:introduction}

In this paper we extend a previous formalization of
\emph{time-independent}, \emph{deterministic} sequential decision
problems~\cite{botta+al2013c} to general sequential decision problems
(general SDPs).


Sequential decision problems are problems in which a decision maker is
required to make a step-by-step sequence of decisions. At each step, the
decision maker selects a \emph{control} upon observing some
\emph{state}.

Time-independent, deterministic SDPs are sequential decision problems in
which the state space (the set of states that can be observed by the
decision maker) and the control space (the set of controls that can be
selected in a given state) do not depend on the specific decision step
and the result of selecting a control in a given state is a unique new
state.

In contrast, general SDPs are sequential decision problems in which both
the state space and the control space can depend on the specific
decision step and the outcome of a step can be a set of new states
(non-deterministic SDPs), a probability distribution of new states
(stochastic SDPs) or, more generally, a monadic structure of states, see
section \ref{section:gsdp}.

Throughout the paper, we use the word ``time'' (and correspondent
phrasings: ``time-independent'', ``time-dependent'', etc.) to denote
a decision step number. In other words, we write ``at time 3 \ldots'' as
a shortcut for ``at the third decision step \ldots''.
The intuition is that decision step 3 takes place \emph{after} decision
steps 0, 1 and 2 and \emph{before} decision steps 4, 5, etc. In certain
decision problems, some physical time -- discrete or, perhaps,
continuous -- might be observable and relevant for decision making. In
these cases, such time becomes a proper component of the state space and
the function that computes a new state from a current state and a
control has to fulfill certain monotonicity conditions.

Sequential decision problems for a finite number of steps, often
called finite horizon SDPs, are in principle well understood. In
standard
textbooks~\cite{cormen+al2001,bertsekas1995,sutton+barto1998}, SDPs
are typically introduced by examples: a few specific problems are
analyzed and dissected and ad-hoc implementations of Bellman's
backwards induction algorithm~\cite{bellman1957} are derived for such
problems.

To the best of our knowledge, no generic algorithm for solving general
sequential decision problems is currently available. This has a number
of disadvantages:

An obvious one is that, in front of a particular instance of an SDP,
be that a variant of knapsack, optimal bracketing, inter-temporal
optimization of social welfare functions or more specific
applications, scientists have to find solution algorithms developed for
similar problems --- backwards induction or non-linear optimization,
for example --- and adapt or re-implement them for their particular
problem.

This is not only time-consuming but also error-prone. For most
practitioners, showing that their ad-hoc implementation actually
delivers optimal solutions is often an insurmountable task.

In this work, we address this problem by using dependent types ---
types that are allowed to ``depend'' on values~\cite{idristutorial}
--- in order to formalise general SDPs, implement a generic version of
Bellman's backwards induction, and obtain a machine-checkable proof
that the implementation is correct.

The use of a dependently-typed language (we have used Idris and are in
the process of developing an equivalent implementation in Agda) is
essential to our approach.  It allows us not only to provide a generic
program, but also a generic proof.  If the users limit themselves to
using the framework by instantiating the problem-dependent elements,
they obtain a concrete program \emph{and} a concrete proof of
correctness.  An error in the instantiation will result in an error in
the proof, and will be signalled by the type checker.

Expert implementors might want to re-implement problem-specific
solution algorithms, e.g., in order to exploit some known properties
of the particular problem at hand. But they would at least be able to
test their solutions against provably correct ones.  Moreover, if they
use the framework, the proof obligations are going to be signalled by
the type checker.

The fact that the specifications, implementations, and proof of
correctness are all expressed in the same programming language is, in
our opinion, the most important advantage of dependently-typed
languages.  Any change in the implementation immediately leads to a
verification against the specification and the proof.  If something is
wrong, the implementation is not executed: the program does not
compile.  This is not the case with pencil-and-paper proofs of
correctness, or with programs extracted from specifications and proofs
using a system such as Coq.

Our approach is similar in spirit to that proposed by de Moor
\cite{de_moor1995} and developed in the \emph{Algebra of Programming}
book \cite{bird+demoor1997}.  There, the specification,
implementation, and proof are all expressed in the relational calculus
of allegories, but they are left at the paper and pencil stage.  For a
discussion of other differences and of the similarities we refer the
reader to our previous paper~\cite{botta+al2013c}, which the present
paper extends.

This extension is presented in two steps. First, we generalize
time-independent (remember that we use the word ``time'' as an alias to
``decision step''; thus, time-independent SDPs are sequential decision
problems in which the state space and the control space do not depend on
the decision step), deterministic decision problems to the case in which
the state and the control spaces can depend on time but the transition
function is still deterministic.
Then, we extend this case to the general one of monadic transition functions.
As it turns out, neither extension is trivial: the requirement of
machine-checkable correctness brings to the light notions and
assumptions which, in informal and semi-formal derivations are often
swept under the rug.

In particular, the extension to the time-dependent case has lead us to
formalize the notions of \emph{viability} and \emph{reachability} of
states. For the deterministic case these notions are more or less
straightforward. But they become more interesting when
non-deterministic and stochastic transition functions are considered
(as outlined in section \ref{section:monadic}).

We believe that these notions would be a good starting point for
building a theory of controllability for the kind of dynamical systems
commonly encountered in climate impact research. These were the systems
originally studied in Ionescu's dissertation~\cite{ionescu2009} and the
monadic case is an extension of the theory presented there for dynamical systems.

In the next section we introduce, informally, the notion of sequential
decision processes and problems. In section \ref{section:base_case} we
summarize the results for the time-independent, deterministic case and
use this as the starting point for the two extensions discussed in
sections \ref{section:time-dependent} and \ref{section:monadic},
respectively.
\section{Sequential decision processes and problems}
\label{section:gsdp}

In a nutshell, a sequential decision process is a process in which a
decision maker is required to take a finite number of decision steps,
one after the other. The process starts in a state \ensuremath{\Varid{x}_{0}} at an initial
step number \ensuremath{\Varid{t}_{0}}.

Here \ensuremath{\Varid{x}_{0}} represents all information available to the decision maker at
\ensuremath{\Varid{t}_{0}}. In a decision process like those underlying models of
international environmental agreements, for instance, \ensuremath{\Varid{x}_{0}} could be a
triple of real numbers representing some estimate of the greenhouse gas
(GHG) concentration in the atmosphere, a measure of a gross domestic
product and, perhaps, the number of years elapsed from some
pre-industrial reference state. In an optimal bracketing problem, \ensuremath{\Varid{x}_{0}}
could be a string of characters representing the ``sizes'' of a list of
``arguments'' which are to be processed pairwise with some associative
binary operation. In all cases, \ensuremath{\Varid{t}_{0}} is the initial value of the
decision step counter.

The control space -- the set of controls (actions, options, choices,
etc.)  available to the decision maker -- can depend both on the initial
step number and state. Upon selecting a control \ensuremath{\Varid{y}_{0}} two events take
place: the system enters a new state \ensuremath{\Varid{x}_{1}} and the decision maker
receives a reward \ensuremath{\Varid{r}_{0}}.

In a deterministic decision problem, a transition function completely
determines the next state \ensuremath{\Varid{x}_{1}} given the time (step number) of the
decision \ensuremath{\Varid{t}_{0}}, the current state \ensuremath{\Varid{x}_{0}}, and the selected control~\ensuremath{\Varid{y}_{0}}.
But, in general, transition functions can return sets of new states
(non-deterministic SDPs), probability distributions over new states
(stochastic SDPs) or, more generally, a monadic structures of states,
as presented by Ionescu~\cite{ionescu2009}.

In general, the reward depends both on the ``old'' state and on the
``new'' state, and on the selected control: in many decision problems,
different controls represent different levels of consumption of
resources (fuel, money, CPU time or memory) or different levels of
restrictions (GHG emission abatements) and are often associated with
costs. Different current and next states often imply different levels of
``running'' costs or benefits (of machinery, avoided climate damages,
\dots) or outcome payoffs.

The intuition of finite horizon SDPs is that the decision maker seeks
controls that maximize the sum of the rewards collected over a finite
number of decision steps.
This characterization of SDPs might appear too narrow (why shouldn't a
decision maker be interested, for instance, in maximizing a product of
rewards?) but it is in fact quite general. For an introduction to SDPs
and concrete examples of state spaces, control spaces, transition- and
reward-functions, see \cite{bertsekas1995,sutton+barto1998}.

In control theory, controls that maximize the sum of the rewards
collected over a finite number of steps are called \emph{optimal}
controls. In practice, optimal controls can only be computed when a
specific initial state is given and for problems in which transitions
are deterministic. What is relevant for decision making -- both in the
deterministic case and in the non-deterministic or stochastic case --
are not controls but \emph{policies}.

Informally, a policy is a function from states to controls: it tells
which control to select when in a given state. Thus, for selecting
controls over \ensuremath{\Varid{n}} steps, a decision maker needs a sequence of \ensuremath{\Varid{n}}
policies, one for each step. Optimal policy sequences are sequences of
policies which cannot be improved by associating different controls to
current and future states.
\section{Time-independent, deterministic problems}
\label{section:base_case}

In a previous paper~\cite{botta+al2013c}, we presented a formalization
of time-independent, deterministic SDPs. For this class of problems,
we introduced an abstract context and derived a generic,
machine-checkable implementation of backwards induction.

In this section we recall the context and the main results from that
paper~\cite{botta+al2013c}. There, we illustrated time-independent,
deterministic SDPs using a simplified version of the ``cylinder''
example originally proposed by Reingold, Nievergelt and
Deo~\cite{reingold+nievergelt+deo1977} and extensively studied by Bird
and de Moor~\cite{bird+demoor1997}.
We use the same example here:

A decision-maker can be in one of five states: \ensuremath{\Varid{a}}, \ensuremath{\Varid{b}}, \ensuremath{\Varid{c}}, \ensuremath{\Varid{d}} or
\ensuremath{\Varid{e}}. In \ensuremath{\Varid{a}}, the decision maker can choose between two controls
(sometimes called ``options'' or ``actions''): move ahead (control
\ensuremath{\Conid{A}}) or move to the right (control \ensuremath{\Conid{R}}). In \ensuremath{\Varid{b}}, \ensuremath{\Varid{c}} and \ensuremath{\Varid{d}} he can
move to the left (\ensuremath{\Conid{L}}), ahead or to the right. In \ensuremath{\Varid{e}} he can move to
the left or go ahead.

\begin{figure}
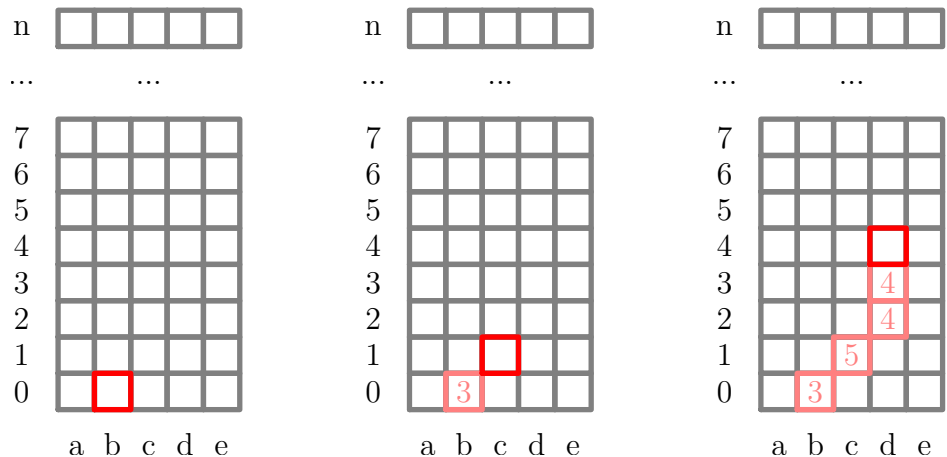

 \begin{asy}
  size(6cm);
  include graph;
  string[] xs = {"a","b","c","d","e"};
  string[] path = {"b","c","d","d","d","c","b","a"};
  string[] cr = {"1","3","5","4","7","8","8","7"};
  int nc = xs.length;
  int nt = path.length;
  real x0 = 0.0;
  real t0 = 0.0;
  real dx = 0.1;
  real dt = 0.1;
  real x;
  real t;
  pair A, B, C, D;
  defaultpen(2);
  for (int j = 0; j < nc; ++j) {
    x = x0 + j * dx;
    label(xs[j], position=(x+dx/2,t0-1.5*dt), align=N);
  }
  for (int i = 0; i < nt; ++i) {
    x = x0;
    t = t0 + i * dt;
    label((string) i, (x-dx,t+dt/2));
    for (int j = 0; j < nc; ++j) {
      x = x0 + j * dx;
      A=(x,t); B=(x+dx,t); C=(x+dx,t+dt); D=(x,t+dt);
      draw(A--B--C--D--A, grey);
    }
  }
  x = x0;
  t = t0 + (0 + nt) * dt + dt/2;
  label("...", (x-dx,t+dt/2));
  label("...", (x+nc*dx/2,t+dt/2));
  x = x0;
  t = t0 + (1 + nt) * dt + dt;
  label("n", (x-dx,t+dt/2));
  for (int j = 0; j < nc; ++j) {
    x = x0 + j * dx;
    A=(x,t); B=(x+dx,t); C=(x+dx,t+dt); D=(x,t+dt);
    draw(A--B--C--D--A, grey);
  }
  int steps = 0;
  for (int i = 0; i < steps - 1; ++i) {
    t = t0 + i * dt;
    int j = search(xs,path[i]);
    x = x0 + j * dx;
    label(cr[j], (x+dx/2,t+dt/2), white);
  }
  if (steps > 0) {
    t = t0 + (steps - 1) * dt;
    int j = search(xs,path[steps - 1]);
    x = x0 + j * dx;
    label(cr[j], (x+dx/2,t+dt/2), white);
  }
  for (int i = 0; i < steps; ++i) {
    t = t0 + i * dt;
    int j = search(xs,path[i]);
    x = x0 + j * dx;
    A=(x,t); B=(x+dx,t); C=(x+dx,t+dt); D=(x,t+dt);
    draw(A--B--C--D--A, lightred);
  }
  t = t0 + steps * dt;
  int j = search(xs,path[steps]);
  x = x0 + j * dx;
  A=(x,t); B=(x+dx,t); C=(x+dx,t+dt); D=(x,t+dt);
  draw(A--B--C--D--A, red);
 \end{asy}
 \hspace*{1cm}
 \begin{asy}
  size(6cm);
  include graph;
  string[] xs = {"a","b","c","d","e"};
  string[] path = {"b","c","d","d","d","c","b","a"};
  string[] cr = {"1","3","5","4","7","8","8","7"};
  int nc = xs.length;
  int nt = path.length;
  real x0 = 0.0;
  real t0 = 0.0;
  real dx = 0.1;
  real dt = 0.1;
  real x;
  real t;
  pair A, B, C, D;
  defaultpen(2);
  for (int j = 0; j < nc; ++j) {
    x = x0 + j * dx;
    label(xs[j], position=(x+dx/2,t0-1.5*dt), align=N);
  }
  for (int i = 0; i < nt; ++i) {
    x = x0;
    t = t0 + i * dt;
    label((string) i, (x-dx,t+dt/2));
    for (int j = 0; j < nc; ++j) {
      x = x0 + j * dx;
      A=(x,t); B=(x+dx,t); C=(x+dx,t+dt); D=(x,t+dt);
      draw(A--B--C--D--A, grey);
    }
  }
  x = x0;
  t = t0 + (0 + nt) * dt + dt/2;
  label("...", (x-dx,t+dt/2));
  label("...", (x+nc*dx/2,t+dt/2));
  x = x0;
  t = t0 + (1 + nt) * dt + dt;
  label("n", (x-dx,t+dt/2));
  for (int j = 0; j < nc; ++j) {
    x = x0 + j * dx;
    A=(x,t); B=(x+dx,t); C=(x+dx,t+dt); D=(x,t+dt);
    draw(A--B--C--D--A, grey);
  }
  int steps = 1;
  for (int i = 0; i < steps - 1; ++i) {
    t = t0 + i * dt;
    int j = search(xs,path[i]);
    x = x0 + j * dx;
    label(cr[j], (x+dx/2,t+dt/2), lightred);
  }
  if (steps > 0) {
    t = t0 + (steps - 1) * dt;
    int j = search(xs,path[steps - 1]);
    x = x0 + j * dx;
    label(cr[j], (x+dx/2,t+dt/2), lightred);
  }
  for (int i = 0; i < steps; ++i) {
    t = t0 + i * dt;
    int j = search(xs,path[i]);
    x = x0 + j * dx;
    A=(x,t); B=(x+dx,t); C=(x+dx,t+dt); D=(x,t+dt);
    draw(A--B--C--D--A, lightred);
  }
  t = t0 + steps * dt;
  int j = search(xs,path[steps]);
  x = x0 + j * dx;
  A=(x,t); B=(x+dx,t); C=(x+dx,t+dt); D=(x,t+dt);
  label("?", (x+dx/2,t+dt/2), white);
  draw(A--B--C--D--A, red);
 \end{asy}
 \hspace*{1cm}
 \begin{asy}
  size(6cm);
  include graph;
  string[] xs = {"a","b","c","d","e"};
  string[] path = {"b","c","d","d","d","c","b","a"};
  string[] cr = {"1","3","5","4","7","8","8","7"};
  int nc = xs.length;
  int nt = path.length;
  real x0 = 0.0;
  real t0 = 0.0;
  real dx = 0.1;
  real dt = 0.1;
  real x;
  real t;
  pair A, B, C, D;
  defaultpen(2);
  for (int j = 0; j < nc; ++j) {
    x = x0 + j * dx;
    label(xs[j], position=(x+dx/2,t0-1.5*dt), align=N);
  }
  for (int i = 0; i < nt; ++i) {
    x = x0;
    t = t0 + i * dt;
    label((string) i, (x-dx,t+dt/2));
    for (int j = 0; j < nc; ++j) {
      x = x0 + j * dx;
      A=(x,t); B=(x+dx,t); C=(x+dx,t+dt); D=(x,t+dt);
      draw(A--B--C--D--A, grey);
    }
  }
  x = x0;
  t = t0 + (0 + nt) * dt + dt/2;
  label("...", (x-dx,t+dt/2));
  label("...", (x+nc*dx/2,t+dt/2));
  x = x0;
  t = t0 + (1 + nt) * dt + dt;
  label("n", (x-dx,t+dt/2));
  for (int j = 0; j < nc; ++j) {
    x = x0 + j * dx;
    A=(x,t); B=(x+dx,t); C=(x+dx,t+dt); D=(x,t+dt);
    draw(A--B--C--D--A, grey);
  }
  int steps = 4;
  for (int i = 0; i < steps - 1; ++i) {
    t = t0 + i * dt;
    int j = search(xs,path[i]);
    x = x0 + j * dx;
    label(cr[j], (x+dx/2,t+dt/2), lightred);
  }
  if (steps > 0) {
    t = t0 + (steps - 1) * dt;
    int j = search(xs,path[steps - 1]);
    x = x0 + j * dx;
    label(cr[j], (x+dx/2,t+dt/2), lightred);
  }
  for (int i = 0; i < steps; ++i) {
    t = t0 + i * dt;
    int j = search(xs,path[i]);
    x = x0 + j * dx;
    A=(x,t); B=(x+dx,t); C=(x+dx,t+dt); D=(x,t+dt);
    draw(A--B--C--D--A, lightred);
  }
  t = t0 + steps * dt;
  int j = search(xs,path[steps]);
  x = x0 + j * dx;
  A=(x,t); B=(x+dx,t); C=(x+dx,t+dt); D=(x,t+dt);
  label("?", (x+dx/2,t+dt/2), white);
  draw(A--B--C--D--A, red);
 \end{asy}
\caption{\small Possible evolutions for the ``cylinder'' problem. Initial state
($b$, left), state and reward after one step ($c$ and 3, middle) and
four steps trajectory and rewards (right). \label{figure:one}}
\end{figure}

Upon selecting a control, the decision maker enters a new state. For
instance, selecting \ensuremath{\Conid{R}} in \ensuremath{\Varid{b}} brings him from \ensuremath{\Varid{b}} to \ensuremath{\Varid{c}}, see Figure
\ref{figure:one}. Thus, each step is characterized by a current state, a
selected control and a new state. A step also yields a reward, for
instance 3 for the transition from \ensuremath{\Varid{b}} to \ensuremath{\Varid{c}} and for control \ensuremath{\Conid{R}}.

The challenge for the decision maker is to make a finite number of
steps, say \ensuremath{\Varid{n}}, by selecting controls that maximize the sum of the
rewards collected.

An example of a possible trajectory and corresponding rewards for the
first four steps is shown on the right of figure \ref{figure:one}. In
this example, the decision maker has so far collected a total reward of
16 by selecting controls according to the sequence \ensuremath{[\mskip1.5mu \Conid{R},\Conid{R},\Conid{A},\Conid{A}\mskip1.5mu]}: \ensuremath{\Conid{R}} in
the first and in the second steps, \ensuremath{\Conid{A}} in the third and in the fourth
steps.

In this problem, the set of possible states $State = \{a,b,c,d,e\}$ is
constant for all steps and the controls available in a state only
depend on that state.
The problem is an instance of a particular class of problems called
time-independent, deterministic SDPs.
In our previous paper~\cite{botta+al2013c} we characterized this class
in terms of four assumptions:
\begin{enumerate}
\item The state space does not depend on the current number of steps.
\item The control space in a given state only depends on that state but
  not on the current number of steps.
\item At each step, the new state depends on the current state and on
  the selected control via a known deterministic function.
\item At each step, the reward is a known function of the current state, of
  the selected control and of the new state.
\end{enumerate}
(Throughout this paper we essentially adopt the notation introduced in
\cite{botta+al2013c}: data types, constructors and \ensuremath{\Conid{Type}}-valued
functions are capitalized, function that return values of a specific
type are lowercased.
We use the mnemonic \ensuremath{\Conid{Spec}} (or \ensuremath{\Varid{spec}}) to denote specifications.
But to improve readability we now use \ensuremath{\Conid{State}} and \ensuremath{\Conid{Ctrl}} (instead of
\ensuremath{\Conid{X}} and \ensuremath{\Conid{Y}}) to denote states and controls.)
The results obtained~\cite{botta+al2013c} for this class of sequential
decision problems can be summarized as follows:
The problems can be formalized in terms of a context containing states
$State$ and controls $Ctrl$ from each state:
\begin{hscode}\SaveRestoreHook
\column{B}{@{}>{\hspre}l<{\hspost}@{}}%
\column{3}{@{}>{\hspre}l<{\hspost}@{}}%
\column{11}{@{}>{\hspre}c<{\hspost}@{}}%
\column{11E}{@{}l@{}}%
\column{14}{@{}>{\hspre}l<{\hspost}@{}}%
\column{E}{@{}>{\hspre}l<{\hspost}@{}}%
\>[3]{}\Conid{State}{}\<[11]%
\>[11]{}\ \mathop{:}\ {}\<[11E]%
\>[14]{}\Conid{Type}{}\<[E]%
\\
\>[3]{}\Conid{Ctrl}{}\<[11]%
\>[11]{}\ \mathop{:}\ {}\<[11E]%
\>[14]{}(\Varid{x}\ \mathop{:}\ \Conid{State})\ \to\ \Conid{Type}{}\<[E]%
\\
\>[3]{}\Varid{step}{}\<[11]%
\>[11]{}\ \mathop{:}\ {}\<[11E]%
\>[14]{}(\Varid{x}\ \mathop{:}\ \Conid{State})\ \to\ (\Varid{y}\ \mathop{:}\ \Conid{Ctrl}\;\Varid{x})\ \to\ \Conid{State}{}\<[E]%
\\
\>[3]{}\Varid{reward}{}\<[11]%
\>[11]{}\ \mathop{:}\ {}\<[11E]%
\>[14]{}(\Varid{x}\ \mathop{:}\ \Conid{State})\ \to\ (\Varid{y}\ \mathop{:}\ \Conid{Ctrl}\;\Varid{x})\ \to\ (\Varid{x'}\ \mathop{:}\ \Conid{State})\ \to\ \mathbb{R}{}\<[E]%
\ColumnHook
\end{hscode}\resethooks
and of the notions of control sequence \ensuremath{\Conid{CtrlSeq}\;\Varid{x}\;\Varid{n}} (from a starting
state \ensuremath{\Varid{x}\ \mathop{:}\ \Conid{State}} and for \ensuremath{\Varid{n}\ \mathop{:}\ \mathbb{N}} steps), value of control sequences
and optimality of control sequences:
\begin{hscode}\SaveRestoreHook
\column{B}{@{}>{\hspre}l<{\hspost}@{}}%
\column{3}{@{}>{\hspre}l<{\hspost}@{}}%
\column{5}{@{}>{\hspre}l<{\hspost}@{}}%
\column{11}{@{}>{\hspre}c<{\hspost}@{}}%
\column{11E}{@{}l@{}}%
\column{14}{@{}>{\hspre}l<{\hspost}@{}}%
\column{E}{@{}>{\hspre}l<{\hspost}@{}}%
\>[3]{}\mathbf{data}\;\Conid{CtrlSeq}\ \mathop{:}\ \Conid{State}\ \to\ \mathbb{N}\ \to\ \Conid{Type}\;\mathbf{where}{}\<[E]%
\\
\>[3]{}\hsindent{2}{}\<[5]%
\>[5]{}\Conid{Nil}{}\<[11]%
\>[11]{}\ \mathop{:}\ {}\<[11E]%
\>[14]{}\Conid{CtrlSeq}\;\Varid{x}\;\Conid{Z}{}\<[E]%
\\
\>[3]{}\hsindent{2}{}\<[5]%
\>[5]{}(\mathbin{::}){}\<[11]%
\>[11]{}\ \mathop{:}\ {}\<[11E]%
\>[14]{}(\Varid{y}\ \mathop{:}\ \Conid{Ctrl}\;\Varid{x})\ \to\ \Conid{CtrlSeq}\;(\Varid{step}\;\Varid{x}\;\Varid{y})\;\Varid{n}\ \to\ \Conid{CtrlSeq}\;\Varid{x}\;(\Conid{S}\;\Varid{n}){}\<[E]%
\ColumnHook
\end{hscode}\resethooks
\begin{hscode}\SaveRestoreHook
\column{B}{@{}>{\hspre}l<{\hspost}@{}}%
\column{3}{@{}>{\hspre}l<{\hspost}@{}}%
\column{14}{@{}>{\hspre}l<{\hspost}@{}}%
\column{25}{@{}>{\hspre}l<{\hspost}@{}}%
\column{36}{@{}>{\hspre}c<{\hspost}@{}}%
\column{36E}{@{}l@{}}%
\column{39}{@{}>{\hspre}l<{\hspost}@{}}%
\column{E}{@{}>{\hspre}l<{\hspost}@{}}%
\>[3]{}\Varid{value}\ \mathop{:}\ \Conid{CtrlSeq}\;\Varid{x}\;\Varid{n}\ \to\ \mathbb{R}{}\<[E]%
\\
\>[3]{}\Varid{value}\;{}\<[14]%
\>[14]{}\{\mskip1.5mu \Varid{n}\mathrel{=}\Conid{Z}\mskip1.5mu\}\;{}\<[25]%
\>[25]{}\anonymous {}\<[36]%
\>[36]{}\mathrel{=}{}\<[36E]%
\>[39]{}\mathrm{0}{}\<[E]%
\\
\>[3]{}\Varid{value}\;\{\mskip1.5mu \Varid{x}\mskip1.5mu\}\;{}\<[14]%
\>[14]{}\{\mskip1.5mu \Varid{n}\mathrel{=}\Conid{S}\;\Varid{m}\mskip1.5mu\}\;{}\<[25]%
\>[25]{}(\Varid{y}\mathbin{::}\Varid{ys}){}\<[36]%
\>[36]{}\mathrel{=}{}\<[36E]%
\>[39]{}\Varid{reward}\;\Varid{x}\;\Varid{y}\;(\Varid{step}\;\Varid{x}\;\Varid{y})\mathbin{+}\Varid{value}\;\Varid{ys}{}\<[E]%
\ColumnHook
\end{hscode}\resethooks
\begin{hscode}\SaveRestoreHook
\column{B}{@{}>{\hspre}l<{\hspost}@{}}%
\column{3}{@{}>{\hspre}l<{\hspost}@{}}%
\column{E}{@{}>{\hspre}l<{\hspost}@{}}%
\>[3]{}\Conid{OptCtrlSeq}\ \mathop{:}\ \Conid{CtrlSeq}\;\Varid{x}\;\Varid{n}\ \to\ \Conid{Type}{}\<[E]%
\\
\>[3]{}\Conid{OptCtrlSeq}\;\{\mskip1.5mu \Varid{x}\mskip1.5mu\}\;\{\mskip1.5mu \Varid{n}\mskip1.5mu\}\;\Varid{ys}\mathrel{=}(\Varid{ys'}\ \mathop{:}\ \Conid{CtrlSeq}\;\Varid{x}\;\Varid{n})\ \to\ \Conid{So}\;(\Varid{value}\;\Varid{ys'}\leq \Varid{value}\;\Varid{ys}){}\<[E]%
\ColumnHook
\end{hscode}\resethooks
In the above formulation, \ensuremath{\Conid{CtrlSeq}\;\Varid{x}\;\Varid{n}} formalizes the notion of a
sequence of controls of length \ensuremath{\Varid{n}} with the first control in \ensuremath{\Conid{Ctrl}\;\Varid{x}}. In other words, if we are given \ensuremath{\Varid{ys}\ \mathop{:}\ \Conid{CtrlSeq}\;\Varid{x}\;\Varid{n}} and we are in
\ensuremath{\Varid{x}}, we know that we can select the first control of \ensuremath{\Varid{ys}}.

The function \ensuremath{\Varid{value}} computes the value of a control sequence. As
explained in the beginning of this section, the challenge for the
decision maker is to select controls that maximize a sum of
rewards. Throughout this paper, we use \ensuremath{\mathbin{+}} to compute the sum. But it
is clear that \ensuremath{\mathbin{+}} does not need to denote standard addition. For
instance, the sum could be ``discounted'' through lower weights for
future rewards.

Thus, a sequence of controls \ensuremath{\Varid{ps}\ \mathop{:}\ \Conid{CtrlSeq}\;\Varid{x}\;\Varid{n}} is optimal iff any other
sequence \ensuremath{\Varid{ps'}\ \mathop{:}\ \Conid{CtrlSeq}\;\Varid{x}\;\Varid{n}} has a value that is smaller or equal to the
value of \ensuremath{\Varid{ps}}. The value of a control sequence of length zero is zero
and the value of a control sequence of length \ensuremath{\Conid{S}\;\Varid{m}} is computed by
adding the reward obtained with the first decision step to the value of
making \ensuremath{\Varid{m}} more decisions with the tail of that sequence.

In the above, the arguments \ensuremath{\Varid{x}} and \ensuremath{\Varid{n}} to \ensuremath{\Conid{CtrlSeq}} in the types of \ensuremath{\Varid{value}} and
$OptCtrlSeq$ occur free. In Idris (as in Haskell), this means that they will be automatically
inserted as implicit arguments. In the definitions of \ensuremath{\Varid{value}} and \ensuremath{\Conid{OptCtrlSeq}},
these implicit arguments are brought into the local scope by adding them to
the pattern match surrounded by curly braces.
We also use the function \ensuremath{\Conid{So}\ \mathop{:}\ \Conid{Bool}\ \to\ \Conid{Type}} for translating between
Booleans and types (the only constructor is \ensuremath{\Conid{Oh}\ \mathop{:}\ \Conid{So}\;\Conid{True}}).

We have shown that one can compute optimal control sequences from
optimal policy sequences. These are policy vectors that maximize the
value function \ensuremath{\Varid{val}} for every state:
\begin{hscode}\SaveRestoreHook
\column{B}{@{}>{\hspre}l<{\hspost}@{}}%
\column{3}{@{}>{\hspre}l<{\hspost}@{}}%
\column{E}{@{}>{\hspre}l<{\hspost}@{}}%
\>[3]{}\Conid{Policy}\ \mathop{:}\ \Conid{Type}{}\<[E]%
\\
\>[3]{}\Conid{Policy}\mathrel{=}(\Varid{x}\ \mathop{:}\ \Conid{State})\ \to\ \Conid{Ctrl}\;\Varid{x}{}\<[E]%
\ColumnHook
\end{hscode}\resethooks
\begin{hscode}\SaveRestoreHook
\column{B}{@{}>{\hspre}l<{\hspost}@{}}%
\column{3}{@{}>{\hspre}l<{\hspost}@{}}%
\column{E}{@{}>{\hspre}l<{\hspost}@{}}%
\>[3]{}\Conid{PolicySeq}\ \mathop{:}\ \mathbb{N}\ \to\ \Conid{Type}{}\<[E]%
\\
\>[3]{}\Conid{PolicySeq}\;\Varid{n}\mathrel{=}\Conid{Vect}\;\Varid{n}\;\Conid{Policy}{}\<[E]%
\ColumnHook
\end{hscode}\resethooks
\begin{hscode}\SaveRestoreHook
\column{B}{@{}>{\hspre}l<{\hspost}@{}}%
\column{3}{@{}>{\hspre}l<{\hspost}@{}}%
\column{5}{@{}>{\hspre}l<{\hspost}@{}}%
\column{9}{@{}>{\hspre}c<{\hspost}@{}}%
\column{9E}{@{}l@{}}%
\column{12}{@{}>{\hspre}l<{\hspost}@{}}%
\column{18}{@{}>{\hspre}l<{\hspost}@{}}%
\column{21}{@{}>{\hspre}l<{\hspost}@{}}%
\column{32}{@{}>{\hspre}c<{\hspost}@{}}%
\column{32E}{@{}l@{}}%
\column{35}{@{}>{\hspre}l<{\hspost}@{}}%
\column{E}{@{}>{\hspre}l<{\hspost}@{}}%
\>[3]{}\Varid{val}\ \mathop{:}\ (\Varid{x}\ \mathop{:}\ \Conid{State})\ \to\ \Conid{PolicySeq}\;\Varid{n}\ \to\ \mathbb{R}{}\<[E]%
\\
\>[3]{}\Varid{val}\;\{\mskip1.5mu \Varid{n}\mathrel{=}\Conid{Z}\mskip1.5mu\}\;{}\<[18]%
\>[18]{}\anonymous \;{}\<[21]%
\>[21]{}\anonymous {}\<[32]%
\>[32]{}\mathrel{=}{}\<[32E]%
\>[35]{}\mathrm{0}{}\<[E]%
\\
\>[3]{}\Varid{val}\;\{\mskip1.5mu \Varid{n}\mathrel{=}\Conid{S}\;\Varid{m}\mskip1.5mu\}\;{}\<[18]%
\>[18]{}\Varid{x}\;{}\<[21]%
\>[21]{}(\Varid{p}\mathbin{::}\Varid{ps}){}\<[32]%
\>[32]{}\mathrel{=}{}\<[32E]%
\>[35]{}\Varid{reward}\;\Varid{x}\;(\Varid{p}\;\Varid{x})\;\Varid{x'}\mathbin{+}\Varid{val}\;\Varid{x'}\;\Varid{ps}\;\mathbf{where}{}\<[E]%
\\
\>[3]{}\hsindent{2}{}\<[5]%
\>[5]{}\Varid{x'}{}\<[9]%
\>[9]{}\ \mathop{:}\ {}\<[9E]%
\>[12]{}\Conid{State}{}\<[E]%
\\
\>[3]{}\hsindent{2}{}\<[5]%
\>[5]{}\Varid{x'}{}\<[9]%
\>[9]{}\mathrel{=}{}\<[9E]%
\>[12]{}\Varid{step}\;\Varid{x}\;(\Varid{p}\;\Varid{x}){}\<[E]%
\ColumnHook
\end{hscode}\resethooks
\begin{hscode}\SaveRestoreHook
\column{B}{@{}>{\hspre}l<{\hspost}@{}}%
\column{3}{@{}>{\hspre}l<{\hspost}@{}}%
\column{E}{@{}>{\hspre}l<{\hspost}@{}}%
\>[3]{}\Conid{OptPolicySeq}\ \mathop{:}\ (\Varid{n}\ \mathop{:}\ \mathbb{N})\ \to\ \Conid{PolicySeq}\;\Varid{n}\ \to\ \Conid{Type}{}\<[E]%
\\
\>[3]{}\Conid{OptPolicySeq}\;\Varid{n}\;\Varid{ps}\mathrel{=}(\Varid{x}\ \mathop{:}\ \Conid{State})\ \to\ (\Varid{ps'}\ \mathop{:}\ \Conid{PolicySeq}\;\Varid{n})\ \to\ \Conid{So}\;(\Varid{val}\;\Varid{x}\;\Varid{ps'}\leq \Varid{val}\;\Varid{x}\;\Varid{ps}){}\<[E]%
\ColumnHook
\end{hscode}\resethooks
We have expressed Bellman's principle of optimality~\cite{bellman1957}
in terms of the notion of optimal extensions of policy sequences

\begin{hscode}\SaveRestoreHook
\column{B}{@{}>{\hspre}l<{\hspost}@{}}%
\column{3}{@{}>{\hspre}l<{\hspost}@{}}%
\column{E}{@{}>{\hspre}l<{\hspost}@{}}%
\>[3]{}\Conid{OptExt}\ \mathop{:}\ \Conid{PolicySeq}\;\Varid{n}\ \to\ \Conid{Policy}\ \to\ \Conid{Type}{}\<[E]%
\\
\>[3]{}\Conid{OptExt}\;\Varid{ps}\;\Varid{p}\mathrel{=}(\Varid{p'}\ \mathop{:}\ \Conid{Policy})\ \to\ (\Varid{x}\ \mathop{:}\ \Conid{State})\ \to\ \Conid{So}\;(\Varid{val}\;\Varid{x}\;(\Varid{p'}\mathbin{::}\Varid{ps})\leq \Varid{val}\;\Varid{x}\;(\Varid{p}\mathbin{::}\Varid{ps})){}\<[E]%
\ColumnHook
\end{hscode}\resethooks
\begin{hscode}\SaveRestoreHook
\column{B}{@{}>{\hspre}l<{\hspost}@{}}%
\column{3}{@{}>{\hspre}l<{\hspost}@{}}%
\column{12}{@{}>{\hspre}c<{\hspost}@{}}%
\column{12E}{@{}l@{}}%
\column{15}{@{}>{\hspre}l<{\hspost}@{}}%
\column{20}{@{}>{\hspre}l<{\hspost}@{}}%
\column{36}{@{}>{\hspre}l<{\hspost}@{}}%
\column{58}{@{}>{\hspre}c<{\hspost}@{}}%
\column{58E}{@{}l@{}}%
\column{E}{@{}>{\hspre}l<{\hspost}@{}}%
\>[3]{}\Conid{Bellman}{}\<[12]%
\>[12]{}\ \mathop{:}\ {}\<[12E]%
\>[15]{}(\Varid{ps}{}\<[20]%
\>[20]{}\ \mathop{:}\ \Conid{PolicySeq}\;\Varid{n}){}\<[36]%
\>[36]{}\ \to\ \Conid{OptPolicySeq}\;\Varid{n}\;\Varid{ps}{}\<[58]%
\>[58]{}\ \to\ {}\<[58E]%
\\
\>[15]{}(\Varid{p}{}\<[20]%
\>[20]{}\ \mathop{:}\ \Conid{Policy}){}\<[36]%
\>[36]{}\ \to\ \Conid{OptExt}\;\Varid{ps}\;\Varid{p}{}\<[58]%
\>[58]{}\ \to\ {}\<[58E]%
\\
\>[15]{}\Conid{OptPolicySeq}\;(\Conid{S}\;\Varid{n})\;(\Varid{p}\mathbin{::}\Varid{ps}){}\<[E]%
\ColumnHook
\end{hscode}\resethooks
and implemented a machine-checkable proof of \ensuremath{\Conid{Bellman}}.
Another machine-checkable proof guarantees that, if one can implement
a function \ensuremath{\Varid{optExt}} that computes an optimal extension of arbitrary policy
sequences
\begin{hscode}\SaveRestoreHook
\column{B}{@{}>{\hspre}l<{\hspost}@{}}%
\column{3}{@{}>{\hspre}l<{\hspost}@{}}%
\column{E}{@{}>{\hspre}l<{\hspost}@{}}%
\>[3]{}\Conid{OptExtLemma}\ \mathop{:}\ (\Varid{ps}\ \mathop{:}\ \Conid{PolicySeq}\;\Varid{n})\ \to\ \Conid{OptExt}\;\Varid{ps}\;(\Varid{optExt}\;\Varid{ps}){}\<[E]%
\ColumnHook
\end{hscode}\resethooks
then
\begin{hscode}\SaveRestoreHook
\column{B}{@{}>{\hspre}l<{\hspost}@{}}%
\column{3}{@{}>{\hspre}l<{\hspost}@{}}%
\column{5}{@{}>{\hspre}l<{\hspost}@{}}%
\column{35}{@{}>{\hspre}c<{\hspost}@{}}%
\column{35E}{@{}l@{}}%
\column{39}{@{}>{\hspre}l<{\hspost}@{}}%
\column{E}{@{}>{\hspre}l<{\hspost}@{}}%
\>[3]{}\Varid{backwardsInduction}\ \mathop{:}\ (\Varid{n}\ \mathop{:}\ \mathbb{N}){}\<[35]%
\>[35]{}\ \to\ {}\<[35E]%
\>[39]{}\Conid{PolicySeq}\;\Varid{n}{}\<[E]%
\\
\>[3]{}\Varid{backwardsInduction}\;\Conid{Z}{}\<[35]%
\>[35]{}\mathrel{=}{}\<[35E]%
\>[39]{}\Conid{Nil}{}\<[E]%
\\
\>[3]{}\Varid{backwardsInduction}\;(\Conid{S}\;\Varid{n}){}\<[35]%
\>[35]{}\mathrel{=}{}\<[35E]%
\>[39]{}(\Varid{optExt}\;\Varid{ps})\mathbin{::}\Varid{ps}\;\mathbf{where}{}\<[E]%
\\
\>[3]{}\hsindent{2}{}\<[5]%
\>[5]{}\Varid{ps}\ \mathop{:}\ \Conid{PolicySeq}\;\Varid{n}{}\<[E]%
\\
\>[3]{}\hsindent{2}{}\<[5]%
\>[5]{}\Varid{ps}\mathrel{=}\Varid{backwardsInduction}\;\Varid{n}{}\<[E]%
\ColumnHook
\end{hscode}\resethooks
yields optimal policy sequences (and, thus, optimal control
sequences) of arbitrary length:
\begin{hscode}\SaveRestoreHook
\column{B}{@{}>{\hspre}l<{\hspost}@{}}%
\column{3}{@{}>{\hspre}l<{\hspost}@{}}%
\column{E}{@{}>{\hspre}l<{\hspost}@{}}%
\>[3]{}\Conid{BackwardsInductionLemma}\ \mathop{:}\ (\Varid{n}\ \mathop{:}\ \mathbb{N})\ \to\ \Conid{OptPolicySeq}\;\Varid{n}\;(\Varid{backwardsInduction}\;\Varid{n}){}\<[E]%
\ColumnHook
\end{hscode}\resethooks
In our previous paper~\cite{botta+al2013c}, we have shown
that it is easy to implement a function that computes the optimal extension of
an arbitrary policy sequence if one can implement
\begin{hscode}\SaveRestoreHook
\column{B}{@{}>{\hspre}l<{\hspost}@{}}%
\column{3}{@{}>{\hspre}l<{\hspost}@{}}%
\column{11}{@{}>{\hspre}c<{\hspost}@{}}%
\column{11E}{@{}l@{}}%
\column{14}{@{}>{\hspre}l<{\hspost}@{}}%
\column{E}{@{}>{\hspre}l<{\hspost}@{}}%
\>[3]{}\Varid{max}{}\<[11]%
\>[11]{}\ \mathop{:}\ {}\<[11E]%
\>[14]{}(\Varid{x}\ \mathop{:}\ \Conid{State})\ \to\ (\Conid{Ctrl}\;\Varid{x}\ \to\ \mathbb{R})\ \to\ \mathbb{R}{}\<[E]%
\\
\>[3]{}\Varid{argmax}{}\<[11]%
\>[11]{}\ \mathop{:}\ {}\<[11E]%
\>[14]{}(\Varid{x}\ \mathop{:}\ \Conid{State})\ \to\ (\Conid{Ctrl}\;\Varid{x}\ \to\ \mathbb{R})\ \to\ \Conid{Ctrl}\;\Varid{x}{}\<[E]%
\ColumnHook
\end{hscode}\resethooks
which fulfill the specifications
\begin{hscode}\SaveRestoreHook
\column{B}{@{}>{\hspre}l<{\hspost}@{}}%
\column{3}{@{}>{\hspre}l<{\hspost}@{}}%
\column{15}{@{}>{\hspre}c<{\hspost}@{}}%
\column{15E}{@{}l@{}}%
\column{18}{@{}>{\hspre}l<{\hspost}@{}}%
\column{E}{@{}>{\hspre}l<{\hspost}@{}}%
\>[3]{}\Conid{MaxSpec}{}\<[15]%
\>[15]{}\ \mathop{:}\ {}\<[15E]%
\>[18]{}\Conid{Type}{}\<[E]%
\\
\>[3]{}\Conid{MaxSpec}{}\<[15]%
\>[15]{}\mathrel{=}{}\<[15E]%
\>[18]{}(\Varid{x}\ \mathop{:}\ \Conid{State})\ \to\ (\Varid{f}\ \mathop{:}\ \Conid{Ctrl}\;\Varid{x}\ \to\ \mathbb{R})\ \to\ (\Varid{y}\ \mathop{:}\ \Conid{Ctrl}\;\Varid{x})\ \to\ {}\<[E]%
\\
\>[18]{}\Conid{So}\;(\Varid{f}\;\Varid{y}\leq \Varid{max}\;\Varid{x}\;\Varid{f}){}\<[E]%
\\
\>[3]{}\Conid{ArgmaxSpec}{}\<[15]%
\>[15]{}\ \mathop{:}\ {}\<[15E]%
\>[18]{}\Conid{Type}{}\<[E]%
\\
\>[3]{}\Conid{ArgmaxSpec}{}\<[15]%
\>[15]{}\mathrel{=}{}\<[15E]%
\>[18]{}(\Varid{x}\ \mathop{:}\ \Conid{State})\ \to\ (\Varid{f}\ \mathop{:}\ \Conid{Ctrl}\;\Varid{x}\ \to\ \mathbb{R})\ \to\ {}\<[E]%
\\
\>[18]{}\Conid{So}\;(\Varid{f}\;(\Varid{argmax}\;\Varid{x}\;\Varid{f})==\Varid{max}\;\Varid{x}\;\Varid{f}){}\<[E]%
\ColumnHook
\end{hscode}\resethooks
When \ensuremath{\Conid{Ctrl}\;\Varid{x}} is finite, such \ensuremath{\Varid{max}} and \ensuremath{\Varid{argmax}} can always be implemented
in a finite number of comparisons.
\section{Time-dependent state spaces}
\label{section:time-dependent}

The results summarized in the previous section are valid under one
implicit assumption: that one can construct control sequences of
arbitrary length from arbitrary initial states.
A sufficient (but not necessary) condition for this is that, for all \ensuremath{\Varid{x}\ \mathop{:}\ \Conid{State}}, the control space \ensuremath{\Conid{Ctrl}\;\Varid{x}} is not empty. As we shall see in a moment,
this assumption is too strong and needs to be refined.

\begin{figure}[hb]
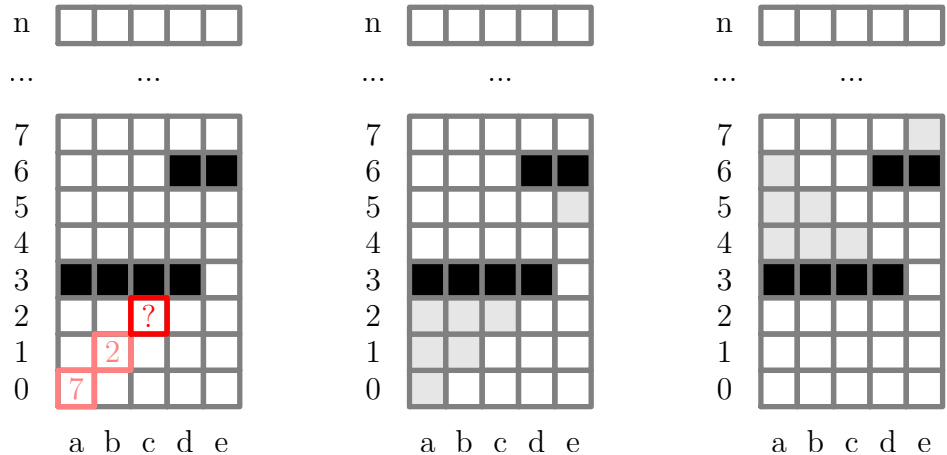

 \begin{asy}
  size(6cm);
  include graph;
  string[] xs = {"a","b","c","d","e"};
  string[] path = {"a","b","c","d","d","d","c","b"};
  string[] cr = {"7","2","7","4","7","8","8","7"};
  int nc = xs.length;
  int nt = path.length;
  real x0 = 0.0;
  real t0 = 0.0;
  real dx = 0.1;
  real dt = 0.1;
  real x;
  real t;
  pair A, B, C, D;
  defaultpen(2);
  for (int j = 0; j < nc; ++j) {
    x = x0 + j * dx;
    label(xs[j], position=(x+dx/2,t0-1.5*dt), align=N);
  }
  for (int i = 0; i < nt; ++i) {
    x = x0;
    t = t0 + i * dt;
    label((string) i, (x-dx,t+dt/2));
    for (int j = 0; j < nc; ++j) {
      x = x0 + j * dx;
      A=(x,t); B=(x+dx,t); C=(x+dx,t+dt); D=(x,t+dt);
      if (i == 3 && j < nc - 1) fill(A--B--C--D--A--cycle);
      if (i == 6 && j > nc - 3) fill(A--B--C--D--A--cycle);
      draw(A--B--C--D--A, grey);
    }
  }
  x = x0;
  t = t0 + (0 + nt) * dt + dt/2;
  label("...", (x-dx,t+dt/2));
  label("...", (x+nc*dx/2,t+dt/2));
  x = x0;
  t = t0 + (1 + nt) * dt + dt;
  label("n", (x-dx,t+dt/2));
  for (int j = 0; j < nc; ++j) {
    x = x0 + j * dx;
    A=(x,t); B=(x+dx,t); C=(x+dx,t+dt); D=(x,t+dt);
    draw(A--B--C--D--A, grey);
  }
  int steps = 2;
  for (int i = 0; i < steps - 1; ++i) {
    t = t0 + i * dt;
    int j = search(xs,path[i]);
    x = x0 + j * dx;
    label(cr[j], (x+dx/2,t+dt/2), lightred);
  }
  if (steps > 0) {
    t = t0 + (steps - 1) * dt;
    int j = search(xs,path[steps - 1]);
    x = x0 + j * dx;
    label(cr[j], (x+dx/2,t+dt/2), lightred);
  }
  for (int i = 0; i < steps; ++i) {
    t = t0 + i * dt;
    int j = search(xs,path[i]);
    x = x0 + j * dx;
    A=(x,t); B=(x+dx,t); C=(x+dx,t+dt); D=(x,t+dt);
    draw(A--B--C--D--A, lightred);
  }
  t = t0 + steps * dt;
  int j = search(xs,path[steps]);
  x = x0 + j * dx;
  A=(x,t); B=(x+dx,t); C=(x+dx,t+dt); D=(x,t+dt);
  label("?", (x+dx/2,t+dt/2), red);
  draw(A--B--C--D--A, red);
 \end{asy}
 \hspace*{1cm}
 \begin{asy}
  size(6cm);
  include graph;
  string[] xs = {"a","b","c","d","e"};
  string[] path = {"a","b","c","d","d","d","c","b"};
  string[] cr = {"7","2","7","4","7","8","8","7"};
  int nc = xs.length;
  int nt = path.length;
  real x0 = 0.0;
  real t0 = 0.0;
  real dx = 0.1;
  real dt = 0.1;
  real x;
  real t;
  pair A, B, C, D;
  defaultpen(2);
  for (int j = 0; j < nc; ++j) {
    x = x0 + j * dx;
    label(xs[j], position=(x+dx/2,t0-1.5*dt), align=N);
  }
  for (int i = 0; i < nt; ++i) {
    x = x0;
    t = t0 + i * dt;
    label((string) i, (x-dx,t+dt/2));
    for (int j = 0; j < nc; ++j) {
      x = x0 + j * dx;
      A=(x,t); B=(x+dx,t); C=(x+dx,t+dt); D=(x,t+dt);
      if (i == 0 && j < nc - 1 - 3) fill(A--B--C--D--A--cycle, lightgrey);
      if (i == 1 && j < nc - 1 - 2) fill(A--B--C--D--A--cycle, lightgrey);
      if (i == 2 && j < nc - 1 - 1) fill(A--B--C--D--A--cycle, lightgrey);
      if (i == 3 && j < nc - 1) fill(A--B--C--D--A--cycle);
      if (i == 5 && j > nc - 3 + 1) fill(A--B--C--D--A--cycle, lightgrey);
      if (i == 6 && j > nc - 3) fill(A--B--C--D--A--cycle);
      draw(A--B--C--D--A, grey);
    }
  }
  x = x0;
  t = t0 + (0 + nt) * dt + dt/2;
  label("...", (x-dx,t+dt/2));
  label("...", (x+nc*dx/2,t+dt/2));
  x = x0;
  t = t0 + (1 + nt) * dt + dt;
  label("n", (x-dx,t+dt/2));
  for (int j = 0; j < nc; ++j) {
    x = x0 + j * dx;
    A=(x,t); B=(x+dx,t); C=(x+dx,t+dt); D=(x,t+dt);
    draw(A--B--C--D--A, grey);
  }
 \end{asy}
 \hspace*{1cm}
 \begin{asy}
  size(6cm);
  include graph;
  string[] xs = {"a","b","c","d","e"};
  string[] path = {"b","c","d","d","d","c","b","a"};
  string[] cr = {"1","3","5","4","7","8","8","7"};
  int nc = xs.length;
  int nt = path.length;
  real x0 = 0.0;
  real t0 = 0.0;
  real dx = 0.1;
  real dt = 0.1;
  real x;
  real t;
  pair A, B, C, D;
  defaultpen(2);
  for (int j = 0; j < nc; ++j) {
    x = x0 + j * dx;
    label(xs[j], position=(x+dx/2,t0-1.5*dt), align=N);
  }
  for (int i = 0; i < nt; ++i) {
    x = x0;
    t = t0 + i * dt;
    label((string) i, (x-dx,t+dt/2));
    for (int j = 0; j < nc; ++j) {
      x = x0 + j * dx;
      A=(x,t); B=(x+dx,t); C=(x+dx,t+dt); D=(x,t+dt);
      if (i == 6 && j < nc - 1 - 3) fill(A--B--C--D--A--cycle, lightgrey);
      if (i == 5 && j < nc - 1 - 2) fill(A--B--C--D--A--cycle, lightgrey);
      if (i == 4 && j < nc - 1 - 1) fill(A--B--C--D--A--cycle, lightgrey);
      if (i == 3 && j < nc - 1) fill(A--B--C--D--A--cycle);
      if (i == 7 && j > nc - 3 + 1) fill(A--B--C--D--A--cycle, lightgrey);
      if (i == 6 && j > nc - 3) fill(A--B--C--D--A--cycle);
      draw(A--B--C--D--A, grey);
    }
  }
  x = x0;
  t = t0 + (0 + nt) * dt + dt/2;
  label("...", (x-dx,t+dt/2));
  label("...", (x+nc*dx/2,t+dt/2));
  x = x0;
  t = t0 + (1 + nt) * dt + dt;
  label("n", (x-dx,t+dt/2));
  for (int j = 0; j < nc; ++j) {
    x = x0 + j * dx;
    A=(x,t); B=(x+dx,t); C=(x+dx,t+dt); D=(x,t+dt);
    draw(A--B--C--D--A, grey);
  }
 \end{asy}
\caption{\small Two steps trajectory starting at state $a$ (left),
  states of limited viability (middle) and unreachable states (right)
  for the ``cylinder'' problem with time-dependent state
  space. \label{figure:two}}
\end{figure}

Consider, again, the cylinder problem. Assume that at a given decision
step, only certain columns are \emph{valid}. For instance, for \ensuremath{\Varid{t}\neq\mathrm{3}}
and \ensuremath{\Varid{t}\neq\mathrm{6}} all states \ensuremath{\Varid{a}} through \ensuremath{\Varid{e}} are valid but at step 3 only \ensuremath{\Varid{e}}
is valid and at step 6 only \ensuremath{\Varid{a}}, \ensuremath{\Varid{b}} and \ensuremath{\Varid{c}} are valid, see figure
\ref{figure:two}.
Similarly, allow the controls available in a given state to depend both
on that state and on the decision step. For instance, from state \ensuremath{\Varid{b}} at
step 0 one might be able to move ahead or to the right. But at step 6
and from the same state, one might only be able to move to the left.

Note that a discrete time (number of decision steps) could be
accounted for in different ways.
One could for instance formalize ``time-dependent'' states as pairs
\ensuremath{(\mathbb{N},\Conid{Type})} or, as we do, by adding an extra \ensuremath{\mathbb{N}} argument.
A study of alternative formalizations of general decision problems is
a very interesting topic but goes well beyond the scope of this
work.
We can provide two ``justifications'' for the formalization proposed
here: that this is (again, to the best of our knowledge) the first
attempt at formalizing such problems generically and that the
formalization via additional \ensuremath{\mathbb{N}} arguments seems natural if one
considers how (non-autonomous) dynamical systems are usually
formalized in the continuous case through systems of differential
equations.

We can easily formalize the context for the time-dependent case by
adding an extra \ensuremath{\mathbb{N}} argument to the declarations of \ensuremath{\Conid{State}} and \ensuremath{\Conid{Ctrl}}
and extending the transition and the reward functions
accordingly (where \ensuremath{\Conid{S}\ \mathop{:}\ \mathbb{N}\ \to\ \mathbb{N}} is the successor function).
%
\begin{hscode}\SaveRestoreHook
\column{B}{@{}>{\hspre}l<{\hspost}@{}}%
\column{3}{@{}>{\hspre}l<{\hspost}@{}}%
\column{11}{@{}>{\hspre}c<{\hspost}@{}}%
\column{11E}{@{}l@{}}%
\column{14}{@{}>{\hspre}l<{\hspost}@{}}%
\column{E}{@{}>{\hspre}l<{\hspost}@{}}%
\>[3]{}\Conid{State}{}\<[11]%
\>[11]{}\ \mathop{:}\ {}\<[11E]%
\>[14]{}(\Varid{t}\ \mathop{:}\ \mathbb{N})\ \to\ \Conid{Type}{}\<[E]%
\\
\>[3]{}\Conid{Ctrl}{}\<[11]%
\>[11]{}\ \mathop{:}\ {}\<[11E]%
\>[14]{}(\Varid{t}\ \mathop{:}\ \mathbb{N})\ \to\ \Conid{State}\;\Varid{t}\ \to\ \Conid{Type}{}\<[E]%
\\
\>[3]{}\Varid{step}{}\<[11]%
\>[11]{}\ \mathop{:}\ {}\<[11E]%
\>[14]{}(\Varid{t}\ \mathop{:}\ \mathbb{N})\ \to\ (\Varid{x}\ \mathop{:}\ \Conid{State}\;\Varid{t})\ \to\ \Conid{Ctrl}\;\Varid{t}\;\Varid{x}\ \to\ \Conid{State}\;(\Conid{S}\;\Varid{t}){}\<[E]%
\\
\>[3]{}\Varid{reward}{}\<[11]%
\>[11]{}\ \mathop{:}\ {}\<[11E]%
\>[14]{}(\Varid{t}\ \mathop{:}\ \mathbb{N})\ \to\ (\Varid{x}\ \mathop{:}\ \Conid{State}\;\Varid{t})\ \to\ \Conid{Ctrl}\;\Varid{t}\;\Varid{x}\ \to\ \Conid{State}\;(\Conid{S}\;\Varid{t})\ \to\ \mathbb{R}{}\<[E]%
\ColumnHook
\end{hscode}\resethooks
In general we will be able to construct control
sequences of a given length from a given initial state only if \ensuremath{\Conid{State}}, \ensuremath{\Conid{Ctrl}}
and \ensuremath{\Varid{step}} satisfy certain compatibility conditions.
For example, assuming that the decision maker can move to the left, go
ahead or move to the right as described in the previous section, there
will be no sequence of more than two controls starting
from \ensuremath{\Varid{a}}, see figure \ref{figure:two} left.
At step \ensuremath{\Varid{t}}, there might be states which are valid but from which only
\ensuremath{\Varid{m}\mathbin{<}\Varid{n}\mathbin{-}\Varid{t}} steps can be done, see figure \ref{figure:two} middle. Conversely,
there might be states which are valid but which cannot be reached from
any initial state, see figure \ref{figure:two} right.

\subsection{Viability}

The time-dependent case makes it clear that, in general, we cannot
assume to be able to construct control sequences of arbitrary length
from arbitrary initial states.
For a given number of steps \ensuremath{\Varid{n}}, we must, at the very least, be able to
distinguish between initial states from which \ensuremath{\Varid{n}} steps can follow and
initial states from which only \ensuremath{\Varid{m}\mathbin{<}\Varid{n}} steps can follow.
Moreover, in building control sequences from initial states from which
\ensuremath{\Varid{n}} steps can actually be made, we may only select controls that bring us
to states from which \ensuremath{\Varid{n}\mathbin{-}\mathrm{1}} steps can be made.
In the example of figure \ref{figure:two} and with \ensuremath{\Varid{b}} as initial state,
for instance, the only control that can be put on the top of a control
sequence of length greater than 2 is \ensuremath{\Conid{R}}. Moves ahead or to the left would lead
to dead ends.

We use the term \emph{viability} to refer to the conditions that \ensuremath{\Conid{State}}, \ensuremath{\Conid{Ctrl}} and
\ensuremath{\Varid{step}} have to satisfy for a sequence of controls of length \ensuremath{\Varid{n}} starting in \ensuremath{\Varid{x}\ \mathop{:}\ \Conid{State}\;\Varid{t}} to exist.
More formally, we say that every state is viable for zero steps (\ensuremath{\Varid{viableSpec0}}) and
that a state \ensuremath{\Varid{x}\ \mathop{:}\ \Conid{State}\;\Varid{t}} is viable for \ensuremath{\Conid{S}\;\Varid{n}} steps if and only if there exists
a command in \ensuremath{\Conid{Ctrl}\;\Varid{t}\;\Varid{x}} which, via \ensuremath{\Varid{step}}, brings \ensuremath{\Varid{x}} into a state which is
viable \ensuremath{\Varid{n}} steps (\ensuremath{\Varid{viableSpec1}} and \ensuremath{\Varid{viableSpec2}}):
%
\begin{hscode}\SaveRestoreHook
\column{B}{@{}>{\hspre}l<{\hspost}@{}}%
\column{3}{@{}>{\hspre}l<{\hspost}@{}}%
\column{16}{@{}>{\hspre}c<{\hspost}@{}}%
\column{16E}{@{}l@{}}%
\column{19}{@{}>{\hspre}l<{\hspost}@{}}%
\column{E}{@{}>{\hspre}l<{\hspost}@{}}%
\>[3]{}\Varid{viable}{}\<[16]%
\>[16]{}\ \mathop{:}\ {}\<[16E]%
\>[19]{}(\Varid{n}\ \mathop{:}\ \mathbb{N})\ \to\ \Conid{State}\;\Varid{t}\ \to\ \Conid{Bool}{}\<[E]%
\\
\>[3]{}\Varid{viableSpec0}{}\<[16]%
\>[16]{}\ \mathop{:}\ {}\<[16E]%
\>[19]{}(\Varid{x}\ \mathop{:}\ \Conid{State}\;\Varid{t})\ \to\ \Conid{Viable}\;\Conid{Z}\;\Varid{x}{}\<[E]%
\\
\>[3]{}\Varid{viableSpec1}{}\<[16]%
\>[16]{}\ \mathop{:}\ {}\<[16E]%
\>[19]{}(\Varid{x}\ \mathop{:}\ \Conid{State}\;\Varid{t})\ \to\ \Conid{Viable}\;(\Conid{S}\;\Varid{n})\;\Varid{x}\ \to\ \Conid{GoodCtrl}\;\Varid{t}\;\Varid{n}\;\Varid{x}{}\<[E]%
\\
\>[3]{}\Varid{viableSpec2}{}\<[16]%
\>[16]{}\ \mathop{:}\ {}\<[16E]%
\>[19]{}(\Varid{x}\ \mathop{:}\ \Conid{State}\;\Varid{t})\ \to\ \Conid{GoodCtrl}\;\Varid{t}\;\Varid{n}\;\Varid{x}\ \to\ \Conid{Viable}\;(\Conid{S}\;\Varid{n})\;\Varid{x}{}\<[E]%
\ColumnHook
\end{hscode}\resethooks
In the above specifications we have introduced \ensuremath{\Conid{Viable}\;\Varid{n}\;\Varid{x}} as a
shorthand for \ensuremath{\Conid{So}\;(\Varid{viable}\;\Varid{n}\;\Varid{x})}. In \ensuremath{\Varid{viableSpec1}} and \ensuremath{\Varid{viableSpec2}} we
use subsets implemented as dependent pairs:
\begin{hscode}\SaveRestoreHook
\column{B}{@{}>{\hspre}l<{\hspost}@{}}%
\column{3}{@{}>{\hspre}l<{\hspost}@{}}%
\column{E}{@{}>{\hspre}l<{\hspost}@{}}%
\>[3]{}\Conid{GoodCtrl}\ \mathop{:}\ (\Varid{t}\ \mathop{:}\ \mathbb{N})\ \to\ (\Varid{n}\ \mathop{:}\ \mathbb{N})\ \to\ \Conid{State}\;\Varid{t}\ \to\ \Conid{Type}{}\<[E]%
\\
\>[3]{}\Conid{GoodCtrl}\;\Varid{t}\;\Varid{n}\;\Varid{x}\mathrel{=}(\Varid{y}\ \mathop{:}\ \Conid{Ctrl}\;\Varid{t}\;\Varid{x}\;*\!\!*\;\Conid{Viable}\;\Varid{n}\;(\Varid{step}\;\Varid{t}\;\Varid{x}\;\Varid{y})){}\<[E]%
\ColumnHook
\end{hscode}\resethooks
These are pairs in which the type of the second element can depend on
the value of the first one. The notation \ensuremath{\Varid{p}\ \mathop{:}\ (\Varid{a}\ \mathop{:}\ \Conid{A}\;*\!\!*\;\Conid{P}\;\Varid{a})}
represents a pair in which the type (\ensuremath{\Conid{P}\;\Varid{a}}) of the second element can
refer to the value (\ensuremath{\Varid{a}}) of the first element, giving a kind of
existential quantification~\cite{idristutorial}. The projection
functions are
\begin{hscode}\SaveRestoreHook
\column{B}{@{}>{\hspre}l<{\hspost}@{}}%
\column{3}{@{}>{\hspre}l<{\hspost}@{}}%
\column{9}{@{}>{\hspre}c<{\hspost}@{}}%
\column{9E}{@{}l@{}}%
\column{12}{@{}>{\hspre}l<{\hspost}@{}}%
\column{E}{@{}>{\hspre}l<{\hspost}@{}}%
\>[3]{}\Varid{outl}{}\<[9]%
\>[9]{}\ \mathop{:}\ {}\<[9E]%
\>[12]{}\{\mskip1.5mu \Conid{A}\ \mathop{:}\ \Conid{Type}\mskip1.5mu\}\ \to\ \{\mskip1.5mu \Conid{P}\ \mathop{:}\ \Conid{A}\ \to\ \Conid{Type}\mskip1.5mu\}\ \to\ (\Varid{a}\ \mathop{:}\ \Conid{A}\;*\!\!*\;\Conid{P}\;\Varid{a})\ \to\ \Conid{A}{}\<[E]%
\\
\>[3]{}\Varid{outr}{}\<[9]%
\>[9]{}\ \mathop{:}\ {}\<[9E]%
\>[12]{}\{\mskip1.5mu \Conid{A}\ \mathop{:}\ \Conid{Type}\mskip1.5mu\}\ \to\ \{\mskip1.5mu \Conid{P}\ \mathop{:}\ \Conid{A}\ \to\ \Conid{Type}\mskip1.5mu\}\ \to\ (\Varid{p}\ \mathop{:}\ (\Varid{a}\ \mathop{:}\ \Conid{A}\;*\!\!*\;\Conid{P}\;\Varid{a}))\ \to\ \Conid{P}\;(\Varid{outl}\;\Varid{p}){}\<[E]%
\ColumnHook
\end{hscode}\resethooks
Thus, in general \ensuremath{(\Varid{a}\ \mathop{:}\ \Conid{A}\;*\!\!*\;\Conid{P}\;\Varid{a})} effectively represents the subset of
\ensuremath{\Conid{A}} whose elements fulfill \ensuremath{\Conid{P}} and our case \ensuremath{\Conid{GoodCtrl}\;\Varid{t}\;\Varid{n}\;\Varid{x}} is the subset
of controls available in \ensuremath{\Varid{x}} at step \ensuremath{\Varid{t}} which lead to next states
which are viable \ensuremath{\Varid{n}} steps.

The declarations of \ensuremath{\Varid{viableSpec0}}, \ensuremath{\Varid{viableSpec1}} and \ensuremath{\Varid{viableSpec2}} are
added to the context.
In implementing an instance of a specific sequential decision problem,
clients are required to define \ensuremath{\Conid{State}}, \ensuremath{\Conid{Ctrl}}, \ensuremath{\Varid{step}}, \ensuremath{\Varid{reward}} and the
\ensuremath{\Varid{viable}} predicate for that problem. In doing so, they have to prove
(or postulate) that their definitions satisfy the above
specifications.

\subsection{Control sequences}

With the notion of viability in place, we can readily extend the notion
of control sequences of section \ref{section:base_case} to the
time-dependent case:
\begin{hscode}\SaveRestoreHook
\column{B}{@{}>{\hspre}l<{\hspost}@{}}%
\column{3}{@{}>{\hspre}l<{\hspost}@{}}%
\column{5}{@{}>{\hspre}l<{\hspost}@{}}%
\column{11}{@{}>{\hspre}c<{\hspost}@{}}%
\column{11E}{@{}l@{}}%
\column{14}{@{}>{\hspre}l<{\hspost}@{}}%
\column{E}{@{}>{\hspre}l<{\hspost}@{}}%
\>[3]{}\mathbf{data}\;\Conid{CtrlSeq}\ \mathop{:}\ (\Varid{x}\ \mathop{:}\ \Conid{State}\;\Varid{t})\ \to\ (\Varid{n}\ \mathop{:}\ \mathbb{N})\ \to\ \Conid{Type}\;\mathbf{where}{}\<[E]%
\\
\>[3]{}\hsindent{2}{}\<[5]%
\>[5]{}\Conid{Nil}{}\<[11]%
\>[11]{}\ \mathop{:}\ {}\<[11E]%
\>[14]{}\Conid{CtrlSeq}\;\Varid{x}\;\Conid{Z}{}\<[E]%
\\
\>[3]{}\hsindent{2}{}\<[5]%
\>[5]{}(\mathbin{::}){}\<[11]%
\>[11]{}\ \mathop{:}\ {}\<[11E]%
\>[14]{}(\Varid{yv}\ \mathop{:}\ \Conid{GoodCtrl}\;\Varid{t}\;\Varid{n}\;\Varid{x})\ \to\ \Conid{CtrlSeq}\;(\Varid{step}\;\Varid{t}\;\Varid{x}\;(\Varid{outl}\;\Varid{yv}))\;\Varid{n}\ \to\ \Conid{CtrlSeq}\;\Varid{x}\;(\Conid{S}\;\Varid{n}){}\<[E]%
\ColumnHook
\end{hscode}\resethooks
\noindent
Notice that now the constructor \ensuremath{\mathbin{::}} (for constructing control
sequences of length \ensuremath{\Conid{S}\;\Varid{n}}) can only be applied to those (implicit !)
\ensuremath{\Varid{x}\ \mathop{:}\ \Conid{State}\;\Varid{t}} for which there exists a ``good'' control \ensuremath{\Varid{y}\mathrel{=}\Varid{outl}\;\Varid{yv}\ \mathop{:}\ \Conid{Ctrl}\;\Varid{t}\;\Varid{x}} such that the
new state \ensuremath{\Varid{step}\;\Varid{t}\;\Varid{x}\;\Varid{y}} is viable \ensuremath{\Varid{n}} steps. The specification
\ensuremath{\Varid{viableSpec2}} ensures us that, in this case, \ensuremath{\Varid{x}} is viable \ensuremath{\Conid{S}\;\Varid{n}}
steps.

The extension of \ensuremath{\Varid{val}}, \ensuremath{\Conid{OptCtrlSeq}} and the proof of optimality of
empty sequences of controls are, as one would expect, straightforward:
\begin{hscode}\SaveRestoreHook
\column{B}{@{}>{\hspre}l<{\hspost}@{}}%
\column{3}{@{}>{\hspre}l<{\hspost}@{}}%
\column{5}{@{}>{\hspre}l<{\hspost}@{}}%
\column{8}{@{}>{\hspre}c<{\hspost}@{}}%
\column{8E}{@{}l@{}}%
\column{9}{@{}>{\hspre}c<{\hspost}@{}}%
\column{9E}{@{}l@{}}%
\column{11}{@{}>{\hspre}l<{\hspost}@{}}%
\column{12}{@{}>{\hspre}l<{\hspost}@{}}%
\column{16}{@{}>{\hspre}l<{\hspost}@{}}%
\column{23}{@{}>{\hspre}l<{\hspost}@{}}%
\column{35}{@{}>{\hspre}c<{\hspost}@{}}%
\column{35E}{@{}l@{}}%
\column{38}{@{}>{\hspre}l<{\hspost}@{}}%
\column{39}{@{}>{\hspre}c<{\hspost}@{}}%
\column{39E}{@{}l@{}}%
\column{42}{@{}>{\hspre}l<{\hspost}@{}}%
\column{E}{@{}>{\hspre}l<{\hspost}@{}}%
\>[3]{}\Varid{val}{}\<[8]%
\>[8]{}\ \mathop{:}\ {}\<[8E]%
\>[11]{}(\Varid{x}\ \mathop{:}\ \Conid{State}\;\Varid{t})\ \to\ (\Varid{n}\ \mathop{:}\ \mathbb{N})\ \to\ \Conid{CtrlSeq}\;\Varid{x}\;\Varid{n}\ \to\ \mathbb{R}{}\<[E]%
\\
\>[3]{}\Varid{val}\;{}\<[12]%
\>[12]{}\anonymous \;{}\<[16]%
\>[16]{}\Conid{Z}\;{}\<[23]%
\>[23]{}\anonymous {}\<[35]%
\>[35]{}\mathrel{=}{}\<[35E]%
\>[38]{}\mathrm{0}{}\<[E]%
\\
\>[3]{}\Varid{val}\;\{\mskip1.5mu \Varid{t}\mskip1.5mu\}\;{}\<[12]%
\>[12]{}\Varid{x}\;{}\<[16]%
\>[16]{}(\Conid{S}\;\Varid{n})\;{}\<[23]%
\>[23]{}(\Varid{yv}\mathbin{::}\Varid{ys}){}\<[35]%
\>[35]{}\mathrel{=}{}\<[35E]%
\>[38]{}\Varid{reward}\;\Varid{t}\;\Varid{x}\;\Varid{y}\;\Varid{x'}\mathbin{+}\Varid{val}\;\Varid{x'}\;\Varid{n}\;\Varid{ys}\;\mathbf{where}{}\<[E]%
\\
\>[3]{}\hsindent{2}{}\<[5]%
\>[5]{}\Varid{y}{}\<[9]%
\>[9]{}\ \mathop{:}\ {}\<[9E]%
\>[12]{}\Conid{Ctrl}\;\Varid{t}\;\Varid{x};{}\<[35]%
\>[35]{}\Varid{y}{}\<[35E]%
\>[39]{}\mathrel{=}{}\<[39E]%
\>[42]{}\Varid{outl}\;\Varid{yv}{}\<[E]%
\\
\>[3]{}\hsindent{2}{}\<[5]%
\>[5]{}\Varid{x'}{}\<[9]%
\>[9]{}\ \mathop{:}\ {}\<[9E]%
\>[12]{}\Conid{State}\;(\Conid{S}\;\Varid{t});~~{}\<[35]%
\>[35]{}\Varid{x'}{}\<[35E]%
\>[39]{}\mathrel{=}{}\<[39E]%
\>[42]{}\Varid{step}\;\Varid{t}\;\Varid{x}\;\Varid{y}{}\<[E]%
\ColumnHook
\end{hscode}\resethooks
\begin{hscode}\SaveRestoreHook
\column{B}{@{}>{\hspre}l<{\hspost}@{}}%
\column{3}{@{}>{\hspre}l<{\hspost}@{}}%
\column{E}{@{}>{\hspre}l<{\hspost}@{}}%
\>[3]{}\Conid{OptCtrlSeq}\ \mathop{:}\ (\Varid{x}\ \mathop{:}\ \Conid{State}\;\Varid{t})\ \to\ (\Varid{n}\ \mathop{:}\ \mathbb{N})\ \to\ \Conid{CtrlSeq}\;\Varid{x}\;\Varid{n}\ \to\ \Conid{Type}{}\<[E]%
\\
\>[3]{}\Conid{OptCtrlSeq}\;\Varid{x}\;\Varid{n}\;\Varid{ys}\mathrel{=}(\Varid{ys'}\ \mathop{:}\ \Conid{CtrlSeq}\;\Varid{x}\;\Varid{n})\ \to\ \Conid{So}\;(\Varid{val}\;\Varid{x}\;\Varid{n}\;\Varid{ys'}\leq \Varid{val}\;\Varid{x}\;\Varid{n}\;\Varid{ys}){}\<[E]%
\ColumnHook
\end{hscode}\resethooks
\begin{hscode}\SaveRestoreHook
\column{B}{@{}>{\hspre}l<{\hspost}@{}}%
\column{3}{@{}>{\hspre}l<{\hspost}@{}}%
\column{26}{@{}>{\hspre}c<{\hspost}@{}}%
\column{26E}{@{}l@{}}%
\column{29}{@{}>{\hspre}l<{\hspost}@{}}%
\column{E}{@{}>{\hspre}l<{\hspost}@{}}%
\>[3]{}\Varid{nilIsOptCtrlSeq}{}\<[26]%
\>[26]{}\ \mathop{:}\ {}\<[26E]%
\>[29]{}(\Varid{x}\ \mathop{:}\ \Conid{State}\;\Varid{t})\ \to\ \Conid{OptCtrlSeq}\;\Varid{x}\;\Conid{Z}\;\Conid{Nil}{}\<[E]%
\\
\>[3]{}\Varid{nilIsOptCtrlSeq}\;\Varid{x}\;\Conid{Nil}{}\<[26]%
\>[26]{}\mathrel{=}{}\<[26E]%
\>[29]{}\Varid{reflexive\char95 Double\char95 lte}\;\mathrm{0}{}\<[E]%
\ColumnHook
\end{hscode}\resethooks

\subsection{Reachability, policy sequences}

In the time-independent case, policies are functions of type \ensuremath{(\Varid{x}\ \mathop{:}\ \Conid{State})\ \to\ \Conid{Ctrl}\;\Varid{x}} and policy sequences are vectors of elements of that type. Given a
policy sequence \ensuremath{\Varid{ps}} and an initial state \ensuremath{\Varid{x}}, one can construct its
corresponding sequence of controls by \ensuremath{\Varid{ctrls}\;\Varid{x}\;\Varid{ps}}:
\begin{hscode}\SaveRestoreHook
\column{B}{@{}>{\hspre}l<{\hspost}@{}}%
\column{3}{@{}>{\hspre}l<{\hspost}@{}}%
\column{21}{@{}>{\hspre}l<{\hspost}@{}}%
\column{E}{@{}>{\hspre}l<{\hspost}@{}}%
\>[3]{}\Varid{ctrl}\ \mathop{:}\ (\Varid{x}\ \mathop{:}\ \Conid{State})\ \to\ \Conid{PolicySeq}\;\Varid{n}\ \to\ \Conid{CtrlSeq}\;\Varid{x}\;\Varid{n}{}\<[E]%
\\
\>[3]{}\Varid{ctrl}\;\Varid{x}\;\Conid{Nil}{}\<[21]%
\>[21]{}\mathrel{=}\Conid{Nil}{}\<[E]%
\\
\>[3]{}\Varid{ctrl}\;\Varid{x}\;(\Varid{p}\mathbin{::}\Varid{ps}){}\<[21]%
\>[21]{}\mathrel{=}\Varid{p}\;\Varid{x}\mathbin{::}\Varid{ctrl}\;(\Varid{step}\;\Varid{x}\;(\Varid{p}\;\Varid{x}))\;\Varid{ps}{}\<[E]%
\ColumnHook
\end{hscode}\resethooks
Thus \ensuremath{\Varid{p}\;\Varid{x}} is of type \ensuremath{\Conid{Ctrl}\;\Varid{x}} which is, in turn, the type of the first
(explicit) argument of the ``cons'' constructor of \ensuremath{\Conid{CtrlSeq}\;\Varid{x}\;\Varid{n}}, see
section \ref{section:base_case}.

As seen above, in the time-dependent case the ``cons'' constructor of
\ensuremath{\Conid{CtrlSeq}\;\Varid{x}\;\Varid{n}} takes as first argument dependent pairs of type \ensuremath{\Conid{GoodCtrl}\;\Varid{t}\;\Varid{n}\;\Varid{x}}.  For sequences of controls to be construable from policy
sequences, policies have to return, in the time-dependent case, values
of this type. Thus, what we want to formalize in the time-dependent case
is the notion of a correspondence between states and sets of controls
that, at a given step \ensuremath{\Varid{t}}, allows us to make a given number of decision
steps \ensuremath{\Varid{n}}. Because of \ensuremath{\Varid{viableSpec1}} we know that such controls exist for
a given \ensuremath{\Varid{x}\ \mathop{:}\ \Conid{State}\;\Varid{t}} if and only if it is viable at least \ensuremath{\Varid{n}} steps. We
use such a requirement to restrict the domain of policies
\begin{hscode}\SaveRestoreHook
\column{B}{@{}>{\hspre}l<{\hspost}@{}}%
\column{3}{@{}>{\hspre}l<{\hspost}@{}}%
\column{19}{@{}>{\hspre}c<{\hspost}@{}}%
\column{19E}{@{}l@{}}%
\column{22}{@{}>{\hspre}l<{\hspost}@{}}%
\column{E}{@{}>{\hspre}l<{\hspost}@{}}%
\>[3]{}\Conid{Policy}\ \mathop{:}\ \mathbb{N}\ \to\ \mathbb{N}\ \to\ \Conid{Type}{}\<[E]%
\\
\>[3]{}\Conid{Policy}\;\Varid{t}\;\Conid{Z}{}\<[19]%
\>[19]{}\mathrel{=}{}\<[19E]%
\>[22]{}\Conid{Unit}{}\<[E]%
\\
\>[3]{}\Conid{Policy}\;\Varid{t}\;(\Conid{S}\;\Varid{n}){}\<[19]%
\>[19]{}\mathrel{=}{}\<[19E]%
\>[22]{}(\Varid{x}\ \mathop{:}\ \Conid{State}\;\Varid{t})\ \to\ \Conid{Viable}\;(\Conid{S}\;\Varid{n})\;\Varid{x}\ \to\ \Conid{GoodCtrl}\;\Varid{t}\;\Varid{n}\;\Varid{x}{}\<[E]%
\ColumnHook
\end{hscode}\resethooks
Here, \ensuremath{\Conid{Unit}} is the singleton type. It is inhabited by a single value
called \ensuremath{()}. Logically, \ensuremath{\Conid{Policy}} states that policies that supports zero
decision steps are trivial values. Policies that support \ensuremath{\Conid{S}\;\Varid{n}} decision
steps starting from states at time \ensuremath{\Varid{t}}, however, are functions that map
states at time \ensuremath{\Varid{t}} which are reachable and viable for \ensuremath{\Conid{S}\;\Varid{n}} steps to
controls leading to states (at time \ensuremath{\Conid{S}\;\Varid{t}}) which are viable for \ensuremath{\Varid{n}}
steps.

Let us go back to the right-hand side of figure \ref{figure:two}. At a
given step, there might be states which are valid but which cannot be
reached. It could be a waste of computational resources to consider such
states, e.g., when constructing optimal extensions inside a backwards
induction step.

We can compute optimal policy sequences more efficiently if we restrict
the domain of our policies to those states which can actually be reached
from the initial states.
We can do this by introducing the notion of \emph{reachability}. We say
that every initial state is reachable (\ensuremath{\Varid{reachableSpec0}}) and that if a state \ensuremath{\Varid{x}\ \mathop{:}\ \Conid{State}\;\Varid{t}} is
reachable, then every control \ensuremath{\Varid{y}\ \mathop{:}\ \Conid{Ctrl}\;\Varid{t}\;\Varid{x}} leads, via \ensuremath{\Varid{step}}, to a reachable
state in \ensuremath{\Conid{State}\;(\Conid{S}\;\Varid{t})} (see \ensuremath{\Varid{reachableSpec1}}). Conversely, if a state \ensuremath{\Varid{x'}\ \mathop{:}\ \Conid{State}\;(\Conid{S}\;\Varid{t})} is
reachable then there exist a state \ensuremath{\Varid{x}\ \mathop{:}\ \Conid{State}\;\Varid{t}} and a control \ensuremath{\Varid{y}\ \mathop{:}\ \Conid{Ctrl}\;\Varid{t}\;\Varid{x}}
such that \ensuremath{\Varid{x}} is reachable and \ensuremath{\Varid{x'}} is equal to \ensuremath{\Varid{step}\;\Varid{t}\;\Varid{x}\;\Varid{y}} (see \ensuremath{\Varid{reachableSpec2}}):
\begin{hscode}\SaveRestoreHook
\column{B}{@{}>{\hspre}l<{\hspost}@{}}%
\column{3}{@{}>{\hspre}l<{\hspost}@{}}%
\column{19}{@{}>{\hspre}c<{\hspost}@{}}%
\column{19E}{@{}l@{}}%
\column{22}{@{}>{\hspre}l<{\hspost}@{}}%
\column{E}{@{}>{\hspre}l<{\hspost}@{}}%
\>[3]{}\Varid{reachable}{}\<[19]%
\>[19]{}\ \mathop{:}\ {}\<[19E]%
\>[22]{}\Conid{State}\;\Varid{t}\ \to\ \Conid{Bool}{}\<[E]%
\\
\>[3]{}\Varid{reachableSpec0}{}\<[19]%
\>[19]{}\ \mathop{:}\ {}\<[19E]%
\>[22]{}(\Varid{x}\ \mathop{:}\ \Conid{State}\;\Conid{Z})\ \to\ \Conid{Reachable}\;\Varid{x}{}\<[E]%
\\
\>[3]{}\Varid{reachableSpec1}{}\<[19]%
\>[19]{}\ \mathop{:}\ {}\<[19E]%
\>[22]{}(\Varid{x}\ \mathop{:}\ \Conid{State}\;\Varid{t})\ \to\ \Conid{Reachable}\;\Varid{x}\ \to\ (\Varid{y}\ \mathop{:}\ \Conid{Ctrl}\;\Varid{t}\;\Varid{x})\ \to\ {}\<[E]%
\\
\>[22]{}\Conid{Reachable}\;(\Varid{step}\;\Varid{t}\;\Varid{x}\;\Varid{y}){}\<[E]%
\\
\>[3]{}\Varid{reachableSpec2}{}\<[19]%
\>[19]{}\ \mathop{:}\ {}\<[19E]%
\>[22]{}(\Varid{x'}\ \mathop{:}\ \Conid{State}\;(\Conid{S}\;\Varid{t}))\ \to\ \Conid{Reachable}\;\Varid{x'}\ \to\ {}\<[E]%
\\
\>[22]{}(\Varid{x}\ \mathop{:}\ \Conid{State}\;\Varid{t}\;*\!\!*\;(\Conid{Reachable}\;\Varid{x},(\Varid{y}\ \mathop{:}\ \Conid{Ctrl}\;\Varid{t}\;\Varid{x}\;*\!\!*\;\Varid{x'}\mathrel{=}\Varid{step}\;\Varid{t}\;\Varid{x}\;\Varid{y}))){}\<[E]%
\ColumnHook
\end{hscode}\resethooks
As for viability, we have introduced \ensuremath{\Conid{Reachable}\;\Varid{x}} as a shorthand for
\ensuremath{\Conid{So}\;(\Varid{reachable}\;\Varid{x})} in the specification of \ensuremath{\Varid{reachable}}. We can now apply
reachability to refine the notion of policy
\begin{hscode}\SaveRestoreHook
\column{B}{@{}>{\hspre}l<{\hspost}@{}}%
\column{3}{@{}>{\hspre}l<{\hspost}@{}}%
\column{19}{@{}>{\hspre}c<{\hspost}@{}}%
\column{19E}{@{}l@{}}%
\column{22}{@{}>{\hspre}l<{\hspost}@{}}%
\column{E}{@{}>{\hspre}l<{\hspost}@{}}%
\>[3]{}\Conid{Policy}\ \mathop{:}\ \mathbb{N}\ \to\ \mathbb{N}\ \to\ \Conid{Type}{}\<[E]%
\\
\>[3]{}\Conid{Policy}\;\Varid{t}\;\Conid{Z}{}\<[19]%
\>[19]{}\mathrel{=}{}\<[19E]%
\>[22]{}\Conid{Unit}{}\<[E]%
\\
\>[3]{}\Conid{Policy}\;\Varid{t}\;(\Conid{S}\;\Varid{n}){}\<[19]%
\>[19]{}\mathrel{=}{}\<[19E]%
\>[22]{}(\Varid{x}\ \mathop{:}\ \Conid{State}\;\Varid{t})\ \to\ \Conid{Reachable}\;\Varid{x}\ \to\ \Conid{Viable}\;(\Conid{S}\;\Varid{n})\;\Varid{x}\ \to\ \Conid{GoodCtrl}\;\Varid{t}\;\Varid{n}\;\Varid{x}{}\<[E]%
\ColumnHook
\end{hscode}\resethooks
and policy sequences:
\begin{hscode}\SaveRestoreHook
\column{B}{@{}>{\hspre}l<{\hspost}@{}}%
\column{3}{@{}>{\hspre}l<{\hspost}@{}}%
\column{5}{@{}>{\hspre}l<{\hspost}@{}}%
\column{11}{@{}>{\hspre}c<{\hspost}@{}}%
\column{11E}{@{}l@{}}%
\column{14}{@{}>{\hspre}l<{\hspost}@{}}%
\column{E}{@{}>{\hspre}l<{\hspost}@{}}%
\>[3]{}\mathbf{data}\;\Conid{PolicySeq}\ \mathop{:}\ \mathbb{N}\ \to\ \mathbb{N}\ \to\ \Conid{Type}\;\mathbf{where}{}\<[E]%
\\
\>[3]{}\hsindent{2}{}\<[5]%
\>[5]{}\Conid{Nil}{}\<[11]%
\>[11]{}\ \mathop{:}\ {}\<[11E]%
\>[14]{}\Conid{PolicySeq}\;\Varid{t}\;\Conid{Z}{}\<[E]%
\\
\>[3]{}\hsindent{2}{}\<[5]%
\>[5]{}(\mathbin{::}){}\<[11]%
\>[11]{}\ \mathop{:}\ {}\<[11E]%
\>[14]{}\Conid{Policy}\;\Varid{t}\;(\Conid{S}\;\Varid{n})\ \to\ \Conid{PolicySeq}\;(\Conid{S}\;\Varid{t})\;\Varid{n}\ \to\ \Conid{PolicySeq}\;\Varid{t}\;(\Conid{S}\;\Varid{n}){}\<[E]%
\ColumnHook
\end{hscode}\resethooks
In contrast to the time-independent case, \ensuremath{\Conid{PolicySeq}} now takes an
additional \ensuremath{\mathbb{N}} argument. This represents the time (number of decision
steps, value of the decision steps counter) at which the first policy of
the sequence can be applied. The previous one-index idiom \ensuremath{(\Conid{S}\;\Varid{n})\;\cdots\;\Varid{n}\;\cdots\;(\Conid{S}\;\Varid{n})} becomes a two-index idiom: \ensuremath{\Varid{t}\;(\Conid{S}\;\Varid{n})\;\cdots\;(\Conid{S}\;\Varid{t})\;\Varid{n}\;\cdots\;\Varid{t}\;(\Conid{S}\;\Varid{n})}. The value function \ensuremath{\Varid{val}} maximized by optimal
policy sequences is:
\begin{hscode}\SaveRestoreHook
\column{B}{@{}>{\hspre}l<{\hspost}@{}}%
\column{3}{@{}>{\hspre}l<{\hspost}@{}}%
\column{5}{@{}>{\hspre}l<{\hspost}@{}}%
\column{8}{@{}>{\hspre}c<{\hspost}@{}}%
\column{8E}{@{}l@{}}%
\column{9}{@{}>{\hspre}c<{\hspost}@{}}%
\column{9E}{@{}l@{}}%
\column{10}{@{}>{\hspre}l<{\hspost}@{}}%
\column{11}{@{}>{\hspre}l<{\hspost}@{}}%
\column{12}{@{}>{\hspre}l<{\hspost}@{}}%
\column{17}{@{}>{\hspre}l<{\hspost}@{}}%
\column{34}{@{}>{\hspre}c<{\hspost}@{}}%
\column{34E}{@{}l@{}}%
\column{36}{@{}>{\hspre}l<{\hspost}@{}}%
\column{37}{@{}>{\hspre}l<{\hspost}@{}}%
\column{40}{@{}>{\hspre}c<{\hspost}@{}}%
\column{40E}{@{}l@{}}%
\column{43}{@{}>{\hspre}l<{\hspost}@{}}%
\column{E}{@{}>{\hspre}l<{\hspost}@{}}%
\>[3]{}\Varid{val}{}\<[8]%
\>[8]{}\ \mathop{:}\ {}\<[8E]%
\>[11]{}(\Varid{t}\ \mathop{:}\ \mathbb{N})\ \to\ (\Varid{n}\ \mathop{:}\ \mathbb{N})\ \to\ {}\<[E]%
\\
\>[11]{}(\Varid{x}\ \mathop{:}\ \Conid{State}\;\Varid{t})\ \to\ (\Varid{r}\ \mathop{:}\ \Conid{Reachable}\;\Varid{x})\ \to\ (\Varid{v}\ \mathop{:}\ \Conid{Viable}\;\Varid{n}\;\Varid{x})\ \to\ {}\<[E]%
\\
\>[11]{}\Conid{PolicySeq}\;\Varid{t}\;\Varid{n}\ \to\ \mathbb{R}{}\<[E]%
\\
\>[3]{}\Varid{val}\;\anonymous \;{}\<[10]%
\>[10]{}\Conid{Z}\;{}\<[17]%
\>[17]{}\anonymous \;\anonymous \;\anonymous \;\anonymous {}\<[34]%
\>[34]{}\mathrel{=}{}\<[34E]%
\>[37]{}\mathrm{0}{}\<[E]%
\\
\>[3]{}\Varid{val}\;\Varid{t}\;{}\<[10]%
\>[10]{}(\Conid{S}\;\Varid{n})\;{}\<[17]%
\>[17]{}\Varid{x}\;\Varid{r}\;\Varid{v}\;(\Varid{p}\mathbin{::}\Varid{ps}){}\<[34]%
\>[34]{}\mathrel{=}{}\<[34E]%
\>[37]{}\Varid{reward}\;\Varid{t}\;\Varid{x}\;\Varid{y}\;\Varid{x'}\mathbin{+}\Varid{val}\;(\Conid{S}\;\Varid{t})\;\Varid{n}\;\Varid{x'}\;\Varid{r'}\;\Varid{v'}\;\Varid{ps}\;\mathbf{where}{}\<[E]%
\\
\>[3]{}\hsindent{2}{}\<[5]%
\>[5]{}\Varid{y}{}\<[9]%
\>[9]{}\ \mathop{:}\ {}\<[9E]%
\>[12]{}\Conid{Ctrl}\;\Varid{t}\;\Varid{x};~~{}\<[36]%
\>[36]{}\Varid{y}{}\<[40]%
\>[40]{}\mathrel{=}{}\<[40E]%
\>[43]{}\Varid{outl}\;(\Varid{p}\;\Varid{x}\;\Varid{r}\;\Varid{v}){}\<[E]%
\\
\>[3]{}\hsindent{2}{}\<[5]%
\>[5]{}\Varid{x'}{}\<[9]%
\>[9]{}\ \mathop{:}\ {}\<[9E]%
\>[12]{}\Conid{State}\;(\Conid{S}\;\Varid{t});~~{}\<[36]%
\>[36]{}\Varid{x'}{}\<[40]%
\>[40]{}\mathrel{=}{}\<[40E]%
\>[43]{}\Varid{step}\;\Varid{t}\;\Varid{x}\;\Varid{y}{}\<[E]%
\\
\>[3]{}\hsindent{2}{}\<[5]%
\>[5]{}\Varid{r'}{}\<[9]%
\>[9]{}\ \mathop{:}\ {}\<[9E]%
\>[12]{}\Conid{Reachable}\;\Varid{x'};~~{}\<[36]%
\>[36]{}\Varid{r'}{}\<[40]%
\>[40]{}\mathrel{=}{}\<[40E]%
\>[43]{}\Varid{reachableSpec1}\;\Varid{x}\;\Varid{r}\;\Varid{y}{}\<[E]%
\\
\>[3]{}\hsindent{2}{}\<[5]%
\>[5]{}\Varid{v'}{}\<[9]%
\>[9]{}\ \mathop{:}\ {}\<[9E]%
\>[12]{}\Conid{Viable}\;\Varid{n}\;\Varid{x'};~~{}\<[36]%
\>[36]{}\Varid{v'}{}\<[40]%
\>[40]{}\mathrel{=}{}\<[40E]%
\>[43]{}\Varid{outr}\;(\Varid{p}\;\Varid{x}\;\Varid{r}\;\Varid{v}){}\<[E]%
\ColumnHook
\end{hscode}\resethooks

\subsection{The full framework}
\label{subsection:full_framework}

With these notions of viability, control sequence, reachability, policy and
policy sequence, the previous~\cite{botta+al2013c} formal framework for
time-independent sequential decision problems can be easily extended to the
time-dependent case.

The notions of optimality of policy sequences, optimal extension of
policy sequences, Bellman's principle of optimality, the generic
implementation of backwards induction and its machine-checkable
correctness can all be derived almost automatically from the
time-independent case.

We do not present the complete framework here (but it is available on GitHub). To give an idea of the
differences between the time-dependent and the time-independent cases,
we compare the proofs of Bellman's principle of optimality.

Consider, first, the time-independent case. As explained in section
\ref{section:base_case}, Bellman's principle of optimality says that if
\ensuremath{\Varid{ps}} is an optimal policy sequence of length \ensuremath{\Varid{n}} and \ensuremath{\Varid{p}} is an optimal
extension of \ensuremath{\Varid{ps}} then \ensuremath{\Varid{p}\mathbin{::}\Varid{ps}} is an optimal policy sequence of length
\ensuremath{\Conid{S}\;\Varid{n}}. In the time-dependent case we need as additional argument the
current number of decision steps \ensuremath{\Varid{t}} and we make the length of the
policy sequence \ensuremath{\Varid{n}} explicit. The other arguments are, as in the
time-independent case, a policy sequence \ensuremath{\Varid{ps}}, a proof of optimality of
\ensuremath{\Varid{ps}}, a policy \ensuremath{\Varid{p}} and a proof that \ensuremath{\Varid{p}} is an optimal extension of \ensuremath{\Varid{ps}}:
\begin{hscode}\SaveRestoreHook
\column{B}{@{}>{\hspre}l<{\hspost}@{}}%
\column{3}{@{}>{\hspre}l<{\hspost}@{}}%
\column{12}{@{}>{\hspre}c<{\hspost}@{}}%
\column{12E}{@{}l@{}}%
\column{15}{@{}>{\hspre}l<{\hspost}@{}}%
\column{42}{@{}>{\hspre}l<{\hspost}@{}}%
\column{70}{@{}>{\hspre}c<{\hspost}@{}}%
\column{70E}{@{}l@{}}%
\column{E}{@{}>{\hspre}l<{\hspost}@{}}%
\>[3]{}\Conid{Bellman}{}\<[12]%
\>[12]{}\ \mathop{:}\ {}\<[12E]%
\>[15]{}(\Varid{t}\ \mathop{:}\ \mathbb{N})\ \to\ (\Varid{n}\ \mathop{:}\ \mathbb{N})\ \to\ {}\<[E]%
\\
\>[15]{}(\Varid{ps}\ \mathop{:}\ \Conid{PolicySeq}\;(\Conid{S}\;\Varid{t})\;\Varid{n}){}\<[42]%
\>[42]{}\ \to\ \Conid{OptPolicySeq}\;(\Conid{S}\;\Varid{t})\;\Varid{n}\;\Varid{ps}{}\<[70]%
\>[70]{}\ \to\ {}\<[70E]%
\\
\>[15]{}(\Varid{p}\ \mathop{:}\ \Conid{Policy}\;\Varid{t}\;(\Conid{S}\;\Varid{n})){}\<[42]%
\>[42]{}\ \to\ \Conid{OptExt}\;\Varid{t}\;\Varid{n}\;\Varid{ps}\;\Varid{p}{}\<[70]%
\>[70]{}\ \to\ {}\<[70E]%
\\
\>[15]{}\Conid{OptPolicySeq}\;\Varid{t}\;(\Conid{S}\;\Varid{n})\;(\Varid{p}\mathbin{::}\Varid{ps}){}\<[E]%
\ColumnHook
\end{hscode}\resethooks
The result is a proof of optimality of \ensuremath{\Varid{p}\mathbin{::}\Varid{ps}}. Notice that the
types of the last 4 arguments of \ensuremath{\Conid{Bellman}} and the type of its result
now depend on \ensuremath{\Varid{t}}.

As discussed in \cite{botta+al2013c}, a proof of \ensuremath{\Conid{Bellman}} can be
derived easily. According to the notion of optimality for policy
sequences of section \ref{section:base_case}, one has to show that
\begin{hscode}\SaveRestoreHook
\column{B}{@{}>{\hspre}l<{\hspost}@{}}%
\column{3}{@{}>{\hspre}l<{\hspost}@{}}%
\column{E}{@{}>{\hspre}l<{\hspost}@{}}%
\>[3]{}\Varid{val}\;\Varid{t}\;(\Conid{S}\;\Varid{n})\;\Varid{x}\;\Varid{r}\;\Varid{v}\;(\Varid{p'}\mathbin{::}\Varid{ps'})\leq \Varid{val}\;\Varid{t}\;(\Conid{S}\;\Varid{n})\;\Varid{x}\;\Varid{r}\;\Varid{v}\;(\Varid{p}\mathbin{::}\Varid{ps}){}\<[E]%
\ColumnHook
\end{hscode}\resethooks
for arbitrary \ensuremath{\Varid{x}\ \mathop{:}\ \Conid{State}} and \ensuremath{(\Varid{p'}\mathbin{::}\Varid{ps'})\ \mathop{:}\ \Conid{PolicySeq}\;\Varid{t}\;(\Conid{S}\;\Varid{n})}. This is
straightforward. Let
\begin{hscode}\SaveRestoreHook
\column{B}{@{}>{\hspre}l<{\hspost}@{}}%
\column{3}{@{}>{\hspre}l<{\hspost}@{}}%
\column{7}{@{}>{\hspre}c<{\hspost}@{}}%
\column{7E}{@{}l@{}}%
\column{10}{@{}>{\hspre}l<{\hspost}@{}}%
\column{43}{@{}>{\hspre}l<{\hspost}@{}}%
\column{47}{@{}>{\hspre}c<{\hspost}@{}}%
\column{47E}{@{}l@{}}%
\column{50}{@{}>{\hspre}l<{\hspost}@{}}%
\column{E}{@{}>{\hspre}l<{\hspost}@{}}%
\>[3]{}\Varid{y}{}\<[7]%
\>[7]{}\mathrel{=}{}\<[7E]%
\>[10]{}\Varid{outl}\;(\Varid{p'}\;\Varid{x}\;\Varid{r}\;\Varid{v});{}\<[43]%
\>[43]{}\Varid{x'}{}\<[47]%
\>[47]{}\mathrel{=}{}\<[47E]%
\>[50]{}\Varid{step}\;\Varid{t}\;\Varid{x}\;\Varid{y};{}\<[E]%
\\
\>[3]{}\Varid{r'}{}\<[7]%
\>[7]{}\mathrel{=}{}\<[7E]%
\>[10]{}\Varid{reachableSpec1}\;\Varid{x}\;\Varid{r}\;\Varid{y};~~{}\<[43]%
\>[43]{}\Varid{v'}{}\<[47]%
\>[47]{}\mathrel{=}{}\<[47E]%
\>[50]{}\Varid{outr}\;(\Varid{p'}\;\Varid{x}\;\Varid{r}\;\Varid{v});{}\<[E]%
\ColumnHook
\end{hscode}\resethooks
then
\begin{hscode}\SaveRestoreHook
\column{B}{@{}>{\hspre}l<{\hspost}@{}}%
\column{3}{@{}>{\hspre}l<{\hspost}@{}}%
\column{5}{@{}>{\hspre}l<{\hspost}@{}}%
\column{E}{@{}>{\hspre}l<{\hspost}@{}}%
\>[3]{}\Varid{val}\;\Varid{t}\;(\Conid{S}\;\Varid{n})\;\Varid{x}\;\Varid{r}\;\Varid{v}\;(\Varid{p'}\mathbin{::}\Varid{ps'}){}\<[E]%
\\
\>[3]{}\hsindent{2}{}\<[5]%
\>[5]{}\mathrel{=}\mbox{\commentbegin  def.\ of \ensuremath{\Varid{val}}  \commentend}{}\<[E]%
\\
\>[3]{}\Varid{reward}\;\Varid{t}\;\Varid{x}\;\Varid{y}\;\Varid{x'}\mathbin{+}\Varid{val}\;(\Conid{S}\;\Varid{t})\;\Varid{n}\;\Varid{x'}\;\Varid{r'}\;\Varid{v'}\;\Varid{ps'}{}\<[E]%
\\
\>[3]{}\hsindent{2}{}\<[5]%
\>[5]{}\leq \mbox{\commentbegin  optimality of \ensuremath{\Varid{ps}}, monotonicity of \ensuremath{\mathbin{+}}  \commentend}{}\<[E]%
\\
\>[3]{}\Varid{reward}\;\Varid{t}\;\Varid{x}\;\Varid{y}\;\Varid{x'}\mathbin{+}\Varid{val}\;(\Conid{S}\;\Varid{t})\;\Varid{n}\;\Varid{x'}\;\Varid{r'}\;\Varid{v'}\;\Varid{ps}{}\<[E]%
\\
\>[3]{}\hsindent{2}{}\<[5]%
\>[5]{}\mathrel{=}\mbox{\commentbegin  def. of \ensuremath{\Varid{val}}  \commentend}{}\<[E]%
\\
\>[3]{}\Varid{val}\;\Varid{t}\;(\Conid{S}\;\Varid{n})\;\Varid{x}\;\Varid{r}\;\Varid{v}\;(\Varid{p'}\mathbin{::}\Varid{ps}){}\<[E]%
\\
\>[3]{}\hsindent{2}{}\<[5]%
\>[5]{}\leq \mbox{\commentbegin  \ensuremath{\Varid{p}} is an optimal extension of \ensuremath{\Varid{ps}}  \commentend}{}\<[E]%
\\
\>[3]{}\Varid{val}\;\Varid{t}\;(\Conid{S}\;\Varid{n})\;\Varid{x}\;\Varid{r}\;\Varid{v}\;(\Varid{p}\mathbin{::}\Varid{ps}){}\<[E]%
\ColumnHook
\end{hscode}\resethooks
We can turn the equational proof into an Idris proof:

\begin{hscode}\SaveRestoreHook
\column{B}{@{}>{\hspre}l<{\hspost}@{}}%
\column{3}{@{}>{\hspre}l<{\hspost}@{}}%
\column{5}{@{}>{\hspre}l<{\hspost}@{}}%
\column{7}{@{}>{\hspre}l<{\hspost}@{}}%
\column{9}{@{}>{\hspre}l<{\hspost}@{}}%
\column{11}{@{}>{\hspre}l<{\hspost}@{}}%
\column{18}{@{}>{\hspre}c<{\hspost}@{}}%
\column{18E}{@{}l@{}}%
\column{21}{@{}>{\hspre}l<{\hspost}@{}}%
\column{45}{@{}>{\hspre}l<{\hspost}@{}}%
\column{49}{@{}>{\hspre}c<{\hspost}@{}}%
\column{49E}{@{}l@{}}%
\column{52}{@{}>{\hspre}l<{\hspost}@{}}%
\column{E}{@{}>{\hspre}l<{\hspost}@{}}%
\>[3]{}\Conid{Bellman}\;\Varid{t}\;\Varid{n}\;\Varid{ps}\;\Varid{ops}\;\Varid{p}\;\Varid{oep}\mathrel{=}{}\<[E]%
\\
\>[3]{}\hsindent{2}{}\<[5]%
\>[5]{}\Varid{opps}\;\mathbf{where}{}\<[E]%
\\
\>[5]{}\hsindent{2}{}\<[7]%
\>[7]{}\Varid{opps}\ \mathop{:}\ \Conid{OptPolicySeq}\;\Varid{t}\;(\Conid{S}\;\Varid{n})\;(\Varid{p}\mathbin{::}\Varid{ps}){}\<[E]%
\\
\>[5]{}\hsindent{2}{}\<[7]%
\>[7]{}\Varid{opps}\;\Conid{Nil}\;\Varid{x}\;\Varid{r}\;\Varid{v}\;\mathbf{impossible}{}\<[E]%
\\
\>[5]{}\hsindent{2}{}\<[7]%
\>[7]{}\Varid{opps}\;(\Varid{p'}\mathbin{::}\Varid{ps'})\;\Varid{x}\;\Varid{r}\;\Varid{v}\mathrel{=}{}\<[E]%
\\
\>[7]{}\hsindent{2}{}\<[9]%
\>[9]{}\Varid{transitive\char95 Double\char95 lte}\;\Varid{step2}\;\Varid{step3}\;\mathbf{where}{}\<[E]%
\\
\>[9]{}\hsindent{2}{}\<[11]%
\>[11]{}\Varid{y}{}\<[18]%
\>[18]{}\ \mathop{:}\ {}\<[18E]%
\>[21]{}\Conid{Ctrl}\;\Varid{t}\;\Varid{x};{}\<[45]%
\>[45]{}\Varid{y}{}\<[49]%
\>[49]{}\mathrel{=}{}\<[49E]%
\>[52]{}\Varid{outl}\;(\Varid{p'}\;\Varid{x}\;\Varid{r}\;\Varid{v}){}\<[E]%
\\
\>[9]{}\hsindent{2}{}\<[11]%
\>[11]{}\Varid{x'}{}\<[18]%
\>[18]{}\ \mathop{:}\ {}\<[18E]%
\>[21]{}\Conid{State}\;(\Conid{S}\;\Varid{t});{}\<[45]%
\>[45]{}\Varid{x'}{}\<[49]%
\>[49]{}\mathrel{=}{}\<[49E]%
\>[52]{}\Varid{step}\;\Varid{t}\;\Varid{x}\;\Varid{y}{}\<[E]%
\\
\>[9]{}\hsindent{2}{}\<[11]%
\>[11]{}\Varid{r'}{}\<[18]%
\>[18]{}\ \mathop{:}\ {}\<[18E]%
\>[21]{}\Conid{Reachable}\;\Varid{x'};~~{}\<[45]%
\>[45]{}\Varid{r'}{}\<[49]%
\>[49]{}\mathrel{=}{}\<[49E]%
\>[52]{}\Varid{reachableSpec1}\;\Varid{x}\;\Varid{r}\;\Varid{y}{}\<[E]%
\\
\>[9]{}\hsindent{2}{}\<[11]%
\>[11]{}\Varid{v'}{}\<[18]%
\>[18]{}\ \mathop{:}\ {}\<[18E]%
\>[21]{}\Conid{Viable}\;\Varid{n}\;\Varid{x'};{}\<[45]%
\>[45]{}\Varid{v'}{}\<[49]%
\>[49]{}\mathrel{=}{}\<[49E]%
\>[52]{}\Varid{outr}\;(\Varid{p'}\;\Varid{x}\;\Varid{r}\;\Varid{v}){}\<[E]%
\\
\>[9]{}\hsindent{2}{}\<[11]%
\>[11]{}\Varid{step1}{}\<[18]%
\>[18]{}\ \mathop{:}\ {}\<[18E]%
\>[21]{}\Conid{So}\;(\Varid{val}\;(\Conid{S}\;\Varid{t})\;\Varid{n}\;\Varid{x'}\;\Varid{r'}\;\Varid{v'}\;\Varid{ps'}\leq \Varid{val}\;(\Conid{S}\;\Varid{t})\;\Varid{n}\;\Varid{x'}\;\Varid{r'}\;\Varid{v'}\;\Varid{ps}){}\<[E]%
\\
\>[9]{}\hsindent{2}{}\<[11]%
\>[11]{}\Varid{step1}{}\<[18]%
\>[18]{}\mathrel{=}{}\<[18E]%
\>[21]{}\Varid{ops}\;\Varid{ps'}\;\Varid{x'}\;\Varid{r'}\;\Varid{v'}{}\<[E]%
\\
\>[9]{}\hsindent{2}{}\<[11]%
\>[11]{}\Varid{step2}{}\<[18]%
\>[18]{}\ \mathop{:}\ {}\<[18E]%
\>[21]{}\Conid{So}\;(\Varid{val}\;\Varid{t}\;(\Conid{S}\;\Varid{n})\;\Varid{x}\;\Varid{r}\;\Varid{v}\;(\Varid{p'}\mathbin{::}\Varid{ps'})\leq \Varid{val}\;\Varid{t}\;(\Conid{S}\;\Varid{n})\;\Varid{x}\;\Varid{r}\;\Varid{v}\;(\Varid{p'}\mathbin{::}\Varid{ps})){}\<[E]%
\\
\>[9]{}\hsindent{2}{}\<[11]%
\>[11]{}\Varid{step2}{}\<[18]%
\>[18]{}\mathrel{=}{}\<[18E]%
\>[21]{}\Varid{monotone\char95 Double\char95 plus\char95 lte}\;(\Varid{reward}\;\Varid{t}\;\Varid{x}\;\Varid{y}\;\Varid{x'})\;\Varid{step1}{}\<[E]%
\\
\>[9]{}\hsindent{2}{}\<[11]%
\>[11]{}\Varid{step3}{}\<[18]%
\>[18]{}\ \mathop{:}\ {}\<[18E]%
\>[21]{}\Conid{So}\;(\Varid{val}\;\Varid{t}\;(\Conid{S}\;\Varid{n})\;\Varid{x}\;\Varid{r}\;\Varid{v}\;(\Varid{p'}\mathbin{::}\Varid{ps})\leq \Varid{val}\;\Varid{t}\;(\Conid{S}\;\Varid{n})\;\Varid{x}\;\Varid{r}\;\Varid{v}\;(\Varid{p}\mathbin{::}\Varid{ps})){}\<[E]%
\\
\>[9]{}\hsindent{2}{}\<[11]%
\>[11]{}\Varid{step3}{}\<[18]%
\>[18]{}\mathrel{=}{}\<[18E]%
\>[21]{}\Varid{oep}\;\Varid{p'}\;\Varid{x}\;\Varid{r}\;\Varid{v}{}\<[E]%
\ColumnHook
\end{hscode}\resethooks
Both the informal and the formal proof require only minor changes from the
proofs for the time-independent case presented in the previous
paper~\cite{botta+al2013c}.
\section{Monadic transition functions}
\label{section:monadic}

As explained in our previous paper~\cite{botta+al2013c}, many sequential
decision problems cannot be described in terms of a deterministic transition
function.

Even for physical systems which are believed to be governed by
deterministic laws, uncertainties might have to be taken into
account. They can arise because of different modelling options,
imperfectly known initial and boundary conditions and phenomenological
closures or through the choice of different approximate solution
methods.

In decision problems in climate impact research, financial markets,
and sports, for instance, uncertainties are the rule rather than the
exception. It would be blatantly unrealistic to assume that we can predict
the impact of, e.g., emission policies over a relevant time horizon in a
perfectly deterministic manner.
Even under the strongest rationality
assumptions -- each player perfectly knows how its cost and benefits
depend on its options and on the options of the other players and has
very strong reasons to assume that the other players enjoy the same kind
of knowledge -- errors, for instance ``fat-finger'' mistakes, can be made.

In systems which are not deterministic, the \emph{kind} of knowledge
which is available to a decision maker can be different in different cases.
Sometimes one is able to assess not only which states can
be obtained by selecting a given control in a given state but also
their probabilities. These systems are called \emph{stochastic}. In
other cases, the decision maker might know the possible outcomes
of a single decision but nothing more. The corresponding systems are
called \emph{non-deterministic}.

The notion of \emph{monadic} systems, originally introduced by
Ionescu~\cite{ionescu2009}, is a simple, yet powerful, way of treating
deterministic, non-deterministic, stochastic and other systems in a uniform
fashion.  It has been developed in the context of climate vulnerability
research, but can be applied to other systems as well. In a nutshell, the idea
is to generalize a generic transition function of type \ensuremath{\alpha\ \to\ \alpha} to
\ensuremath{\alpha\ \to\ \Conid{M}\;\alpha} where \ensuremath{\Conid{M}} is a monad.

For \ensuremath{\Conid{M}\mathrel{=}\Conid{Id}}, \ensuremath{\Conid{M}\mathrel{=}\Conid{List}} and \ensuremath{\Conid{M}\mathrel{=}\Conid{SimpleProb}}, one recovers the
deterministic, the non-deterministic and the stochastic cases. As in
\cite{ionescu2009}, we use \ensuremath{\Conid{SimpleProb}\;\alpha} to formalize the notion of
finite probability distributions (a probability distribution with finite
support) on \ensuremath{\alpha}, for instance:
\begin{hscode}\SaveRestoreHook
\column{B}{@{}>{\hspre}l<{\hspost}@{}}%
\column{3}{@{}>{\hspre}l<{\hspost}@{}}%
\column{5}{@{}>{\hspre}l<{\hspost}@{}}%
\column{19}{@{}>{\hspre}c<{\hspost}@{}}%
\column{19E}{@{}l@{}}%
\column{22}{@{}>{\hspre}l<{\hspost}@{}}%
\column{E}{@{}>{\hspre}l<{\hspost}@{}}%
\>[3]{}\mathbf{data}\;\Conid{SimpleProb}\ \mathop{:}\ \Conid{Type}\ \to\ \Conid{Type}\;\mathbf{where}{}\<[E]%
\\
\>[3]{}\hsindent{2}{}\<[5]%
\>[5]{}\Conid{MkSimpleProb}{}\<[19]%
\>[19]{}\ \mathop{:}\ {}\<[19E]%
\>[22]{}(\Varid{as}\ \mathop{:}\ \Conid{Vect}\;\Varid{n}\;\alpha)\ \to\ {}\<[E]%
\\
\>[22]{}(\Varid{ps}\ \mathop{:}\ \Conid{Vect}\;\Varid{n}\;\mathbb{R})\ \to\ {}\<[E]%
\\
\>[22]{}(\Varid{k}\ \mathop{:}\ \Conid{Fin}\;\Varid{n}\ \to\ \Conid{So}\;(\Varid{index}\;\Varid{k}\;\Varid{ps}\geq \mathrm{0.0}))\ \to\ {}\<[E]%
\\
\>[22]{}\Varid{sum}\;\Varid{ps}\mathrel{=}\mathrm{1.0}\ \to\ {}\<[E]%
\\
\>[22]{}\Conid{SimpleProb}\;\alpha{}\<[E]%
\ColumnHook
\end{hscode}\resethooks
We write \ensuremath{\Conid{M}\ \mathop{:}\ \Conid{Type}\ \to\ \Conid{Type}} for a monad and \ensuremath{{\Varid{fmap}\!}}, \ensuremath{{\Varid{ret}}} and \ensuremath{\bind} for its
\ensuremath{\Varid{fmap}}, \ensuremath{\Varid{return}} and \ensuremath{\Varid{bind}} operators:
\begin{hscode}\SaveRestoreHook
\column{B}{@{}>{\hspre}l<{\hspost}@{}}%
\column{3}{@{}>{\hspre}l<{\hspost}@{}}%
\column{12}{@{}>{\hspre}c<{\hspost}@{}}%
\column{12E}{@{}l@{}}%
\column{15}{@{}>{\hspre}l<{\hspost}@{}}%
\column{E}{@{}>{\hspre}l<{\hspost}@{}}%
\>[3]{}{\Varid{fmap}\!}{}\<[12]%
\>[12]{}\ \mathop{:}\ {}\<[12E]%
\>[15]{}(\alpha\ \to\ \beta)\ \to\ \Conid{M}\;\alpha\ \to\ \Conid{M}\;\beta{}\<[E]%
\\
\>[3]{}{\Varid{ret}}{}\<[12]%
\>[12]{}\ \mathop{:}\ {}\<[12E]%
\>[15]{}\alpha\ \to\ \Conid{M}\;\alpha{}\<[E]%
\\
\>[3]{}(\bind){}\<[12]%
\>[12]{}\ \mathop{:}\ {}\<[12E]%
\>[15]{}\Conid{M}\;\alpha\ \to\ (\alpha\ \to\ \Conid{M}\;\beta)\ \to\ \Conid{M}\;\beta{}\<[E]%
\ColumnHook
\end{hscode}\resethooks
\ensuremath{{\Varid{fmap}\!}} is required to preserve identity and composition that is, every
monad is a functor:
\begin{hscode}\SaveRestoreHook
\column{B}{@{}>{\hspre}l<{\hspost}@{}}%
\column{3}{@{}>{\hspre}l<{\hspost}@{}}%
\column{17}{@{}>{\hspre}c<{\hspost}@{}}%
\column{17E}{@{}l@{}}%
\column{20}{@{}>{\hspre}l<{\hspost}@{}}%
\column{E}{@{}>{\hspre}l<{\hspost}@{}}%
\>[3]{}\Varid{functorSpec1}{}\<[17]%
\>[17]{}\ \mathop{:}\ {}\<[17E]%
\>[20]{}{\Varid{fmap}\!}\;\Varid{id}\mathrel{=}\Varid{id}{}\<[E]%
\\
\>[3]{}\Varid{functorSpec2}{}\<[17]%
\>[17]{}\ \mathop{:}\ {}\<[17E]%
\>[20]{}{\Varid{fmap}\!}\;(\Varid{f}\mathbin{\circ}\Varid{g})\mathrel{=}({\Varid{fmap}\!}\;\Varid{f})\mathbin{\circ}({\Varid{fmap}\!}\;\Varid{g}){}\<[E]%
\ColumnHook
\end{hscode}\resethooks
\ensuremath{{\Varid{ret}}} and \ensuremath{\bind} are required to fulfill the monad laws
\begin{hscode}\SaveRestoreHook
\column{B}{@{}>{\hspre}l<{\hspost}@{}}%
\column{3}{@{}>{\hspre}l<{\hspost}@{}}%
\column{15}{@{}>{\hspre}c<{\hspost}@{}}%
\column{15E}{@{}l@{}}%
\column{18}{@{}>{\hspre}l<{\hspost}@{}}%
\column{E}{@{}>{\hspre}l<{\hspost}@{}}%
\>[3]{}\Varid{monadSpec1}{}\<[15]%
\>[15]{}\ \mathop{:}\ {}\<[15E]%
\>[18]{}({\Varid{fmap}\!}\;\Varid{f})\mathbin{\circ}{\Varid{ret}}\mathrel{=}{\Varid{ret}}\mathbin{\circ}\Varid{f}{}\<[E]%
\\
\>[3]{}\Varid{monadSpec2}{}\<[15]%
\>[15]{}\ \mathop{:}\ {}\<[15E]%
\>[18]{}({\Varid{ret}}\;\Varid{a})\;\bind\;\Varid{f}\mathrel{=}\Varid{f}\;\Varid{a}{}\<[E]%
\\
\>[3]{}\Varid{monadSpec3}{}\<[15]%
\>[15]{}\ \mathop{:}\ {}\<[15E]%
\>[18]{}\Varid{ma}\;\bind\;{\Varid{ret}}\mathrel{=}\Varid{ma}{}\<[E]%
\\
\>[3]{}\Varid{monadSpec4}{}\<[15]%
\>[15]{}\ \mathop{:}\ {}\<[15E]%
\>[18]{}\{\mskip1.5mu \Varid{f}\ \mathop{:}\ \Varid{a}\ \to\ \Conid{M}\;\Varid{b}\mskip1.5mu\}\ \to\ \{\mskip1.5mu \Varid{g}\ \mathop{:}\ \Varid{b}\ \to\ \Conid{M}\;\Varid{c}\mskip1.5mu\}\ \to\ {}\<[E]%
\\
\>[18]{}(\Varid{ma}\;\bind\;\Varid{f})\;\bind\;\Varid{g}\quad\mathrel{=}\quad\Varid{ma}\;\bind\;(\lambda \Varid{a}\Rightarrow (\Varid{f}\;\Varid{a})\;\bind\;\Varid{g}){}\<[E]%
\ColumnHook
\end{hscode}\resethooks
For the stochastic case (\ensuremath{\Conid{M}\mathrel{=}\Conid{SimpleProb}}), for example, \ensuremath{\bind} encodes
the total probability law and the monadic laws have natural
interpretations in terms of conditional probabilities and concentrated
probabilities.

As it turns out, the monadic laws are not necessary for computing
optimal policy sequences. But, as we will see in section
\ref{subsection:trajectories}, they do play a crucial role for computing
possible state-control trajectories from (optimal or non optimal)
sequences of policies. For stochastic decision problems, for instance,
and a specific policy sequence, \ensuremath{\bind} makes it possible to compute a
probability distribution over all possible trajectories which can be
realized by selecting controls according to that sequence. This is
important, e.g. in policy advice, for providing a better understanding of
the (possible) implications of adopting certain policies.

We can apply the approach developed for monadic dynamical systems to
sequential decision problems to extend the transition function
to the time-dependent monadic case
\begin{hscode}\SaveRestoreHook
\column{B}{@{}>{\hspre}l<{\hspost}@{}}%
\column{3}{@{}>{\hspre}l<{\hspost}@{}}%
\column{E}{@{}>{\hspre}l<{\hspost}@{}}%
\>[3]{}\Varid{step}\ \mathop{:}\ (\Varid{t}\ \mathop{:}\ \mathbb{N})\ \to\ (\Varid{x}\ \mathop{:}\ \Conid{State}\;\Varid{t})\ \to\ \Conid{Ctrl}\;\Varid{t}\;\Varid{x}\ \to\ \Conid{M}\;(\Conid{State}\;(\Conid{S}\;\Varid{t})){}\<[E]%
\ColumnHook
\end{hscode}\resethooks
As it turns out, extending the time-dependent formulation to the monadic
case is almost straightforward and we do not present the full details
here. There are, however, a few important aspects that need to be taken
into account.
We discuss four aspects of the monadic extension in the next four
sections.

\subsection{Monadic containers}
\label{subsection:monadic_containers}

For our application not all monads make sense.
We generalize from the deterministic case where there is just one
possible state to some form of container of possible next states.
A monadic container has, in addition to the monadic interface, a
membership test:
\begin{hscode}\SaveRestoreHook
\column{B}{@{}>{\hspre}l<{\hspost}@{}}%
\column{3}{@{}>{\hspre}l<{\hspost}@{}}%
\column{E}{@{}>{\hspre}l<{\hspost}@{}}%
\>[3]{}(\in)\ \mathop{:}\ \alpha\ \to\ \Conid{M}\;\alpha\ \to\ \Conid{Bool}{}\<[E]%
\ColumnHook
\end{hscode}\resethooks
For the generalization of \ensuremath{\Varid{viable}} we also require the predicate
\ensuremath{{\Varid{areAllTrue}}} defined on \ensuremath{\Conid{M}}-structures of Booleans
\begin{hscode}\SaveRestoreHook
\column{B}{@{}>{\hspre}l<{\hspost}@{}}%
\column{3}{@{}>{\hspre}l<{\hspost}@{}}%
\column{E}{@{}>{\hspre}l<{\hspost}@{}}%
\>[3]{}{\Varid{areAllTrue}}\ \mathop{:}\ \Conid{M}\;\Conid{Bool}\ \to\ \Conid{Bool}{}\<[E]%
\ColumnHook
\end{hscode}\resethooks
The idea is that \ensuremath{{\Varid{areAllTrue}}\;\Varid{mb}} is true if and only if all Boolean
values contained in \ensuremath{\Varid{mb}} are true.
We express this by requiring the following specification
\begin{hscode}\SaveRestoreHook
\column{B}{@{}>{\hspre}l<{\hspost}@{}}%
\column{3}{@{}>{\hspre}l<{\hspost}@{}}%
\column{11}{@{}>{\hspre}c<{\hspost}@{}}%
\column{11E}{@{}l@{}}%
\column{14}{@{}>{\hspre}l<{\hspost}@{}}%
\column{E}{@{}>{\hspre}l<{\hspost}@{}}%
\>[3]{}{\Varid{areAllTrueSpec}}{}\<[11]%
\>[11]{}\ \mathop{:}\ {}\<[11E]%
\>[14]{}(\Varid{b}\ \mathop{:}\ \Conid{Bool})\ \to\ \Conid{So}\;({\Varid{areAllTrue}}\;({\Varid{ret}}\;\Varid{b})==\Varid{b}){}\<[E]%
\\
\>[3]{}{\Varid{isInAreAllTrueSpec}}{}\<[11]%
\>[11]{}\ \mathop{:}\ {}\<[11E]%
\>[14]{}(\Varid{mx}\ \mathop{:}\ \Conid{M}\;\alpha)\ \to\ (\Varid{p}\ \mathop{:}\ \alpha\ \to\ \Conid{Bool})\ \to\ {}\<[E]%
\\
\>[14]{}\Conid{So}\;({\Varid{areAllTrue}}\;({\Varid{fmap}\!}\;\Varid{p}\;\Varid{mx}))\ \to\ {}\<[E]%
\\
\>[14]{}(\Varid{x}\ \mathop{:}\ \alpha)\ \to\ \Conid{So}\;(\Varid{x}\in\Varid{mx})\ \to\ \Conid{So}\;(\Varid{p}\;\Varid{x}){}\<[E]%
\ColumnHook
\end{hscode}\resethooks
It is enough to require this for the special case of \ensuremath{\alpha} equal
to \ensuremath{\Conid{State}\;(\Conid{S}\;\Varid{t})}.

A key property of the monadic containers is that if we map a function
\ensuremath{\Varid{f}} over a container \ensuremath{\Varid{ma}}, \ensuremath{\Varid{f}} will only be used on values in the
subset of \ensuremath{\alpha} which are in \ensuremath{\Varid{ma}}.
We model the subset as \ensuremath{(\Varid{a}\ \mathop{:}\ \alpha\;*\!\!*\;\Conid{So}\;(\Varid{a}\in\Varid{ma}))} and we
formalise the key property by requiring a function \ensuremath{\Varid{toSub}} which takes
any \ensuremath{\Varid{a}\ \mathop{:}\ \alpha} in the container into the subset:
\begin{hscode}\SaveRestoreHook
\column{B}{@{}>{\hspre}l<{\hspost}@{}}%
\column{3}{@{}>{\hspre}l<{\hspost}@{}}%
\column{14}{@{}>{\hspre}c<{\hspost}@{}}%
\column{14E}{@{}l@{}}%
\column{17}{@{}>{\hspre}l<{\hspost}@{}}%
\column{E}{@{}>{\hspre}l<{\hspost}@{}}%
\>[3]{}\Varid{toSub}{}\<[14]%
\>[14]{}\ \mathop{:}\ {}\<[14E]%
\>[17]{}(\Varid{ma}\ \mathop{:}\ \Conid{M}\;\alpha)\ \to\ \Conid{M}\;(\Varid{a}\ \mathop{:}\ \alpha\;*\!\!*\;\Conid{So}\;(\Varid{a}\in\Varid{ma})){}\<[E]%
\\
\>[3]{}\Varid{toSubSpec}{}\<[14]%
\>[14]{}\ \mathop{:}\ {}\<[14E]%
\>[17]{}(\Varid{ma}\ \mathop{:}\ \Conid{M}\;\alpha)\ \to\ {\Varid{fmap}\!}\;\Varid{outl}\;(\Varid{toSub}\;\Varid{ma})\mathrel{=}\Varid{ma}{}\<[E]%
\ColumnHook
\end{hscode}\resethooks
The specification requires \ensuremath{\Varid{toSub}} to be a tagged identity function.
For the cases mentioned above (\ensuremath{\Conid{M}\mathrel{=}\Conid{Id}}, \ensuremath{\Conid{List}} and \ensuremath{\Conid{SimpleProb}}) this
is easily implemented.

\subsection{Viability, reachability}
\label{subsection:viable_reachable}

In section \ref{section:time-dependent} we said that a state \ensuremath{\Varid{x}\ \mathop{:}\ \Conid{State}\;\Varid{t}}
is viable for \ensuremath{\Conid{S}\;\Varid{n}} steps if an only if there exists a control \ensuremath{\Varid{y}\ \mathop{:}\ \Conid{Ctrl}\;\Varid{t}\;\Varid{x}} such that \ensuremath{\Varid{step}\;\Varid{t}\;\Varid{x}\;\Varid{y}} is viable \ensuremath{\Varid{n}} steps.

As explained above, monadic extensions are introduced to generalize the
notion of deterministic transition function. For \ensuremath{\Conid{M}\mathrel{=}\Conid{SimpleProb}}, for
instance, \ensuremath{\Varid{step}\;\Varid{t}\;\Varid{x}\;\Varid{y}} is a probability distribution on \ensuremath{\Conid{State}\;(\Conid{S}\;\Varid{t})}. Its
support represents the set of states that can be reached in one step
from \ensuremath{\Varid{x}} by selecting the control \ensuremath{\Varid{y}}.

According to this interpretation, \ensuremath{\Conid{M}} is a monadic container and the
states in \ensuremath{\Varid{step}\;\Varid{t}\;\Varid{x}\;\Varid{y}} are \emph{possible} states at step \ensuremath{\Conid{S}\;\Varid{t}}. For \ensuremath{\Varid{x}\ \mathop{:}\ \Conid{State}\;\Varid{t}} to be viable for \ensuremath{\Conid{S}\;\Varid{n}} steps, there must exist a control \ensuremath{\Varid{y}\ \mathop{:}\ \Conid{Ctrl}\;\Varid{t}\;\Varid{x}} such that all next states which are possible are viable for \ensuremath{\Varid{n}}
steps. We call such a control a \emph{feasible} control
\begin{hscode}\SaveRestoreHook
\column{B}{@{}>{\hspre}l<{\hspost}@{}}%
\column{3}{@{}>{\hspre}l<{\hspost}@{}}%
\column{14}{@{}>{\hspre}c<{\hspost}@{}}%
\column{14E}{@{}l@{}}%
\column{17}{@{}>{\hspre}l<{\hspost}@{}}%
\column{E}{@{}>{\hspre}l<{\hspost}@{}}%
\>[3]{}\Varid{viable}{}\<[14]%
\>[14]{}\ \mathop{:}\ {}\<[14E]%
\>[17]{}(\Varid{n}\ \mathop{:}\ \mathbb{N})\ \to\ \Conid{State}\;\Varid{t}\ \to\ \Conid{Bool}{}\<[E]%
\\
\>[3]{}{\Varid{feasible}}{}\<[14]%
\>[14]{}\ \mathop{:}\ {}\<[14E]%
\>[17]{}(\Varid{n}\ \mathop{:}\ \mathbb{N})\ \to\ (\Varid{x}\ \mathop{:}\ \Conid{State}\;\Varid{t})\ \to\ \Conid{Ctrl}\;\Varid{t}\;\Varid{x}\ \to\ \Conid{Bool}{}\<[E]%
\\
\>[3]{}{\Varid{feasible}}\;\{\mskip1.5mu \Varid{t}\mskip1.5mu\}\;\Varid{n}\;\Varid{x}\;\Varid{y}\mathrel{=}{\Varid{areAllTrue}}\;({\Varid{fmap}\!}\;(\Varid{viable}\;\Varid{n})\;(\Varid{step}\;\Varid{t}\;\Varid{x}\;\Varid{y})){}\<[E]%
\ColumnHook
\end{hscode}\resethooks
With the notion of feasibility in place, we can extend the specification
of \ensuremath{\Varid{viable}} to the monadic case
\begin{hscode}\SaveRestoreHook
\column{B}{@{}>{\hspre}l<{\hspost}@{}}%
\column{3}{@{}>{\hspre}l<{\hspost}@{}}%
\column{16}{@{}>{\hspre}c<{\hspost}@{}}%
\column{16E}{@{}l@{}}%
\column{19}{@{}>{\hspre}l<{\hspost}@{}}%
\column{E}{@{}>{\hspre}l<{\hspost}@{}}%
\>[3]{}\Varid{viableSpec0}{}\<[16]%
\>[16]{}\ \mathop{:}\ {}\<[16E]%
\>[19]{}(\Varid{x}\ \mathop{:}\ \Conid{State}\;\Varid{t})\ \to\ \Conid{Viable}\;\Conid{Z}\;\Varid{x}{}\<[E]%
\\
\>[3]{}\Varid{viableSpec1}{}\<[16]%
\>[16]{}\ \mathop{:}\ {}\<[16E]%
\>[19]{}(\Varid{x}\ \mathop{:}\ \Conid{State}\;\Varid{t})\ \to\ \Conid{Viable}\;(\Conid{S}\;\Varid{n})\;\Varid{x}\ \to\ \Conid{GoodCtrl}\;\Varid{t}\;\Varid{n}\;\Varid{x}{}\<[E]%
\\
\>[3]{}\Varid{viableSpec2}{}\<[16]%
\>[16]{}\ \mathop{:}\ {}\<[16E]%
\>[19]{}(\Varid{x}\ \mathop{:}\ \Conid{State}\;\Varid{t})\ \to\ \Conid{GoodCtrl}\;\Varid{t}\;\Varid{n}\;\Varid{x}\ \to\ \Conid{Viable}\;(\Conid{S}\;\Varid{n})\;\Varid{x}{}\<[E]%
\ColumnHook
\end{hscode}\resethooks
As in the time-dependent case, \ensuremath{\Conid{Viable}\;\Varid{n}\;\Varid{x}} as a shorthand for \ensuremath{\Conid{So}\;(\Varid{viable}\;\Varid{n}\;\Varid{x})} and
\begin{hscode}\SaveRestoreHook
\column{B}{@{}>{\hspre}l<{\hspost}@{}}%
\column{3}{@{}>{\hspre}l<{\hspost}@{}}%
\column{E}{@{}>{\hspre}l<{\hspost}@{}}%
\>[3]{}\Conid{GoodCtrl}\ \mathop{:}\ (\Varid{t}\ \mathop{:}\ \mathbb{N})\ \to\ (\Varid{n}\ \mathop{:}\ \mathbb{N})\ \to\ \Conid{State}\;\Varid{t}\ \to\ \Conid{Type}{}\<[E]%
\\
\>[3]{}\Conid{GoodCtrl}\;\Varid{t}\;\Varid{n}\;\Varid{x}\mathrel{=}(\Varid{y}\ \mathop{:}\ \Conid{Ctrl}\;\Varid{t}\;\Varid{x}\;*\!\!*\;{\Conid{Feasible}}\;\Varid{n}\;\Varid{x}\;\Varid{y}){}\<[E]%
\ColumnHook
\end{hscode}\resethooks
Here, \ensuremath{{\Conid{Feasible}}\;\Varid{n}\;\Varid{x}\;\Varid{y}} is a shorthand for \ensuremath{\Conid{So}\;({\Varid{feasible}}\;\Varid{n}\;\Varid{x}\;\Varid{y})}.

The notion of reachability introduced in section
\ref{section:time-dependent} can be extended to the monadic case
straightforwardly: every initial state is reachable. If a state \ensuremath{\Varid{x}\ \mathop{:}\ \Conid{State}\;\Varid{t}} is reachable, every control \ensuremath{\Varid{y}\ \mathop{:}\ \Conid{Ctrl}\;\Varid{t}\;\Varid{x}} leads, via \ensuremath{\Varid{step}} to an
\ensuremath{\Conid{M}}-structure of reachable states. Conversely, if a state \ensuremath{\Varid{x'}\ \mathop{:}\ \Conid{State}\;(\Conid{S}\;\Varid{t})}
is reachable then there exist a state \ensuremath{\Varid{x}\ \mathop{:}\ \Conid{State}\;\Varid{t}} and a control \ensuremath{\Varid{y}\ \mathop{:}\ \Conid{Ctrl}\;\Varid{t}\;\Varid{x}} such that \ensuremath{\Varid{x}} is reachable and \ensuremath{\Varid{x'}} is in the \ensuremath{\Conid{M}}-structure \ensuremath{\Varid{step}\;\Varid{t}\;\Varid{x}\;\Varid{y}}:
\begin{hscode}\SaveRestoreHook
\column{B}{@{}>{\hspre}l<{\hspost}@{}}%
\column{3}{@{}>{\hspre}l<{\hspost}@{}}%
\column{19}{@{}>{\hspre}c<{\hspost}@{}}%
\column{19E}{@{}l@{}}%
\column{22}{@{}>{\hspre}l<{\hspost}@{}}%
\column{E}{@{}>{\hspre}l<{\hspost}@{}}%
\>[3]{}\Varid{reachable}{}\<[19]%
\>[19]{}\ \mathop{:}\ {}\<[19E]%
\>[22]{}\Conid{State}\;\Varid{t}\ \to\ \Conid{Bool}{}\<[E]%
\\[\blanklineskip]%
\\
\\[\blanklineskip]%
\>[3]{}\Varid{reachableSpec0}{}\<[19]%
\>[19]{}\ \mathop{:}\ {}\<[19E]%
\>[22]{}(\Varid{x}\ \mathop{:}\ \Conid{State}\;\Conid{Z})\ \to\ \Conid{Reachable}\;\Varid{x}{}\<[E]%
\ColumnHook
\end{hscode}\resethooks
\begin{hscode}\SaveRestoreHook
\column{B}{@{}>{\hspre}l<{\hspost}@{}}%
\column{3}{@{}>{\hspre}l<{\hspost}@{}}%
\column{19}{@{}>{\hspre}c<{\hspost}@{}}%
\column{19E}{@{}l@{}}%
\column{22}{@{}>{\hspre}l<{\hspost}@{}}%
\column{E}{@{}>{\hspre}l<{\hspost}@{}}%
\>[3]{}\Varid{reachableSpec1}{}\<[19]%
\>[19]{}\ \mathop{:}\ {}\<[19E]%
\>[22]{}(\Varid{x}\ \mathop{:}\ \Conid{State}\;\Varid{t})\ \to\ \Conid{Reachable}\;\Varid{x}\ \to\ (\Varid{y}\ \mathop{:}\ \Conid{Ctrl}\;\Varid{t}\;\Varid{x})\ \to\ {}\<[E]%
\\
\>[22]{}(\Varid{x'}\ \mathop{:}\ \Conid{State}\;(\Conid{S}\;\Varid{t}))\ \to\ \Conid{So}\;(\Varid{x'}\in\Varid{step}\;\Varid{t}\;\Varid{x}\;\Varid{y})\ \to\ \Conid{Reachable}\;\Varid{x'}{}\<[E]%
\ColumnHook
\end{hscode}\resethooks
\begin{hscode}\SaveRestoreHook
\column{B}{@{}>{\hspre}l<{\hspost}@{}}%
\column{3}{@{}>{\hspre}l<{\hspost}@{}}%
\column{19}{@{}>{\hspre}c<{\hspost}@{}}%
\column{19E}{@{}l@{}}%
\column{22}{@{}>{\hspre}l<{\hspost}@{}}%
\column{E}{@{}>{\hspre}l<{\hspost}@{}}%
\>[3]{}\Varid{reachableSpec2}{}\<[19]%
\>[19]{}\ \mathop{:}\ {}\<[19E]%
\>[22]{}(\Varid{x'}\ \mathop{:}\ \Conid{State}\;(\Conid{S}\;\Varid{t}))\ \to\ \Conid{Reachable}\;\Varid{x'}\ \to\ {}\<[E]%
\\
\>[22]{}(\Varid{x}\ \mathop{:}\ \Conid{State}\;\Varid{t}\;*\!\!*\;(\Conid{Reachable}\;\Varid{x},(\Varid{y}\ \mathop{:}\ \Conid{Ctrl}\;\Varid{t}\;\Varid{x}\;*\!\!*\;\Conid{So}\;(\Varid{x'}\in\Varid{step}\;\Varid{t}\;\Varid{x}\;\Varid{y})))){}\<[E]%
\ColumnHook
\end{hscode}\resethooks
As in the time-dependent case, \ensuremath{\Conid{Reachable}\;\Varid{x}} is a shorthand for \ensuremath{\Conid{So}\;(\Varid{reachable}\;\Varid{x})}.







\subsection{Aggregation measure}
\label{subsection:agg}

In the monadic case, the notions of policy and policy sequence are the
same as in the deterministic case. The notion of value of a policy
sequence, however, requires some attention.

In the deterministic case, the value of selecting controls according to
the policy sequence \ensuremath{(\Varid{p}\mathbin{::}\Varid{ps})} of length \ensuremath{\Conid{S}\;\Varid{n}} when in state \ensuremath{\Varid{x}} at step
\ensuremath{\Varid{t}} is given by
\begin{hscode}\SaveRestoreHook
\column{B}{@{}>{\hspre}l<{\hspost}@{}}%
\column{3}{@{}>{\hspre}l<{\hspost}@{}}%
\column{E}{@{}>{\hspre}l<{\hspost}@{}}%
\>[3]{}\Varid{reward}\;\Varid{t}\;\Varid{x}\;\Varid{y}\;\Varid{x'}\mathbin{+}\Varid{val}\;(\Conid{S}\;\Varid{t})\;\Varid{n}\;\Varid{x'}\;\Varid{r'}\;\Varid{v'}\;\Varid{ps}{}\<[E]%
\ColumnHook
\end{hscode}\resethooks
where \ensuremath{\Varid{y}} is the control selected by \ensuremath{\Varid{p}}, \ensuremath{\Varid{x'}\mathrel{=}\Varid{step}\;\Varid{t}\;\Varid{x}\;\Varid{y}} and \ensuremath{\Varid{r'}} and
\ensuremath{\Varid{v'}} are proofs that \ensuremath{\Varid{x'}} is reachable and viable for \ensuremath{\Varid{n}} steps. In the
monadic case, \ensuremath{\Varid{step}} returns an \ensuremath{\Conid{M}}-structure of states. In general, for
each possible state in \ensuremath{\Varid{step}\;\Varid{t}\;\Varid{x}\;\Varid{y}} there will be a corresponding value
of the above sum.

As shown by Ionescu~\cite{ionescu2009}, one can easily extend the
notion of the value of a policy sequence to the monadic case if one
has a way of \emph{measuring} (or aggregating) an \ensuremath{\Conid{M}}-structure of
\ensuremath{\mathbb{R}} satisfying a monotonicity condition:
\begin{hscode}\SaveRestoreHook
\column{B}{@{}>{\hspre}l<{\hspost}@{}}%
\column{3}{@{}>{\hspre}l<{\hspost}@{}}%
\column{13}{@{}>{\hspre}c<{\hspost}@{}}%
\column{13E}{@{}l@{}}%
\column{16}{@{}>{\hspre}l<{\hspost}@{}}%
\column{E}{@{}>{\hspre}l<{\hspost}@{}}%
\>[3]{}{\Varid{meas}}{}\<[13]%
\>[13]{}\ \mathop{:}\ {}\<[13E]%
\>[16]{}\Conid{M}\;\mathbb{R}\ \to\ \mathbb{R}{}\<[E]%
\\
\>[3]{}{\Varid{measMon}}{}\<[13]%
\>[13]{}\ \mathop{:}\ {}\<[13E]%
\>[16]{}(\Varid{f}\ \mathop{:}\ \Conid{State}\;\Varid{t}\ \to\ \mathbb{R})\ \to\ (\Varid{g}\ \mathop{:}\ \Conid{State}\;\Varid{t}\ \to\ \mathbb{R})\ \to\ {}\<[E]%
\\
\>[16]{}((\Varid{x}\ \mathop{:}\ \Conid{State}\;\Varid{t})\ \to\ \Conid{So}\;(\Varid{f}\;\Varid{x}\leq \Varid{g}\;\Varid{x}))\ \to\ {}\<[E]%
\\
\>[16]{}(\Varid{mx}\ \mathop{:}\ \Conid{M}\;(\Conid{State}\;\Varid{t}))\ \to\ \Conid{So}\;({\Varid{meas}}\;({\Varid{fmap}\!}\;\Varid{f}\;\Varid{mx})\leq {\Varid{meas}}\;({\Varid{fmap}\!}\;\Varid{g}\;\Varid{mx})){}\<[E]%
\ColumnHook
\end{hscode}\resethooks
With \ensuremath{{\Varid{meas}}}, the value of a policy sequence in the monadic case can be
easily computed
\begin{hscode}\SaveRestoreHook
\column{B}{@{}>{\hspre}l<{\hspost}@{}}%
\column{3}{@{}>{\hspre}l<{\hspost}@{}}%
\column{5}{@{}>{\hspre}l<{\hspost}@{}}%
\column{6}{@{}>{\hspre}l<{\hspost}@{}}%
\column{9}{@{}>{\hspre}c<{\hspost}@{}}%
\column{9E}{@{}l@{}}%
\column{10}{@{}>{\hspre}c<{\hspost}@{}}%
\column{10E}{@{}l@{}}%
\column{11}{@{}>{\hspre}l<{\hspost}@{}}%
\column{12}{@{}>{\hspre}l<{\hspost}@{}}%
\column{13}{@{}>{\hspre}l<{\hspost}@{}}%
\column{18}{@{}>{\hspre}l<{\hspost}@{}}%
\column{35}{@{}>{\hspre}c<{\hspost}@{}}%
\column{35E}{@{}l@{}}%
\column{36}{@{}>{\hspre}l<{\hspost}@{}}%
\column{38}{@{}>{\hspre}l<{\hspost}@{}}%
\column{39}{@{}>{\hspre}l<{\hspost}@{}}%
\column{44}{@{}>{\hspre}c<{\hspost}@{}}%
\column{44E}{@{}l@{}}%
\column{47}{@{}>{\hspre}l<{\hspost}@{}}%
\column{E}{@{}>{\hspre}l<{\hspost}@{}}%
\>[3]{}\Conid{Mval}{}\<[9]%
\>[9]{}\ \mathop{:}\ {}\<[9E]%
\>[12]{}(\Varid{t}\ \mathop{:}\ \mathbb{N})\ \to\ (\Varid{n}\ \mathop{:}\ \mathbb{N})\ \to\ {}\<[E]%
\\
\>[12]{}(\Varid{x}\ \mathop{:}\ \Conid{State}\;\Varid{t})\ \to\ (\Varid{r}\ \mathop{:}\ \Conid{Reachable}\;\Varid{x})\ \to\ (\Varid{v}\ \mathop{:}\ \Conid{Viable}\;\Varid{n}\;\Varid{x})\ \to\ {}\<[E]%
\\
\>[12]{}\Conid{PolicySeq}\;\Varid{t}\;\Varid{n}\ \to\ \mathbb{R}{}\<[E]%
\\
\>[3]{}\Conid{Mval}\;\anonymous \;{}\<[11]%
\>[11]{}\Conid{Z}\;{}\<[18]%
\>[18]{}\anonymous \;\anonymous \;\anonymous \;\anonymous {}\<[35]%
\>[35]{}\mathrel{=}{}\<[35E]%
\>[38]{}\mathrm{0}{}\<[E]%
\\
\>[3]{}\Conid{Mval}\;\Varid{t}\;{}\<[11]%
\>[11]{}(\Conid{S}\;\Varid{n})\;{}\<[18]%
\>[18]{}\Varid{x}\;\Varid{r}\;\Varid{v}\;(\Varid{p}\mathbin{::}\Varid{ps}){}\<[35]%
\>[35]{}\mathrel{=}{}\<[35E]%
\>[38]{}{\Varid{meas}}\;({\Varid{fmap}\!}\;\Varid{f}\;(\Varid{toSub}\;\Varid{mx'}))\;\mathbf{where}{}\<[E]%
\\
\>[3]{}\hsindent{2}{}\<[5]%
\>[5]{}\Varid{y}{}\<[10]%
\>[10]{}\ \mathop{:}\ {}\<[10E]%
\>[13]{}\Conid{Ctrl}\;\Varid{t}\;\Varid{x};{}\<[39]%
\>[39]{}\Varid{y}{}\<[44]%
\>[44]{}\mathrel{=}{}\<[44E]%
\>[47]{}\Varid{outl}\;(\Varid{p}\;\Varid{x}\;\Varid{r}\;\Varid{v}){}\<[E]%
\\
\>[3]{}\hsindent{2}{}\<[5]%
\>[5]{}\Varid{mx'}{}\<[10]%
\>[10]{}\ \mathop{:}\ {}\<[10E]%
\>[13]{}\Conid{M}\;(\Conid{State}\;(\Conid{S}\;\Varid{t}));~{}\<[39]%
\>[39]{}\Varid{mx'}{}\<[44]%
\>[44]{}\mathrel{=}{}\<[44E]%
\>[47]{}\Varid{step}\;\Varid{t}\;\Varid{x}\;\Varid{y}{}\<[E]%
\\
\>[3]{}\hsindent{2}{}\<[5]%
\>[5]{}\Varid{f}\ \mathop{:}\ (\Varid{x'}\ \mathop{:}\ \Conid{State}\;(\Conid{S}\;\Varid{t})\;*\!\!*\;\Conid{So}\;(\Varid{x'}\in\Varid{mx'}))\ \to\ \mathbb{R}{}\<[E]%
\\
\>[3]{}\hsindent{2}{}\<[5]%
\>[5]{}\Varid{f}\;(\Varid{x'}\;*\!\!*\;\Varid{x'ins})\mathrel{=}\Varid{reward}\;\Varid{t}\;\Varid{x}\;\Varid{y}\;\Varid{x'}\mathbin{+}\Conid{Mval}\;(\Conid{S}\;\Varid{t})\;\Varid{n}\;\Varid{x'}\;\Varid{r'}\;\Varid{v'}\;\Varid{ps}\;\mathbf{where}{}\<[E]%
\\
\>[5]{}\hsindent{1}{}\<[6]%
\>[6]{}\Varid{r'}{}\<[10]%
\>[10]{}\ \mathop{:}\ {}\<[10E]%
\>[13]{}\Conid{Reachable}\;\Varid{x'};~{}\<[36]%
\>[36]{}\Varid{r'}\mathrel{=}\Varid{reachableSpec1}\;\Varid{x}\;\Varid{r}\;\Varid{y}\;\Varid{x'}\;\Varid{x'ins}{}\<[E]%
\\
\>[5]{}\hsindent{1}{}\<[6]%
\>[6]{}\Varid{v'}{}\<[10]%
\>[10]{}\ \mathop{:}\ {}\<[10E]%
\>[13]{}\Conid{Viable}\;\Varid{n}\;\Varid{x'};{}\<[36]%
\>[36]{}\Varid{v'}\mathrel{=}{\Varid{isInAreAllTrueSpec}}\;\Varid{mx'}\;(\Varid{viable}\;\Varid{n})\;(\Varid{outr}\;(\Varid{p}\;\Varid{x}\;\Varid{r}\;\Varid{v}))\;\Varid{x'}\;\Varid{x'ins}{}\<[E]%
\ColumnHook
\end{hscode}\resethooks
Notice that, in the implementation of \ensuremath{\Varid{f}}, we (can) call \ensuremath{\Conid{Mval}} only for
those values of \ensuremath{\Varid{x'}} which are provably reachable (\ensuremath{\Varid{r'}}) and viable for
\ensuremath{\Varid{n}} steps (\ensuremath{\Varid{v'}}). Using \ensuremath{\Varid{r}}, \ensuremath{\Varid{v}} and \ensuremath{\Varid{outr}\;(\Varid{p}\;\Varid{x}\;\Varid{r}\;\Varid{v})}, it is easy to
compute \ensuremath{\Varid{r'}} and \ensuremath{\Varid{v'}} for \ensuremath{\Varid{x'}} in \ensuremath{\Varid{mx'}}.

The monotonicity condition for \ensuremath{{\Varid{meas}}} plays a crucial role in proving Bellman's
principle of optimality. The principle itself is formulated as in the
deterministic case, see section \ref{subsection:full_framework}. But now, proving
\begin{hscode}\SaveRestoreHook
\column{B}{@{}>{\hspre}l<{\hspost}@{}}%
\column{3}{@{}>{\hspre}l<{\hspost}@{}}%
\column{E}{@{}>{\hspre}l<{\hspost}@{}}%
\>[3]{}\Conid{Mval}\;\Varid{t}\;(\Conid{S}\;\Varid{n})\;\Varid{x}\;\Varid{r}\;\Varid{v}\;(\Varid{p'}\mathbin{::}\Varid{ps'})\leq \Conid{Mval}\;\Varid{t}\;(\Conid{S}\;\Varid{n})\;\Varid{x}\;\Varid{r}\;\Varid{v}\;(\Varid{p'}\mathbin{::}\Varid{ps}){}\<[E]%
\ColumnHook
\end{hscode}\resethooks
requires proving that
\begin{hscode}\SaveRestoreHook
\column{B}{@{}>{\hspre}l<{\hspost}@{}}%
\column{3}{@{}>{\hspre}l<{\hspost}@{}}%
\column{E}{@{}>{\hspre}l<{\hspost}@{}}%
\>[3]{}{\Varid{meas}}\;({\Varid{fmap}\!}\;\Varid{f}\;(\Varid{step}\;\Varid{t}\;\Varid{x}\;\Varid{y}))\leq {\Varid{meas}}\;({\Varid{fmap}\!}\;\Varid{g}\;(\Varid{step}\;\Varid{t}\;\Varid{x}\;\Varid{y})){}\<[E]%
\ColumnHook
\end{hscode}\resethooks
where \ensuremath{\Varid{f}} and \ensuremath{\Varid{g}} are the functions that map \ensuremath{\Varid{x'}} in \ensuremath{\Varid{mx'}} to
\begin{hscode}\SaveRestoreHook
\column{B}{@{}>{\hspre}l<{\hspost}@{}}%
\column{3}{@{}>{\hspre}l<{\hspost}@{}}%
\column{E}{@{}>{\hspre}l<{\hspost}@{}}%
\>[3]{}\Varid{reward}\;\Varid{t}\;\Varid{x}\;\Varid{y}\;\Varid{x'}\mathbin{+}\Conid{Mval}\;(\Conid{S}\;\Varid{t})\;\Varid{n}\;\Varid{x'}\;\Varid{r'}\;\Varid{v'}\;\Varid{ps'}{}\<[E]%
\ColumnHook
\end{hscode}\resethooks
and
\begin{hscode}\SaveRestoreHook
\column{B}{@{}>{\hspre}l<{\hspost}@{}}%
\column{3}{@{}>{\hspre}l<{\hspost}@{}}%
\column{E}{@{}>{\hspre}l<{\hspost}@{}}%
\>[3]{}\Varid{reward}\;\Varid{t}\;\Varid{x}\;\Varid{y}\;\Varid{x'}\mathbin{+}\Conid{Mval}\;(\Conid{S}\;\Varid{t})\;\Varid{n}\;\Varid{x'}\;\Varid{r'}\;\Varid{v'}\;\Varid{ps}{}\<[E]%
\ColumnHook
\end{hscode}\resethooks
respectively (and \ensuremath{\Varid{r'}}, \ensuremath{\Varid{v'}} are reachability and viability proofs for
\ensuremath{\Varid{x'}}, as above). We can use optimality of \ensuremath{\Varid{ps}} and monotonicity of \ensuremath{\mathbin{+}}
as in the deterministic case to infer that the first sum is not bigger
than the second one for arbitrary \ensuremath{\Varid{x'}}. The monotonicity condition
guarantees that inequality of measured values follows.

A final remark on \ensuremath{{\Varid{meas}}}: in standard textbooks, stochastic sequential
decision problems are often tackled by assuming \ensuremath{{\Varid{meas}}} to be the
function that computes (at step \ensuremath{\Varid{t}}, state \ensuremath{\Varid{x}}, for a policy sequence \ensuremath{\Varid{p}\mathbin{::}\Varid{ps}}, etc.) the expected value of \ensuremath{{\Varid{fmap}\!}\;\Varid{f}\;(\Varid{step}\;\Varid{t}\;\Varid{x}\;\Varid{y})} where \ensuremath{\Varid{y}} is
the control selected by \ensuremath{\Varid{p}} at step \ensuremath{\Varid{t}} and \ensuremath{\Varid{f}} is defined as above.
Our framework allows clients to apply whatever aggregation best suits
their specific application domain as long as it fulfills the
monotonicity requirement. This holds for arbitrary \ensuremath{\Conid{M}}, not just for the
stochastic case.

\subsection{Trajectories}
\label{subsection:trajectories}

Dropping the assumption of determinism has an important implication on
sequential decision problems: the notion of control sequences (and,
therefore, of optimal control sequences) becomes, in a certain sense,
void. What in the non-deterministic and stochastic cases do matter are just
policies and policy sequences.

Intuitively, this is easy to understand. If the evolution of a system is
not deterministic, it makes very little sense to ask for the best
actions for the future. The best action at step \ensuremath{\Varid{t}\mathbin{+}\Varid{n}} will depend on
the state that will be reached after \ensuremath{\Varid{n}} steps. That state is not known
in advance. Thus -- for non-deterministic systems -- only policies are
relevant: if we have an optimal policy for step \ensuremath{\Varid{t}\mathbin{+}\Varid{n}}, we know all we
need to optimally select controls at that step.

On a more formal level, the implication of dropping the assumption of
determinism is that the notions of control sequence and policy sequence
become roughly equivalent.
Consider, for example, a non-deterministic case (\ensuremath{\Conid{M}\mathrel{=}\Conid{List}}) and an
initial state \ensuremath{\Varid{x}_{0}\ \mathop{:}\ \Conid{State}\;\mathrm{0}}.  Assume we have a rule
\begin{hscode}\SaveRestoreHook
\column{B}{@{}>{\hspre}l<{\hspost}@{}}%
\column{3}{@{}>{\hspre}l<{\hspost}@{}}%
\column{E}{@{}>{\hspre}l<{\hspost}@{}}%
\>[3]{}\Varid{p}_{0}\ \mathop{:}\ (\Varid{x}\ \mathop{:}\ \Conid{State}\;\mathrm{0})\ \to\ \Conid{Ctrl}\;\mathrm{0}\;\Varid{x}{}\<[E]%
\ColumnHook
\end{hscode}\resethooks
which allows us to select a control \ensuremath{\Varid{y}_{0}\mathrel{=}\Varid{p}_{0}\;\Varid{x}_{0}} at step 0. We can then compute
the singleton list consisting of the dependent pair \ensuremath{(\Varid{x}_{0}\;*\!\!*\;\Varid{y}_{0})}:
\begin{hscode}\SaveRestoreHook
\column{B}{@{}>{\hspre}l<{\hspost}@{}}%
\column{3}{@{}>{\hspre}l<{\hspost}@{}}%
\column{9}{@{}>{\hspre}c<{\hspost}@{}}%
\column{9E}{@{}l@{}}%
\column{12}{@{}>{\hspre}l<{\hspost}@{}}%
\column{E}{@{}>{\hspre}l<{\hspost}@{}}%
\>[3]{}\Varid{mxy}_{0}{}\<[9]%
\>[9]{}\ \mathop{:}\ {}\<[9E]%
\>[12]{}\Conid{M}\;(\Varid{x}\ \mathop{:}\ \Conid{State}\;\mathrm{0}\;*\!\!*\;\Conid{Ctrl}\;\mathrm{0}\;\Varid{x}){}\<[E]%
\\
\>[3]{}\Varid{mxy}_{0}{}\<[9]%
\>[9]{}\mathrel{=}{}\<[9E]%
\>[12]{}{\Varid{ret}}\;(\Varid{x}_{0}\;*\!\!*\;\Varid{y}_{0}){}\<[E]%
\ColumnHook
\end{hscode}\resethooks
Via the transition function, \ensuremath{\Varid{mxy}_{0}} yields a list of possible future
states \ensuremath{\Varid{mx}_{1}}:
\begin{hscode}\SaveRestoreHook
\column{B}{@{}>{\hspre}l<{\hspost}@{}}%
\column{3}{@{}>{\hspre}l<{\hspost}@{}}%
\column{8}{@{}>{\hspre}c<{\hspost}@{}}%
\column{8E}{@{}l@{}}%
\column{11}{@{}>{\hspre}l<{\hspost}@{}}%
\column{E}{@{}>{\hspre}l<{\hspost}@{}}%
\>[3]{}\Varid{mx}_{1}{}\<[8]%
\>[8]{}\ \mathop{:}\ {}\<[8E]%
\>[11]{}\Conid{M}\;(\Conid{State}\;\mathrm{1}){}\<[E]%
\\
\>[3]{}\Varid{mx}_{1}{}\<[8]%
\>[8]{}\mathrel{=}{}\<[8E]%
\>[11]{}\Varid{step}\;\mathrm{0}\;\Varid{x}_{0}\;\Varid{y}_{0}{}\<[E]%
\ColumnHook
\end{hscode}\resethooks
Thus, after one step and in contrast to the deterministic case, we do
not have just one set of controls to choose from. Instead, we have as
many sets as there are elements in \ensuremath{\Varid{mx}_{1}}. Because the controls available
in a given state depend, in general, on that state, we do not have a
uniformly valid rule for joining all such control spaces into a single
one to select a new control from. But again, if we have a rule for selecting
controls at step 1
\begin{hscode}\SaveRestoreHook
\column{B}{@{}>{\hspre}l<{\hspost}@{}}%
\column{3}{@{}>{\hspre}l<{\hspost}@{}}%
\column{7}{@{}>{\hspre}c<{\hspost}@{}}%
\column{7E}{@{}l@{}}%
\column{10}{@{}>{\hspre}l<{\hspost}@{}}%
\column{E}{@{}>{\hspre}l<{\hspost}@{}}%
\>[3]{}\Varid{p}_{1}{}\<[7]%
\>[7]{}\ \mathop{:}\ {}\<[7E]%
\>[10]{}(\Varid{x}\ \mathop{:}\ \Conid{State}\;\mathrm{1})\ \to\ \Conid{Ctrl}\;\mathrm{1}\;\Varid{x}{}\<[E]%
\ColumnHook
\end{hscode}\resethooks
we can pair each state in \ensuremath{\Varid{mx}_{1}} with its corresponding control and
compute a list of state-control dependent pairs
\begin{hscode}\SaveRestoreHook
\column{B}{@{}>{\hspre}l<{\hspost}@{}}%
\column{3}{@{}>{\hspre}l<{\hspost}@{}}%
\column{5}{@{}>{\hspre}l<{\hspost}@{}}%
\column{9}{@{}>{\hspre}c<{\hspost}@{}}%
\column{9E}{@{}l@{}}%
\column{11}{@{}>{\hspre}c<{\hspost}@{}}%
\column{11E}{@{}l@{}}%
\column{12}{@{}>{\hspre}l<{\hspost}@{}}%
\column{14}{@{}>{\hspre}l<{\hspost}@{}}%
\column{E}{@{}>{\hspre}l<{\hspost}@{}}%
\>[3]{}\Varid{mxy}_{1}{}\<[9]%
\>[9]{}\ \mathop{:}\ {}\<[9E]%
\>[12]{}\Conid{M}\;(\Varid{x}\ \mathop{:}\ \Conid{State}\;\mathrm{1}\;*\!\!*\;\Conid{Ctrl}\;\mathrm{1}\;\Varid{x}){}\<[E]%
\\
\>[3]{}\Varid{mxy}_{1}{}\<[9]%
\>[9]{}\mathrel{=}{}\<[9E]%
\>[12]{}{\Varid{fmap}\!}\;\Varid{f}\;\Varid{mx}_{1}\;\mathbf{where}{}\<[E]%
\\
\>[3]{}\hsindent{2}{}\<[5]%
\>[5]{}\Varid{f}{}\<[11]%
\>[11]{}\ \mathop{:}\ {}\<[11E]%
\>[14]{}(\Varid{x}\ \mathop{:}\ \Conid{State}\;\mathrm{1})\ \to\ (\Varid{x}\ \mathop{:}\ \Conid{State}\;\mathrm{1}\;*\!\!*\;\Conid{Ctrl}\;\mathrm{1}\;\Varid{x}){}\<[E]%
\\
\>[3]{}\hsindent{2}{}\<[5]%
\>[5]{}\Varid{f}\;\Varid{x}_{1}{}\<[11]%
\>[11]{}\mathrel{=}{}\<[11E]%
\>[14]{}(\Varid{x}_{1}\;*\!\!*\;\Varid{p}_{1}\;\Varid{x}_{1}){}\<[E]%
\ColumnHook
\end{hscode}\resethooks
In general, if we have a rule \ensuremath{\Varid{p}}
\begin{hscode}\SaveRestoreHook
\column{B}{@{}>{\hspre}l<{\hspost}@{}}%
\column{3}{@{}>{\hspre}l<{\hspost}@{}}%
\column{E}{@{}>{\hspre}l<{\hspost}@{}}%
\>[3]{}\Varid{p}\ \mathop{:}\ (\Varid{t}\ \mathop{:}\ \mathbb{N})\ \to\ (\Varid{x}\ \mathop{:}\ \Conid{State}\;\Varid{t})\ \to\ \Conid{Ctrl}\;\Varid{t}\;\Varid{x}{}\<[E]%
\ColumnHook
\end{hscode}\resethooks
and an initial state \ensuremath{\Varid{x}_{0}}, we can compute lists of state-control pairs
\ensuremath{\Varid{mxy}\;\Varid{t}} for arbitrary \ensuremath{\Varid{t}}
\begin{hscode}\SaveRestoreHook
\column{B}{@{}>{\hspre}l<{\hspost}@{}}%
\column{3}{@{}>{\hspre}l<{\hspost}@{}}%
\column{5}{@{}>{\hspre}l<{\hspost}@{}}%
\column{7}{@{}>{\hspre}l<{\hspost}@{}}%
\column{13}{@{}>{\hspre}c<{\hspost}@{}}%
\column{13E}{@{}l@{}}%
\column{14}{@{}>{\hspre}l<{\hspost}@{}}%
\column{16}{@{}>{\hspre}l<{\hspost}@{}}%
\column{E}{@{}>{\hspre}l<{\hspost}@{}}%
\>[3]{}\Varid{mxy}\ \mathop{:}\ (\Varid{t}\ \mathop{:}\ \mathbb{N})\ \to\ \Conid{M}\;(\Varid{x}\ \mathop{:}\ \Conid{State}\;\Varid{t}\;*\!\!*\;\Conid{Ctrl}\;\Varid{t}\;\Varid{x}){}\<[E]%
\\
\>[3]{}\Varid{mxy}\;\Conid{Z}{}\<[14]%
\>[14]{}\mathrel{=}{\Varid{ret}}\;(\Varid{x}_{0}\;*\!\!*\;\Varid{p}\;\Conid{Z}\;\Varid{x}_{0}){}\<[E]%
\\
\>[3]{}\Varid{mxy}\;(\Conid{S}\;\Varid{t}){}\<[14]%
\>[14]{}\mathrel{=}(\Varid{mxy}\;\Varid{t})\;\bind\;\Varid{g}\;\mathbf{where}{}\<[E]%
\\
\>[3]{}\hsindent{2}{}\<[5]%
\>[5]{}\Varid{g}\ \mathop{:}\ (\Varid{x}\ \mathop{:}\ \Conid{State}\;\Varid{t}\;*\!\!*\;\Conid{Ctrl}\;\Varid{t}\;\Varid{x})\ \to\ \Conid{M}\;(\Varid{x}\ \mathop{:}\ \Conid{State}\;(\Conid{S}\;\Varid{t})\;*\!\!*\;\Conid{Ctrl}\;(\Conid{S}\;\Varid{t})\;\Varid{x}){}\<[E]%
\\
\>[3]{}\hsindent{2}{}\<[5]%
\>[5]{}\Varid{g}\;(\Varid{xt}\;*\!\!*\;\Varid{yt})\mathrel{=}{\Varid{fmap}\!}\;\Varid{f}\;(\Varid{step}\;\Varid{t}\;\Varid{xt}\;\Varid{yt})\;\mathbf{where}{}\<[E]%
\\
\>[5]{}\hsindent{2}{}\<[7]%
\>[7]{}\Varid{f}{}\<[13]%
\>[13]{}\ \mathop{:}\ {}\<[13E]%
\>[16]{}(\Varid{x}\ \mathop{:}\ \Conid{State}\;(\Conid{S}\;\Varid{t}))\ \to\ (\Varid{x}\ \mathop{:}\ \Conid{State}\;(\Conid{S}\;\Varid{t})\;*\!\!*\;\Conid{Ctrl}\;(\Conid{S}\;\Varid{t})\;\Varid{x}){}\<[E]%
\\
\>[5]{}\hsindent{2}{}\<[7]%
\>[7]{}\Varid{f}\;\Varid{xt}{}\<[13]%
\>[13]{}\mathrel{=}{}\<[13E]%
\>[16]{}(\Varid{xt}\;*\!\!*\;\Varid{p}\;(\Conid{S}\;\Varid{t})\;\Varid{xt}){}\<[E]%
\ColumnHook
\end{hscode}\resethooks
For a given \ensuremath{\Varid{t}}, \ensuremath{\Varid{mxy}\;\Varid{t}} is a list of state-control pairs. It contains
all states and controls which can be reached in \ensuremath{\Varid{t}} steps from \ensuremath{\Varid{x}_{0}} by
selecting controls according to \ensuremath{\Varid{p}\;\mathrm{0}} \dots \ensuremath{\Varid{p}\;\Varid{t}}. We can see \ensuremath{\Varid{mxy}\;\Varid{t}} as
a list-based, possibly redundant representation of a subset of the graph
of \ensuremath{\Varid{p}\;\Varid{t}}.

We can take a somewhat orthogonal view and compute, for each element in
\ensuremath{\Varid{mxy}\;\Varid{t}}, the sequence of state-control pairs of length \ensuremath{\Varid{t}} leading to
that element from \ensuremath{(\Varid{x}_{0}\;*\!\!*\;\Varid{p}_{0}\;\Varid{x}_{0})}. What we obtain is a list of
sequences. Formally:
\begin{hscode}\SaveRestoreHook
\column{B}{@{}>{\hspre}l<{\hspost}@{}}%
\column{3}{@{}>{\hspre}l<{\hspost}@{}}%
\column{5}{@{}>{\hspre}l<{\hspost}@{}}%
\column{11}{@{}>{\hspre}c<{\hspost}@{}}%
\column{11E}{@{}l@{}}%
\column{14}{@{}>{\hspre}l<{\hspost}@{}}%
\column{E}{@{}>{\hspre}l<{\hspost}@{}}%
\>[3]{}\mathbf{data}\;\Conid{StateCtrlSeq}\ \mathop{:}\ (\Varid{t}\ \mathop{:}\ \mathbb{N})\ \to\ (\Varid{n}\ \mathop{:}\ \mathbb{N})\ \to\ \Conid{Type}\;\mathbf{where}{}\<[E]%
\\
\>[3]{}\hsindent{2}{}\<[5]%
\>[5]{}\Conid{Nil}{}\<[11]%
\>[11]{}\ \mathop{:}\ {}\<[11E]%
\>[14]{}(\Varid{x}\ \mathop{:}\ \Conid{State}\;\Varid{t})\ \to\ \Conid{StateCtrlSeq}\;\Varid{t}\;\Conid{Z}{}\<[E]%
\\
\>[3]{}\hsindent{2}{}\<[5]%
\>[5]{}(\mathbin{::}){}\<[11]%
\>[11]{}\ \mathop{:}\ {}\<[11E]%
\>[14]{}(\Varid{x}\ \mathop{:}\ \Conid{State}\;\Varid{t}\;*\!\!*\;\Conid{Ctrl}\;\Varid{t}\;\Varid{x})\ \to\ \Conid{StateCtrlSeq}\;(\Conid{S}\;\Varid{t})\;\Varid{n}\ \to\ \Conid{StateCtrlSeq}\;\Varid{t}\;(\Conid{S}\;\Varid{n}){}\<[E]%
\ColumnHook
\end{hscode}\resethooks
\begin{hscode}\SaveRestoreHook
\column{B}{@{}>{\hspre}l<{\hspost}@{}}%
\column{3}{@{}>{\hspre}l<{\hspost}@{}}%
\column{5}{@{}>{\hspre}l<{\hspost}@{}}%
\column{7}{@{}>{\hspre}l<{\hspost}@{}}%
\column{10}{@{}>{\hspre}l<{\hspost}@{}}%
\column{11}{@{}>{\hspre}l<{\hspost}@{}}%
\column{17}{@{}>{\hspre}c<{\hspost}@{}}%
\column{17E}{@{}l@{}}%
\column{19}{@{}>{\hspre}l<{\hspost}@{}}%
\column{20}{@{}>{\hspre}l<{\hspost}@{}}%
\column{26}{@{}>{\hspre}l<{\hspost}@{}}%
\column{29}{@{}>{\hspre}l<{\hspost}@{}}%
\column{32}{@{}>{\hspre}l<{\hspost}@{}}%
\column{35}{@{}>{\hspre}l<{\hspost}@{}}%
\column{36}{@{}>{\hspre}l<{\hspost}@{}}%
\column{38}{@{}>{\hspre}l<{\hspost}@{}}%
\column{43}{@{}>{\hspre}l<{\hspost}@{}}%
\column{48}{@{}>{\hspre}l<{\hspost}@{}}%
\column{E}{@{}>{\hspre}l<{\hspost}@{}}%
\>[3]{}\Varid{stateCtrlTrj}{}\<[17]%
\>[17]{}\ \mathop{:}\ {}\<[17E]%
\>[20]{}(\Varid{t}\ \mathop{:}\ \mathbb{N})\ \to\ (\Varid{n}\ \mathop{:}\ \mathbb{N})\ \to\ {}\<[E]%
\\
\>[20]{}(\Varid{x}\ \mathop{:}\ \Conid{State}\;\Varid{t})\ \to\ (\Varid{r}\ \mathop{:}\ \Conid{Reachable}\;\Varid{x})\ \to\ (\Varid{v}\ \mathop{:}\ \Conid{Viable}\;\Varid{n}\;\Varid{x})\ \to\ {}\<[E]%
\\
\>[20]{}(\Varid{ps}\ \mathop{:}\ \Conid{PolicySeq}\;\Varid{t}\;\Varid{n})\ \to\ \Conid{M}\;(\Conid{StateCtrlSeq}\;\Varid{t}\;\Varid{n}){}\<[E]%
\\
\>[3]{}\Varid{stateCtrlTrj}\;\anonymous \;{}\<[19]%
\>[19]{}\Conid{Z}\;{}\<[26]%
\>[26]{}\Varid{x}\;{}\<[29]%
\>[29]{}\anonymous \;{}\<[32]%
\>[32]{}\anonymous \;{}\<[35]%
\>[35]{}\anonymous {}\<[48]%
\>[48]{}\mathrel{=}{\Varid{ret}}\;(\Conid{Nil}\;\Varid{x}){}\<[E]%
\\
\>[3]{}\Varid{stateCtrlTrj}\;\Varid{t}\;{}\<[19]%
\>[19]{}(\Conid{S}\;\Varid{n})\;{}\<[26]%
\>[26]{}\Varid{x}\;{}\<[29]%
\>[29]{}\Varid{r}\;{}\<[32]%
\>[32]{}\Varid{v}\;{}\<[35]%
\>[35]{}(\Varid{p}\mathbin{::}\Varid{ps'}){}\<[48]%
\>[48]{}\mathrel{=}{\Varid{fmap}\!}\;\Varid{prepend}\;(\Varid{toSub}\;\Varid{mx'}\;\bind\;\Varid{f})\;\mathbf{where}{}\<[E]%
\\
\>[3]{}\hsindent{2}{}\<[5]%
\>[5]{}\Varid{y}{}\<[10]%
\>[10]{}\ \mathop{:}\ \Conid{Ctrl}\;\Varid{t}\;\Varid{x};{}\<[38]%
\>[38]{}\Varid{y}{}\<[43]%
\>[43]{}\mathrel{=}\Varid{outl}\;(\Varid{p}\;\Varid{x}\;\Varid{r}\;\Varid{v}){}\<[E]%
\\
\>[3]{}\hsindent{2}{}\<[5]%
\>[5]{}\Varid{mx'}{}\<[10]%
\>[10]{}\ \mathop{:}\ \Conid{M}\;(\Conid{State}\;(\Conid{S}\;\Varid{t}));~{}\<[38]%
\>[38]{}\Varid{mx'}{}\<[43]%
\>[43]{}\mathrel{=}\Varid{step}\;\Varid{t}\;\Varid{x}\;\Varid{y}{}\<[E]%
\\
\>[3]{}\hsindent{2}{}\<[5]%
\>[5]{}\Varid{prepend}\ \mathop{:}\ \Conid{StateCtrlSeq}\;(\Conid{S}\;\Varid{t})\;\Varid{n}\ \to\ \Conid{StateCtrlSeq}\;\Varid{t}\;(\Conid{S}\;\Varid{n}){}\<[E]%
\\
\>[3]{}\hsindent{2}{}\<[5]%
\>[5]{}\Varid{prepend}\;\Varid{xys}\mathrel{=}(\Varid{x}\;*\!\!*\;\Varid{y})\mathbin{::}\Varid{xys}{}\<[E]%
\\
\>[3]{}\hsindent{2}{}\<[5]%
\>[5]{}\Varid{f}\ \mathop{:}\ (\Varid{x'}\ \mathop{:}\ \Conid{State}\;(\Conid{S}\;\Varid{t})\;*\!\!*\;\Conid{So}\;(\Varid{x'}\in\Varid{mx'}))\ \to\ \Conid{M}\;(\Conid{StateCtrlSeq}\;(\Conid{S}\;\Varid{t})\;\Varid{n}){}\<[E]%
\\
\>[3]{}\hsindent{2}{}\<[5]%
\>[5]{}\Varid{f}\;(\Varid{x'}\;*\!\!*\;\Varid{x'}\!\Varid{inmx'})\mathrel{=}\Varid{stateCtrlTrj}\;(\Conid{S}\;\Varid{t})\;\Varid{n}\;\Varid{x'}\;\Varid{r'}\;\Varid{v'}\;\Varid{ps'}\;\mathbf{where}{}\<[E]%
\\
\>[5]{}\hsindent{2}{}\<[7]%
\>[7]{}\!\!\Varid{r'}{}\<[11]%
\>[11]{}\ \mathop{:}\ \Conid{Reachable}\;\Varid{x'};~{}\<[36]%
\>[36]{}\!\Varid{r'}\mathrel{=}\Varid{reachableSpec1}\;\Varid{x}\;\Varid{r}\;\Varid{y}\;\Varid{x'}\;\Varid{x'}\!\Varid{inmx'}{}\<[E]%
\\
\>[5]{}\hsindent{2}{}\<[7]%
\>[7]{}\!\!\Varid{v'}{}\<[11]%
\>[11]{}\ \mathop{:}\ \Conid{Viable}\;\Varid{n}\;\Varid{x'};{}\<[36]%
\>[36]{}\!\Varid{v'}\mathrel{=}{\Varid{isInAreAllTrueSpec}}\;\Varid{mx'}\;(\Varid{viable}\;\Varid{n})\;(\Varid{outr}\;(\Varid{p}\;\Varid{x}\;\Varid{r}\;\Varid{v}))\;\Varid{x'}\;\Varid{x'}\!\Varid{inmx'}{}\<[E]%
\ColumnHook
\end{hscode}\resethooks
For an initial state \ensuremath{\Varid{x}\ \mathop{:}\ \Conid{State}\;\mathrm{0}} which is viable for \ensuremath{\Varid{n}} steps,
and a policy sequence \ensuremath{\Varid{ps}}, \ensuremath{\Varid{stateCtrlTrj}} provides a complete and
detailed information about all possible state-control sequences of
length \ensuremath{\Varid{n}} which can be obtained by selecting controls according to
\ensuremath{\Varid{ps}}.

For \ensuremath{\Conid{M}\mathrel{=}\Conid{List}}, this information is a list of state-control
sequences. For \ensuremath{\Conid{M}\mathrel{=}\Conid{SimpleProb}} it is a probability distribution of
sequences. In general, it is an \ensuremath{\Conid{M}}-structure of state-control sequences.

If \ensuremath{\Varid{ps}} is an optimal policy sequence, we can search \ensuremath{\Varid{stateCtrlTrj}} for
different best-case scenarios, assess their \ensuremath{\Conid{Mval}}-impacts or, perhaps identify
policies which are near optimal but easier to implement than optimal ones.

\section{Conclusions and outlook}
\label{section:conclusions_outlook}

We have presented a dependently typed, generic framework for
finite-horizon sequential decision problems.
These include problems in which the state space and the control space
can depend on the current decision step and the outcome of a step can be
a set of new states (non-deterministic SDPs) a probability distribution
of new states (stochastic SDPs) or, more generally, a monadic structure
of states.

The framework supports the specification and the solution of specific
SDPs that is, the computation of optimal
controls for an arbitrary (but finite) number of decision steps \ensuremath{\Varid{n}} and
starting from initial states which are viable for \ensuremath{\Varid{n}} steps, through
instantiation of an abstract context.

Users of the framework are expected to implement their
problem-specific context. This is done by specifying the ``bare''
problem \ensuremath{\Conid{State}}, \ensuremath{\Conid{Ctrl}}, \ensuremath{\Conid{M}}, \ensuremath{\Varid{step}}, \ensuremath{\Varid{reward}}; the basic container
monad functionalities \ensuremath{{\Varid{fmap}\!}}, \ensuremath{\Conid{MisIn}}, \ensuremath{{\Varid{areAllTrue}}}, \ensuremath{\Varid{toSub}} and
their specification \ensuremath{{\Varid{isInAreAllTrueSpec}}}; the measure \ensuremath{{\Varid{meas}}} and its
specification \ensuremath{{\Varid{measMon}}}; the viability and reachability functions
\ensuremath{\Varid{viable}} and \ensuremath{\Varid{reachable}} with their specification \ensuremath{\Varid{viableSpec0}},
\ensuremath{\Varid{viableSpec1}}, \ensuremath{\Varid{viableSpec2}}, \ensuremath{\Varid{reachableSpec0}}, \ensuremath{\Varid{reachableSpec1}},
\ensuremath{\Varid{reachableSpec2}} and the maximization functions \ensuremath{\Varid{max}} and \ensuremath{\Varid{argmax}}
with their specification \ensuremath{\Varid{maxSpec}} and \ensuremath{\Varid{argmaxSpec}} (used in the
implementation of \ensuremath{\Varid{optExtension}}).

The generic backwards induction algorithm provides them with an optimal
sequence of policies for their specific problem. The framework's design
is based on a clear cut separation between proofs and computations. Thus,
for example, \ensuremath{\Varid{backwardsInduction}} returns a bare policy sequence, not a
policy sequence paired with an optimality proof. Instead, the optimality
proof is implemented as a separate lemma
\begin{hscode}\SaveRestoreHook
\column{B}{@{}>{\hspre}l<{\hspost}@{}}%
\column{3}{@{}>{\hspre}l<{\hspost}@{}}%
\column{30}{@{}>{\hspre}l<{\hspost}@{}}%
\column{E}{@{}>{\hspre}l<{\hspost}@{}}%
\>[3]{}\Conid{BackwardsInductionLemma}\ \mathop{:}\ {}\<[30]%
\>[30]{}(\Varid{t}\ \mathop{:}\ \mathbb{N})\ \to\ (\Varid{n}\ \mathop{:}\ \mathbb{N})\ \to\ {}\<[E]%
\\
\>[30]{}\Conid{OptPolicySeq}\;\Varid{t}\;\Varid{n}\;(\Varid{backwardsInduction}\;\Varid{t}\;\Varid{n}){}\<[E]%
\ColumnHook
\end{hscode}\resethooks
This approach supports an incremental approach towards correctness: If
no guarantee of optimality is required, users do not need to implement
the full context. In this case, some of the above specifications, those
of \ensuremath{\Varid{max}} and \ensuremath{\Varid{argmax}}, for instance, do not need to be implemented.

We understand our contribution as a first step towards building a
software infrastructure for computing provably optimal policies for
general SDPs and we have made the essential components of such
infrastructure publicly available on GitHub. The repository is
\url{https://github.com/nicolabotta/SeqDecProbs} and the code for this
paper is in
\href{https://github.com/nicolabotta/SeqDecProbs/tree/master/manuscripts/2014.LMCS/code}{\texttt{tree/master/manuscripts/2014.LMCS/code}}.

This paper is part of a longer series exploring the use of dependent
types for scientific computing \cite{ionescujansson2013DTPinSciComp}
including the interplay between testing and proving
\cite{ionescujansson:LIPIcs:2013:3899}.
We have developed parts of the library code in Agda (as well as in
Idris) to explore the stronger module system and we have noticed that
several notions could benefit from using the relational algebra
framework (called AoPA) built up in \cite{MuKoJansson2009AoPA}.
Rewriting more of the code in AoPA style is future work.

\subsection{Generic tabulation}

The policies computed by backwards induction are provably optimal but
backwards induction itself is often computationally intractable. For the
cases in which \ensuremath{\Conid{State}\;\Varid{t}} is finite, the framework provides a tabulated
version of the algorithm which is linear in the number of decision steps
(not presented here but available on Github).

The tabulated version is still generic but does not come with a
machine-checkable proof of correctness. Nevertheless, users can apply the slow
but provably correct algorithm to ``small'' problems and use these results to
validate the fast version. Or they can use the tabulated version for
production code and switch back to the safe implementation for verification.

\subsection{Viability and reachability defaults}

As seen in the previous sections, in order to apply the framework to a
specific problem, a user has to implement a problem-specific viability
predicate
\begin{hscode}\SaveRestoreHook
\column{B}{@{}>{\hspre}l<{\hspost}@{}}%
\column{3}{@{}>{\hspre}l<{\hspost}@{}}%
\column{E}{@{}>{\hspre}l<{\hspost}@{}}%
\>[3]{}\Varid{viable}\ \mathop{:}\ (\Varid{n}\ \mathop{:}\ \mathbb{N})\ \to\ \Conid{State}\;\Varid{t}\ \to\ \Conid{Bool}{}\<[E]%
\ColumnHook
\end{hscode}\resethooks
Attempts at computing optimal policies of length \ensuremath{\Varid{n}} from initial
states which are not viable for at least \ensuremath{\Varid{n}} steps are then detected
by the type checker and rejected. This guarantees that no exceptions
will occur at run time. In other words: the framework will reject
problems which are not well-posed and will provide provably optimal
solutions for well-posed problems.

As seen in section \ref{subsection:viable_reachable}, for this to work,
\ensuremath{\Varid{viable}} has to be consistent with the problem specific controls \ensuremath{\Conid{Ctrl}} and
transition function \ensuremath{\Varid{step}} that is, it has to fulfill:
\begin{hscode}\SaveRestoreHook
\column{B}{@{}>{\hspre}l<{\hspost}@{}}%
\column{3}{@{}>{\hspre}l<{\hspost}@{}}%
\column{16}{@{}>{\hspre}c<{\hspost}@{}}%
\column{16E}{@{}l@{}}%
\column{19}{@{}>{\hspre}l<{\hspost}@{}}%
\column{E}{@{}>{\hspre}l<{\hspost}@{}}%
\>[3]{}\Varid{viableSpec0}{}\<[16]%
\>[16]{}\ \mathop{:}\ {}\<[16E]%
\>[19]{}(\Varid{x}\ \mathop{:}\ \Conid{State}\;\Varid{t})\ \to\ \Conid{Viable}\;\Conid{Z}\;\Varid{x}{}\<[E]%
\\
\>[3]{}\Varid{viableSpec1}{}\<[16]%
\>[16]{}\ \mathop{:}\ {}\<[16E]%
\>[19]{}(\Varid{x}\ \mathop{:}\ \Conid{State}\;\Varid{t})\ \to\ \Conid{Viable}\;(\Conid{S}\;\Varid{n})\;\Varid{x}\ \to\ \Conid{GoodCtrl}\;\Varid{t}\;\Varid{n}\;\Varid{x}{}\<[E]%
\\
\>[3]{}\Varid{viableSpec2}{}\<[16]%
\>[16]{}\ \mathop{:}\ {}\<[16E]%
\>[19]{}(\Varid{x}\ \mathop{:}\ \Conid{State}\;\Varid{t})\ \to\ \Conid{GoodCtrl}\;\Varid{t}\;\Varid{n}\;\Varid{x}\ \to\ \Conid{Viable}\;(\Conid{S}\;\Varid{n})\;\Varid{x}{}\<[E]%
\ColumnHook
\end{hscode}\resethooks
where
\begin{hscode}\SaveRestoreHook
\column{B}{@{}>{\hspre}l<{\hspost}@{}}%
\column{3}{@{}>{\hspre}l<{\hspost}@{}}%
\column{E}{@{}>{\hspre}l<{\hspost}@{}}%
\>[3]{}\Conid{GoodCtrl}\ \mathop{:}\ (\Varid{t}\ \mathop{:}\ \mathbb{N})\ \to\ (\Varid{n}\ \mathop{:}\ \mathbb{N})\ \to\ \Conid{State}\;\Varid{t}\ \to\ \Conid{Type}{}\<[E]%
\\
\>[3]{}\Conid{GoodCtrl}\;\Varid{t}\;\Varid{n}\;\Varid{x}\mathrel{=}(\Varid{y}\ \mathop{:}\ \Conid{Ctrl}\;\Varid{t}\;\Varid{x}\;*\!\!*\;{\Conid{Feasible}}\;\Varid{n}\;\Varid{x}\;\Varid{y}){}\<[E]%
\ColumnHook
\end{hscode}\resethooks
and \ensuremath{{\Conid{Feasible}}\;\Varid{n}\;\Varid{x}\;\Varid{y}} is a shorthand for \ensuremath{\Conid{So}\;({\Varid{feasible}}\;\Varid{n}\;\Varid{x}\;\Varid{y})}:
\begin{hscode}\SaveRestoreHook
\column{B}{@{}>{\hspre}l<{\hspost}@{}}%
\column{3}{@{}>{\hspre}l<{\hspost}@{}}%
\column{16}{@{}>{\hspre}c<{\hspost}@{}}%
\column{16E}{@{}l@{}}%
\column{19}{@{}>{\hspre}l<{\hspost}@{}}%
\column{E}{@{}>{\hspre}l<{\hspost}@{}}%
\>[3]{}{\Varid{feasible}}{}\<[16]%
\>[16]{}\ \mathop{:}\ {}\<[16E]%
\>[19]{}(\Varid{n}\ \mathop{:}\ \mathbb{N})\ \to\ (\Varid{x}\ \mathop{:}\ \Conid{State}\;\Varid{t})\ \to\ \Conid{Ctrl}\;\Varid{t}\;\Varid{x}\ \to\ \Conid{Bool}{}\<[E]%
\\
\>[3]{}{\Varid{feasible}}\;\{\mskip1.5mu \Varid{t}\mskip1.5mu\}\;\Varid{n}\;\Varid{x}\;\Varid{y}\mathrel{=}{\Varid{areAllTrue}}\;({\Varid{fmap}\!}\;(\Varid{viable}\;\Varid{n})\;(\Varid{step}\;\Varid{t}\;\Varid{x}\;\Varid{y})){}\<[E]%
\ColumnHook
\end{hscode}\resethooks
Again, users are responsible for implementing or (if they feel confident
to do so) postulating the specification.

For problems in which the control state \ensuremath{\Conid{Ctrl}\;\Varid{t}\;\Varid{x}} is finite for every \ensuremath{\Varid{t}}
and \ensuremath{\Varid{x}}, the framework provides a default implementation of
\ensuremath{\Varid{viable}}. This is based on the notion of successors:
\begin{hscode}\SaveRestoreHook
\column{B}{@{}>{\hspre}l<{\hspost}@{}}%
\column{3}{@{}>{\hspre}l<{\hspost}@{}}%
\column{15}{@{}>{\hspre}c<{\hspost}@{}}%
\column{15E}{@{}l@{}}%
\column{18}{@{}>{\hspre}l<{\hspost}@{}}%
\column{E}{@{}>{\hspre}l<{\hspost}@{}}%
\>[3]{}\Varid{succs}{}\<[15]%
\>[15]{}\ \mathop{:}\ {}\<[15E]%
\>[18]{}\Conid{State}\;\Varid{t}\ \to\ (\Varid{n}\ \mathop{:}\ \mathbb{N}\;*\!\!*\;\Conid{Vect}\;\Varid{n}\;(\Conid{M}\;(\Conid{State}\;(\Conid{S}\;\Varid{t})))){}\<[E]%
\\
\>[3]{}\Varid{succsSpec1}{}\<[15]%
\>[15]{}\ \mathop{:}\ {}\<[15E]%
\>[18]{}(\Varid{x}\ \mathop{:}\ \Conid{State}\;\Varid{t})\ \to\ (\Varid{y}\ \mathop{:}\ \Conid{Ctrl}\;\Varid{t}\;\Varid{x})\ \to\ \Conid{So}\;((\Varid{step}\;\Varid{t}\;\Varid{x}\;\Varid{y})\mathbin{`\Varid{isIn}`}(\Varid{succs}\;\Varid{x})){}\<[E]%
\\
\>[3]{}\Varid{succsSpec2}{}\<[15]%
\>[15]{}\ \mathop{:}\ {}\<[15E]%
\>[18]{}(\Varid{x}\ \mathop{:}\ \Conid{State}\;\Varid{t})\ \to\ (\Varid{mx'}\ \mathop{:}\ \Conid{M}\;(\Conid{State}\;(\Conid{S}\;\Varid{t})))\ \to\ {}\<[E]%
\\
\>[18]{}\Conid{So}\;(\Varid{mx'}\mathbin{`\Varid{isIn}`}(\Varid{succs}\;\Varid{x}))\ \to\ (\Varid{y}\ \mathop{:}\ \Conid{Ctrl}\;\Varid{t}\;\Varid{x}\;*\!\!*\;\Varid{mx'}\mathrel{=}\Varid{step}\;\Varid{t}\;\Varid{x}\;\Varid{y}){}\<[E]%
\ColumnHook
\end{hscode}\resethooks
Users can still provide their own implementation of \ensuremath{\Varid{viable}}.
Alternatively, they can implement \ensuremath{\Varid{succs}} (and \ensuremath{\Varid{succsSpec1}},
\ensuremath{\Varid{succsSpec2}}) and rely on the default implementation of \ensuremath{\Varid{viable}}
provided by the framework. This is
\begin{hscode}\SaveRestoreHook
\column{B}{@{}>{\hspre}l<{\hspost}@{}}%
\column{3}{@{}>{\hspre}l<{\hspost}@{}}%
\column{17}{@{}>{\hspre}l<{\hspost}@{}}%
\column{20}{@{}>{\hspre}c<{\hspost}@{}}%
\column{20E}{@{}l@{}}%
\column{23}{@{}>{\hspre}l<{\hspost}@{}}%
\column{E}{@{}>{\hspre}l<{\hspost}@{}}%
\>[3]{}\Varid{viable}\;\Conid{Z}\;{}\<[17]%
\>[17]{}\anonymous {}\<[20]%
\>[20]{}\mathrel{=}{}\<[20E]%
\>[23]{}\Conid{True}{}\<[E]%
\\
\>[3]{}\Varid{viable}\;(\Conid{S}\;\Varid{n})\;{}\<[17]%
\>[17]{}\Varid{x}{}\<[20]%
\>[20]{}\mathrel{=}{}\<[20E]%
\>[23]{}\Varid{isAnyBy}\;(\lambda \Varid{mx}\Rightarrow {\Varid{areAllTrue}}\;({\Varid{fmap}\!}\;(\Varid{viable}\;\Varid{n})\;\Varid{mx}))\;(\Varid{succs}\;\Varid{x}){}\<[E]%
\ColumnHook
\end{hscode}\resethooks
and can be shown to fulfill \ensuremath{\Varid{viableSpec0}}, \ensuremath{\Varid{viableSpec1}} and
\ensuremath{\Varid{viableSpec2}}. In a similar way, the framework supports the
implementation of \ensuremath{\Varid{reachable}} with a default based on the notion of
predecessor.

\subsection{Outlook}

In developing the framework presented in this paper, we have
implemented a number of variations of the ``cylinder'' problem
discussed in sections \ref{section:base_case} and
\ref{section:time-dependent} and a simple ``Knapsack'' problem
(see
\href{https://github.com/nicolabotta/SeqDecProbs/blob/master/manuscripts/2014.LMCS/code/DynamicProgramming/S1206\_CylinderExample1.lidr}{\texttt{CylinderExample1}},
\href{https://github.com/nicolabotta/SeqDecProbs/blob/master/manuscripts/2014.LMCS/code/DynamicProgramming/S1206\_CylinderExample4.lidr}{\texttt{CylinderExample4}}
and
\href{https://github.com/nicolabotta/SeqDecProbs/blob/master/manuscripts/2014.LMCS/code/DynamicProgramming/S1106\_KnapsackExample.lidr}{\texttt{KnapsackExample}}
in the GitHub repository).
Since Feb. 2016, we have implemented 5 new examples of computations of
optimal policy sequences for sequential decision problems. These are
published in an extended framework, see
\texttt{frameworks/14-/SeqDecProbsExample1-5}, also in the GitHub repository.

A look at the dependencies of the examples shows that a lot of
infrastructure had to be built in order to specify and solve even these
toy problems. In fact, the theory presented here is rather succinct and
most of the libraries of our framework have been developed in order to
implement examples.

This is not completely surprising: dependently types languages -- Idris
in particular -- are in their infancy. They lack even the most
elementary verified libraries. Thus, for instance, in order to specify
\texttt{SeqDecProbsExample5}, we had to provide, among others, our own implementation of
non-negative rational numbers, of finite types, of verified filtering
operations and of bounded natural numbers.

Of course, dissecting one of the examples would not require a detailed
understanding of all its dependencies. But it would require a careful
discussion of notions (e.g., of finiteness of propositional data types)
that would go well beyond the scope of our contribution. Thus, we plan a
follow-up article focused on application examples. In particular, we
want to apply the framework to study optimal emission policies in a
competitive multi-agent game under threshold uncertainty in the context
of climate impact research. This is the application domain that has
motivated the development of the framework in the very beginning.

The work presented here naturally raises a number of questions. A first
one is related with the notion of reward function
\begin{hscode}\SaveRestoreHook
\column{B}{@{}>{\hspre}l<{\hspost}@{}}%
\column{3}{@{}>{\hspre}l<{\hspost}@{}}%
\column{E}{@{}>{\hspre}l<{\hspost}@{}}%
\>[3]{}\Varid{reward}\ \mathop{:}\ (\Varid{t}\ \mathop{:}\ \mathbb{N})\ \to\ (\Varid{x}\ \mathop{:}\ \Conid{State}\;\Varid{t})\ \to\ \Conid{Ctrl}\;\Varid{t}\;\Varid{x}\ \to\ \Conid{State}\;(\Conid{S}\;\Varid{t})\ \to\ \mathbb{R}{}\<[E]%
\ColumnHook
\end{hscode}\resethooks
As mentioned in our previous paper~\cite{botta+al2013c}, we have taken
\ensuremath{\Varid{reward}} to return values of type \ensuremath{\mathbb{R}} but it is clear that this assumption
can be weakened. A natural question here is what specification the
return type of \ensuremath{\Varid{reward}} has to fulfill for the framework to be implementable.

A second question is directly related with the notion of viability
discussed above.  According to this notion, a necessary (and sufficient)
condition for a state \ensuremath{\Varid{x}\ \mathop{:}\ \Conid{State}\;\Varid{t}} to be viable \ensuremath{\Conid{S}\;\Varid{n}} steps is that there
exists a control \ensuremath{\Varid{y}\ \mathop{:}\ \Conid{Ctrl}\;\Varid{t}\;\Varid{x}} such that all states in \ensuremath{\Varid{step}\;\Varid{t}\;\Varid{x}\;\Varid{y}} are
viable \ensuremath{\Varid{n}} steps.

Remember that \ensuremath{\Varid{step}\;\Varid{t}\;\Varid{x}\;\Varid{y}} is an M-structure of values of type \ensuremath{\Conid{State}\;(\Conid{S}\;\Varid{t})}. In the stochastic cases, \ensuremath{\Varid{step}\;\Varid{t}\;\Varid{x}\;\Varid{y}} is a probability
distribution. Its support is the set of all states that can be reached
from \ensuremath{\Varid{x}} with non-zero probability by selecting \ensuremath{\Varid{y}}. Our notion of
viability requires all such states to be viable \ensuremath{\Varid{n}} steps no matter how
small their probabilities might actually be. It is clear that under
our notion of viability small perturbations of a perfectly
deterministic transition function can easily turn a well-posed problem
into an ill-posed one. The question here is whether there is a natural
way of weakening the viability notion that allows one to preserve
well-posedness in the limit for vanishing probabilities of non-viable
states.

Another question comes from the notion of aggregation measure introduced
in section \ref{subsection:agg}. As mentioned there, in the stochastic
case \ensuremath{\Varid{meas}} is often taken to be the expected value function. Can we
construct other suitable aggregation measures?  What is their impact on
optimal policy selection?

Finally, the formalization presented here has been implemented on the
top of our own extensions of the Idris standard library. Beside
application (framework) specific software components --- e.g., for
implementing the context of SDPs or tabulated
versions of backwards induction --- we have implemented data structures
to represent bounded natural numbers, finite probability distributions,
and setoids.
We also have implemented a number of operations on data structures of
the standard library, e.g., point-wise modifiers for functions and
vectors and filter operations with guaranteed non-empty output.

From a software engineering perspective, an interesting question is
how to organize such extensions in a software layer which is
independent of specific applications (in our case the components that
implement the framework for SDPs) but still not
part of the standard library.
To answer this question we certainly need a better understanding of the
scope of different constructs for structuring programs: modules,
parameter blocks, records and type classes.

For instance it is clear that from our viewpoint -- that of the
developers of the framework -- the specifications \ensuremath{(\in)},
\ensuremath{{\Varid{isInAreAllTrueSpec}}} of section \ref{section:monadic}
demand a more polymorphic formulation. On the other hand, these
functions are specifications: in order to apply the framework, users
have to implement them. How can we avoid over-specification while at the
same time minimizing the requirements we put on the users?

Tackling such questions would obviously go beyond the scope of this
paper and must be deferred to future work.


\vspace*{5mm}
\noindent
\thanks{%
  \textbf{Acknowledments:}
The authors thank the reviewers, whose comments have lead to significant
improvements of the original manuscript.  The work presented in this
paper heavily relies on free software, among others on hugs, GHC, vi,
the GCC compiler, Emacs, \LaTeX\ and on the FreeBSD and Debian GNU/Linux
operating systems. It is our pleasure to thank all developers of these
excellent products.  }

\bibliographystyle{alpha}
\bibliography{references,dtp}


\end{document}